\documentclass[12pt,letterpaper]{article}

\usepackage{epsfig}

\begin{document}


\begin{center}
{\LARGE\bf Logarithmic link smearing for full QCD}
\end{center}
\vspace{4pt}

\begin{center}
{\large\bf Stephan D\"urr} 
\\[4pt]
{\sl Universit\"at Bern, Institut f\"ur theoretische Physik,
Sidlerstr.\,5, CH-3012 Bern, Switzerland}\\
\end{center}
\vspace{4pt}

\begin{abstract}
\noindent
A Lie-algebra based recipe for smoothing gauge links in lattice field
theory is presented, building on the matrix logarithm.
With or without hypercubic nesting, this LOG/HYL smearing yields fat
links which are differentiable w.r.t.\ the original ones.
This is essential for defining UV-filtered (``fat link'') fermion actions which
may be simulated with a HMC-type algorithm.
The effect of this smearing on the distribution of plaquettes and on the
residual mass of tree-level $O(a)$-improved clover fermions in quenched QCD is
studied.
\end{abstract}
\vspace{4pt}


\newcommand{\pad}{\partial}
\newcommand{\psl}{\partial\!\!\!/}
\newcommand{\hqu}{\hbar}
\newcommand{\ovr}{\over}
\newcommand{\til}{\tilde}
\newcommand{\pri}{^\prime}
\renewcommand{\dag}{^\dagger}
\newcommand{\<}{\langle}
\renewcommand{\>}{\rangle}
\newcommand{\gaf}{\gamma_5}
\newcommand{\lap}{\triangle}
\newcommand{\dal}{{\sqcap\!\!\!\!\sqcup}}
\newcommand{\trc}{\mathrm{tr}}
\newcommand{\Mpi}{M_\pi}
\newcommand{\Fpi}{F_\pi}

\newcommand{\al}{\alpha}
\newcommand{\be}{\beta}
\newcommand{\ga}{\gamma}
\newcommand{\de}{\delta}
\newcommand{\ep}{\epsilon}
\newcommand{\ve}{\varepsilon}
\newcommand{\ze}{\zeta}
\newcommand{\et}{\eta}
\renewcommand{\th}{\theta}
\newcommand{\vt}{\vartheta}
\newcommand{\io}{\iota}
\newcommand{\ka}{\kappa}
\newcommand{\la}{\lambda}
\newcommand{\rh}{\rho}
\newcommand{\vr}{\varrho}
\newcommand{\si}{\sigma}
\newcommand{\ta}{\tau}
\newcommand{\ph}{\phi}
\newcommand{\vp}{\varphi}
\newcommand{\ch}{\chi}
\newcommand{\ps}{\psi}
\newcommand{\om}{\omega}

\newcommand{\psb}{\overline{\psi}}
\newcommand{\etb}{\overline{\eta}}
\newcommand{\psd}{\psi^{\dagger}}
\newcommand{\etd}{\eta^{\dagger}}
\newcommand{\qh}{{\hat q}}
\newcommand{\kh}{{\hat k}}

\newcommand{\bdm}{\begin{displaymath}}
\newcommand{\edm}{\end{displaymath}}
\newcommand{\bea}{\begin{eqnarray}}
\newcommand{\eea}{\end{eqnarray}}
\newcommand{\beq}{\begin{equation}}
\newcommand{\eeq}{\end{equation}}

\newcommand{\mr}{\mathrm}
\newcommand{\mb}{\mathbf}
\newcommand{\Nf}{{N_{\!f}}}
\newcommand{\Nc}{{N_{\!c}}}
\newcommand{\ri}{\mr{i}}
\newcommand{\DW}{D_\mr{W}}
\newcommand{\DN}{D_\mr{N}}
\newcommand{\MeV}{\,\mr{MeV}}
\newcommand{\GeV}{\,\mr{GeV}}
\newcommand{\fm}{\,\mr{fm}}
\newcommand{\MSB}{\overline{\mr{MS}}}

\hyphenation{topo-lo-gi-cal simu-la-tion theo-re-ti-cal mini-mum con-tinu-um}


\section{Introduction}


In lattice field theory one is generally interested in taking the continuum
limit where the lattice spacing $a$ tends to zero.
In other words, one calculates the ratio $R(a)$ of two physical masses or
matrix elements and considers the limit where the correlation length in lattice
units diverges, $\xi/a\to\infty$.
Since the physical box volume is supposed to be roughly constant, the total
number of variables (and hence the CPU time needed) grows with a large power
of $a^{-1}$.

To speed up the computation of the continuum ratio two concepts have proven
important.
The first one is known as Symanzik improvement \cite{Symanzik:1983dc}.
Here one augments both the action and the observable by irrelevant terms.
Upon tuning some coefficients it may be achieved that the continuum value is
approached at a quadratic rate, $R(a)=R(0)+\mr{const}\,(a/r_0)^2+O(a^3)$,
rather than linearly, $R(a)=R(0)+\mr{const}\,a/r_0+O(a^2)$, where $r_0$ is a
fixed length.
Hence, if the Symanzik scaling window (the regime where the ``const'' term
dominates) starts at $a\!\simeq\!0.1\fm$ and one wishes to cover three quarters
of the variable in which one extrapolates linearly, the finest lattice will
have $a\simeq0.05\fm$ with improvement versus $a\simeq0.025\fm$ without.

The second ingredient in todays state-of-the-art simulations is some damping of
ultra-violet (UV) fluctuations, that is of unphysical excitations at the scale
of the cut-off $a^{-1}$.
To maintain locality, such damping is achieved via locally averaging field
variables.
In case of a gauge theory one replaces the link $U_\mu(x)$, the
parallel transporter from $x\!+\!\hat\mu$ to $x$, by $U_\mu'(x)$.
The latter is some average of paths, attached to the same endpoints, which stay
within a local neighborhood of $x$ and $x\!+\!\hat\mu$.
From a formal viewpoint such smearing amounts to another change of the action
by irrelevant terms; thus changing ``const'' in the extrapolation law, but not
the power (i.e.\ the Symanzik class is unchanged).
Data suggest that some mild smearing enlarges the scaling window and reduces
the ``const'' (in absolute magnitude) in the extrapolation law
\cite{pioneers,Durr:2005ik}.

Specifically for lattice QCD it has been observed that these two strategies,
when applied together, enhance each others effectiveness.
With staggered quarks this has been demonstrated in a somewhat indirect manner.
Upon combining a Naik term with a specific UV-filtering known as ``AsqTad''
smearing the MILC collaboration has been able to simulate large volumes with
standard lattice spacings and fairly light pion masses \cite{Bernard:2006wx}.
For fat-link clover quarks \cite{pioneers} the enhancement has been
investigated in detail in the quenched approximation \cite{CDH}, and there is a
similar program for large-scale dynamical simulations with almost-realistic
quark masses \cite{Durr:2007ef}.

What remains is an engineering issue.
With the smearing being part of the action or operator definition, it seems
natural to keep the parameter $\al$ and the iteration level $n_\mr{iter}$
unchanged as the lattice spacing is reduced.
This way locality is guaranteed, and $(\al,n_\mr{iter})$ represents a handle to
influence the overall cost, while the ratio $R(0)$ is, ultimately, unchanged.
Of course one may ask what is the ``optimal'' choice of $(\al,n_\mr{iter})$, in
terms of CPU time, to achieve a predefined accuracy of $R(0)$.
In practice, however, one is interested in the continuum limit for a number of
quantities, and often this list enlarges as the simulation program progresses.
Thus, staying away from either extreme choice (i.e.\ no or excessive smearing)
will often be sufficient.

For theories with dynamical fermions an additional engineering constraint
emerges.
There is a single algorithm which scales, at fixed bare parameters, almost
linearly with the box volume, known as hybrid Monte Carlo algorithm (HMC), with
subvarieties called PHMC, RHMC \cite{Clark:2006wq}.
This algorithm demands that the fermion action reacts smoothly to a change of
the field of gauge variables $U_\mu(x)$.
For fat-link actions this requires that the smeared link $U_\mu'(x)$ be
differentiable with respect to $U_\mu(x)$, which amounts to a constraint on
the smearing recipe.

Historically, the first smoothing introduced has been APE smearing \cite{APE}
\bea
U_\mu^\mr{APE}(x)&=&P_{SU(3)}
\Big\{
(1-\al)U_\mu(x)+{\al\ovr2(d\!-\!1)}\sum_{\pm\nu\neq\mu}
U_\nu(x)U_\mu(x\!+\!\hat\nu)U_\nu\dag(x\!+\!\hat\mu)
\Big\}
\\
&=&P_{SU(3)}
\Big\{
(1-\al)I+{\al\ovr2(d\!-\!1)}\sum_{\pm\nu\neq\mu}
U_\nu(x)U_\mu(x\!+\!\hat\nu)U_\nu\dag(x\!+\!\hat\mu)U_\mu\dag(x)
\Big\}\,
U_\mu(x)
\label{def_APE}
\eea
with $P_G$ denoting the projection to the gauge group $G$.
The rewriting in the second line is based on $P_{SU(3)}\{AU\}=P_{SU(3)}\{A\}U$,
valid for the combination of an arbitrary matrix $A$ and a special unitary $U$.
Since the $U(1)$ projection [cf.\ (\ref{proj1}, \ref{proj2}) below for details]
creates a headache in the HMC force, Morningstar and Peardon invented the
``stout'' smearing
\cite{EXP}
\beq
U_\mu^\mr{EXP}(x)=
\exp\Big(
{\al\ovr2}
\sum_{\pm\nu\neq\mu}
\Big\{
[U_\nu(x)U_\mu(x\!+\!\hat\nu)U_\nu\dag(x\!+\!\hat\mu)U_\mu\dag(x)-\mr{h.c.}]
-{1\ovr3}\mr{Tr}[.]
\Big\}
\Big)\,
U_\mu(x)
\label{def_EXP}
\eeq
subsequently dubbed EXP, which, by design, yields differentiable fat links.
However, it turns out that (\ref{def_EXP}) is less effective in damping
extremal plaquettes than (\ref{def_APE}).
In Ref.\ \cite{n-APE} a modified nAPE smearing has been introduced where the
projection is to $U(3)$ only.
The goal of this article is to test yet another smearing which yields an
$SU(3)$ valued differentiable link, with an efficient damping of the UV
fluctuations, such that it could be used in full QCD.

This article is organized as follows.
The next two sections specify the new LOG smearing and show how it can be
used together with the hypercubic nesting trick, defining the HYL smearing.
Sections 4 and 5 investigate the impact on selected observables, in particular
the distribution of plaquettes and the residual mass of fat-clover fermions.
In the latter case evidence is given that $\Mpi\simeq160\MeV$ can be reached at
standard $\be$-values in $O(a)$-improved quenched QCD without hitting the
so-called ``exceptional configuration'' problem.
Sections 6 and 7 sketch how the LOG/HYL smearing is included in a HMC approach
to full QCD and how the matrix logarithm may be used to define a new gauge
action and the pertinent (gluonic) topological charge density.
After a summary, three appendices give technical details.


\section{Logarithmic link smearing}


The re-written form (\ref{def_APE}) of the APE smearing contains the factor
$P_{SU(3)}\{.\}$ by which the original link $U_\mu(x)$ is multiplied.
Upon sending the parameter $\al\!\to\!0$ this prefactor is smoothly deformed
into unity.
If one wishes to stay in the group, the backprojection is needed, since the
weighted arithmetic average of the identity and six (in 4D) ``blades'' (closed
staples) is not a group element any more.
Given the standard definition of the fractional power of a matrix
\beq
U^\al=\exp(\al\log(U))
\label{frac_pow}
\eeq
one starts wondering whether it would make sense to define the smeared link as
\beq
U_\mu'(x)=
\exp\Big(
{\al\ovr2(d\!-\!1)}\sum_{\pm\nu\neq\mu}
\log[U_\nu(x)U_\mu(x\!+\!\hat\nu)U_\nu\dag(x\!+\!\hat\mu)U_\mu\dag(x)]
\Big)\,
U_\mu(x)
\label{def_1}
\eeq
and on a sufficiently smooth gauge background the rationale for this choice is
as follows.
The product $U_\nu(x)U_\mu(x\!+\!\hat\nu)U_\nu\dag(x\!+\!\hat\mu)U_\mu\dag(x)$
is special unitary, its logarithm thus anti-hermitean and traceless.
So is the sum, and this means that the exponential defines again a special
unitary matrix which is to be multiplied onto the original link.
Hence one stays in $SU(3)$.

The problem is that the argument has a loophole: On an arbitrary gauge
configuration the logarithm is \emph{not} traceless, it is just traceless
modulo $2\pi\ri$.
Upon averaging this logarithm with the remaining five (in 4D) contributions,
which we assume to be traceless, one gets an arbitrary trace.
As a result, the exponential is still unitary, but its determinant is not 1,
so one leaves the gauge group.
While, even on a rough configuration,
$\mr{Tr}\log[U_\nu(x)U_\mu(x\!+\!\hat\nu)U_\nu\dag(x\!+\!\hat\mu)U_\mu\dag(x)]
\neq0$ is rare, if one wishes to stay in $SU(3)$, this possibility needs to be
accounted for.

Obviously, by restricting the sum to traceless contributions, the goal of
staying in the gauge group is reached.
The most straightforward option is to define the smearing through
\beq
U_\mu''(x)=
\exp\Big(
{\al\ovr2(d\!-\!1)}\sum_{\pm\nu\neq\mu}
\Big\{
\log[U_\nu(x)U_\mu(x\!+\!\hat\nu)U_\nu\dag(x\!+\!\hat\mu)U_\mu\dag(x)]
-{1\ovr3}\mr{Tr}\log[.]
\Big\}
\Big)\,
U_\mu(x)
\label{def_2}
\eeq
where, as mentioned before, the new term $-{1\ovr3}\mr{Tr}\log[.]$ is almost
always zero.
In tangent space the logarithm of the 4-link product may be decomposed as
$\log[.]=\sum_{a=1}^8\ri\xi^aT^a+\ri\xi^9I$ with $\xi^a\!\in\!\mb{R}$ for
$a\!=\!1,...,8$ and $\xi^9\!\in\!\{0,\pm2\pi/3\}$.
Here, $T^a\!=\!\la^a/2$ with $\la^a$ the Gell-Mann matrices.
In (\ref{def_2}) the 9th (radial) component is simply projected away, and
after this correction, the six contributions to the tangent space are again
subject to an arithmetic average.

A more general approach may allow for unequal weights of the six staples,
depending on whether their logarithm is traceless or not.
Hence, a more sophisticated version is
\beq
U_\mu'''(x)=
\exp\Big(
{\al\ovr2(d\!-\!1)}\sum_{\pm\nu\neq\mu}c_{\pm\nu}
\Big\{
\log[U_\nu(x)U_\mu(x\!+\!\hat\nu)U_\nu\dag(x\!+\!\hat\mu)U_\mu\dag(x)]
-{1\ovr3}\mr{Tr}\log[.]
\Big\}
\Big)\,
U_\mu(x)
\label{def_3}
\eeq
where the coefficient $c_{\pm\nu}$ is a function of
$\log[U_\nu(x)U_\mu(x\!+\!\hat\nu)U_\nu\dag(x\!+\!\hat\mu)U_\mu\dag(x)]$,
for instance $c_\nu=+1$ if $\mr{Tr}\log[.]=0$ and $c_\nu=-0.5$ if
$\mr{Tr}\log[.]=\pm2\pi\ri$.
Still, in view of the application as an ingredient in a HMC, this definition
seems less attractive, since it entails a rather complicated force.


Finally, one may choose to invoke a non-principal logarithm.
In fact, thinking in terms of the eigenvalues $\la_i$ of the original matrix
(which we assume to be special unitary) it is natural to attribute a possible
occurrence of $\mr{Tr}\log[.]=+2\pi\ri$ to the $\la_i$ with the largest
imaginary part of $\log(\la_i)$ (and ditto to the one with the smallest
imaginary part of $\log(\la_i)$ for $\mr{Tr}\log[.]=-2\pi\ri$).
Accordingly, a ``trace-free'' matrix logarithm may be defined which basically
shifts the logarithm of one eigenvalue of the original matrix by $\pm2\pi\ri$
in those cases where the principal logarithm has non-zero trace.
With this function at hand, we define
\beq
U_\mu^\mr{LOG}(x)=
\exp\Big(
{\al\ovr2(d\!-\!1)}\sum_{\pm\nu\neq\mu}
\mr{tf\/log}[U_\nu(x)U_\mu(x\!+\!\hat\nu)U_\nu\dag(x\!+\!\hat\mu)U_\mu\dag(x)]
\Big)\,
U_\mu(x)
\label{def_4}
\eeq
and below, whenever referring to the LOG recipe without specification, the
recipe (\ref{def_4}) will be meant.
This ``trace-free'' logarithm seems most interesting, because in full QCD this
version amounts to a smooth adaptation of the definition to the actual gauge
field, which has a good chance to avoid large HMC forces.
For implementation details of all varieties see App.\,A.

The construct (\ref{def_4}) retains all symmetry properties of the original
link, in particular the behavior under gauge transformations, charge
conjugation, reflections, and permutations of the coordinate axes.
It produces a link in $SU(3)$, with an obvious generalization to $SU(\Nc)$, and
the new $U_\mu^\mr{LOG}(x)$ is differentiable with respect to the original
$U_\mu(x)$.
The normalization of the parameter $\al$, which determines the weight of the
fluctuation, has been chosen such that in leading order perturbation theory the
new LOG recipe (\ref{def_4}) agrees with the APE recipe (\ref{def_APE}).
The same holds true for the stout/EXP recipe (\ref{def_EXP}) if an extra
factor $1/(2(d\!-\!1))$ is included \cite{CDH}.


\section{Hypercubic nesting trick}


To tame the noise in QCD observables, one would like to iterate the smearing,
while keeping the delocalizing effect minimal.
A nice strategy (inspired by the fixed-point action approach and the pertinent
``perfect smearing'') was presented in \cite{HYP}.
In this original form the hypercubic nesting trick uses the APE smearing
(\ref{def_APE}) as core recipe, giving
\bea
\bar V_{\mu,\nu\rh}(x)&=&P_{SU(3)}
\Big\{\,
(1\!-\!\al_3)I+{\al_3\ovr2}\sum_{\pm\si\neq\mu,\nu,\rh}
U_{\si}(x)\,
U_{\mu}(x\!+\!\hat\si)\,
U_{\si}\dag(x\!+\!\hat\mu)\,
U_\mu\dag(x)\,
\Big\}\,U_\mu(x)
\nonumber\\
\til V_{\mu,\nu}(x)&=&P_{SU(3)}
\Big\{\,
(1\!-\!\al_2)I+{\al_2\ovr4}\;\sum_{\pm\rh\neq\mu,\nu}
\bar V_{\rh,\mu\nu}(x)\,
\bar V_{\mu,\nu\rh}(x\!+\!\hat\rh)\,
\bar V_{\rh,\mu\nu}\dag(x\!+\!\hat\mu)\,
U_\mu\dag(x)\,
\Big\}\,U_\mu(x)
\nonumber\\
U_\mu^\mr{HYP}(x)&=&P_{SU(3)}
\Big\{\,
(1\!-\!\al_1)I+{\al_1\ovr6}\;\;\sum_{\pm\nu\neq\mu}
\til V_{\nu,\mu}(x)\,
\til V_{\mu,\nu}(x\!+\!\hat\nu)\,
\til V_{\nu,\mu}\dag(x\!+\!\hat\mu)\,
U_\mu\dag(x)\,
\Big\}\,U_\mu(x)
\label{def_HYP}
\eea
where we stick to the notation of \cite{HYP} in which $\al_1$ denotes the
fluctuation weight in the \emph{last} step.
This has been generalized to the case of ``stout/EXP'' smearing \cite{CDH}
and in complete analogy
\bea
\bar V_{\mu,\nu\rh} (x)\!&\!=\!&\!\exp
\Big({\al_3\ovr2}\sum_{\pm\si\neq\mu,\nu,\rh}\mr{tf\/log}\big[
U_{\si}(x)\,
U_{\mu}(x\!+\!\hat\si)\,
U_{\si}\dag(x\!+\!\hat\mu)\,
U_\mu\dag(x)\,\big]
\Big)U_\mu(x)
\nonumber\\
\til V_{\mu,\nu} (x)\!&\!=\!&\!\exp
\Big({\al_2\ovr4}\sum_{\pm\rh\neq\mu,\nu}\mr{tf\/log}\big[
\bar V_{\rh,\mu\nu}(x)\,
\bar V_{\mu,\nu\rh}(x\!+\!\hat\rh)\,
\bar V_{\rh,\mu\nu}\dag(x\!+\!\hat\mu)\,
U_\mu\dag(x)\,\big]
\Big)U_\mu(x)
\nonumber\\
U_\mu^\mr{HYL}(x)\!&\!=\!&\!\exp
\Big({\al_1\ovr6}\sum_{\pm\nu\neq\mu}\mr{tf\/log}\big[
\til V_{\nu,\mu}(x)\,
\til V_{\mu,\nu}(x\!+\!\hat\nu)\,
\til V_{\nu,\mu}\dag(x\!+\!\hat\mu)\,
U_\mu\dag(x)\,\big]
\Big)U_\mu(x)
\label{def_HYL}
\eea
is the hypercubically nested version of LOG smearing.
With these formulae given, all aspects of logarithmic link smearing have been
specified and we are ready for a numerical investigation.


\section{Effect on selected gluonic observables}


The LOG/HYL smearing (\ref{def_4}, \ref{def_HYL}) may be used both in gluonic
observables (e.g.\ for a fat-link topological charge or Polyakov loop) and in
fermionic observables (e.g.\ for a fat-link Dirac operator).
This section is devoted to the effect on one quantity in the pure gauge theory,
the distribution of the plaquette.
The comparison between the different smearing recipes will be organized both in
an unfair manner (sticking to the perturbative equivalence rule described in
Sect.\,2) and in a fair way (using optimized parameters in each recipe).

\begin{figure}[t]
\epsfig{file=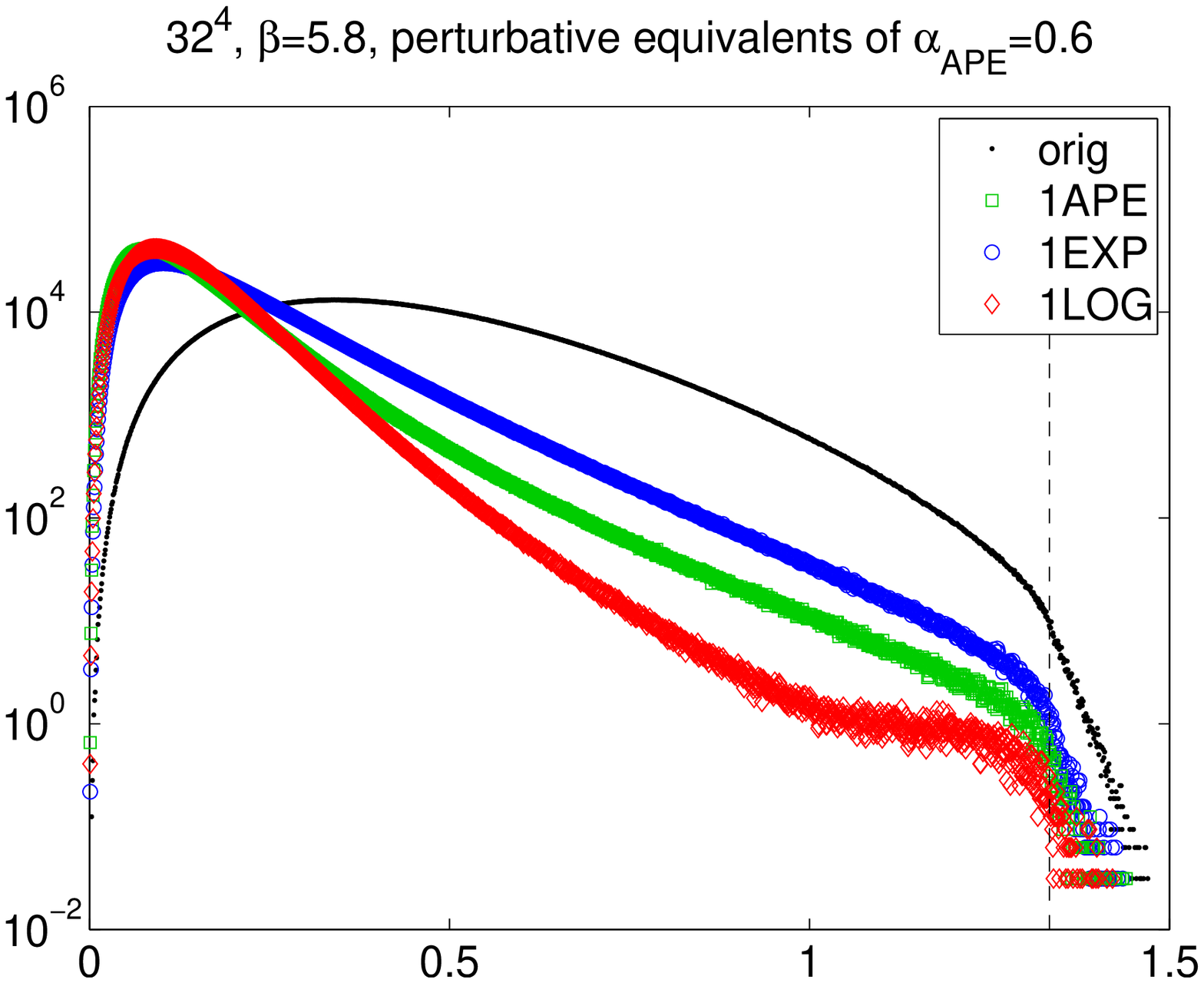,width=8.2cm}\hfill
\epsfig{file=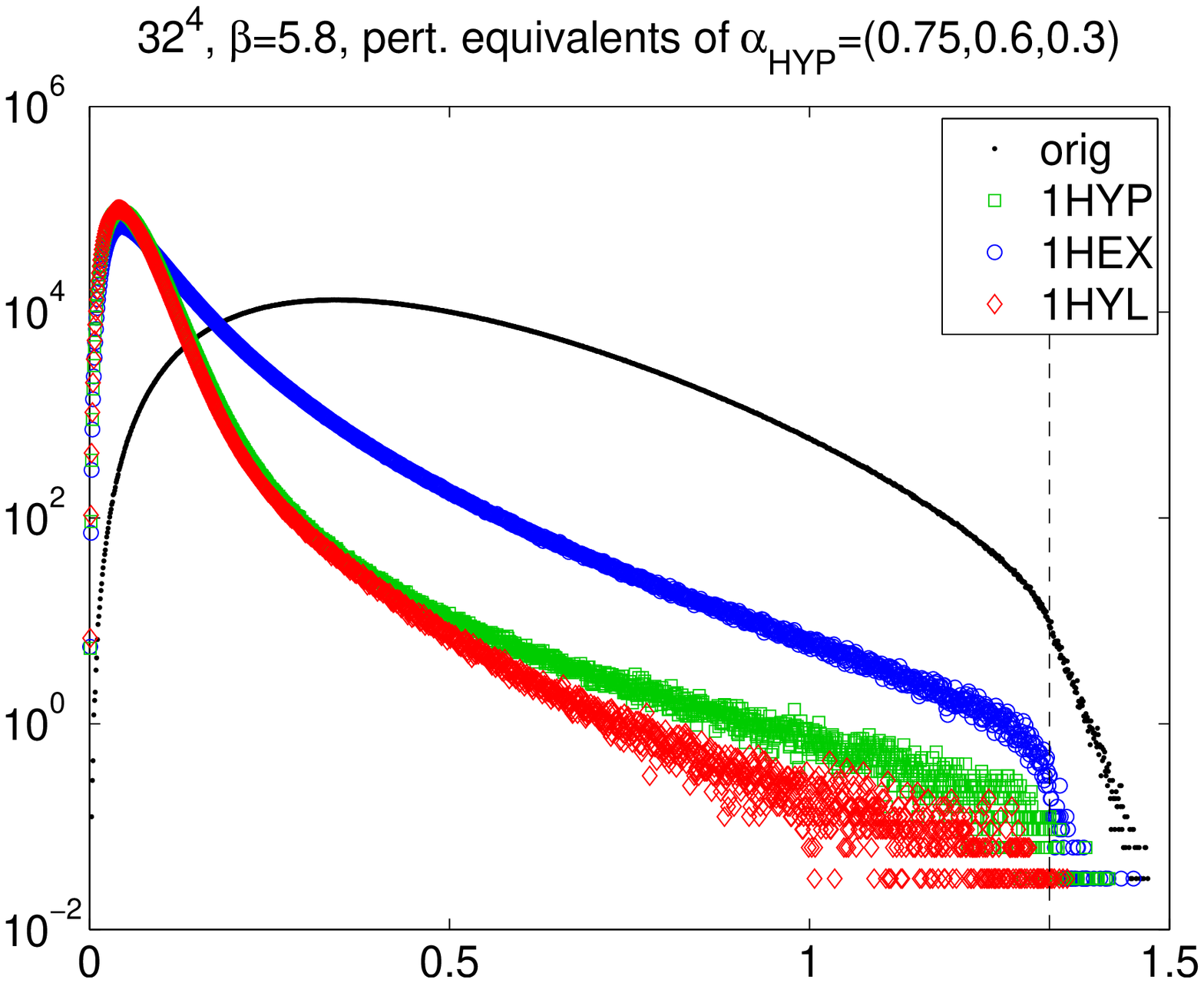,width=8.2cm}\\
\epsfig{file=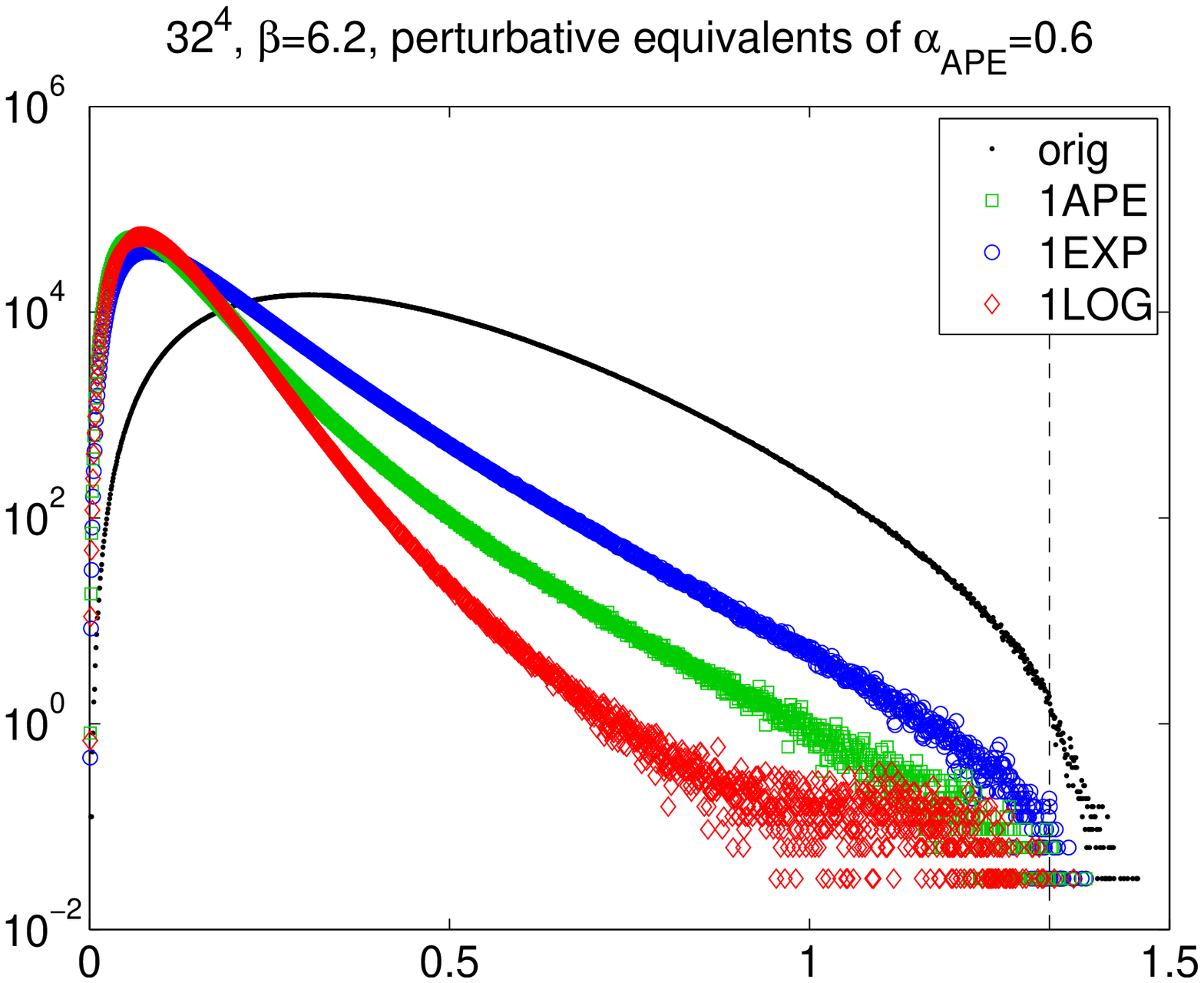,width=8.2cm}\hfill
\epsfig{file=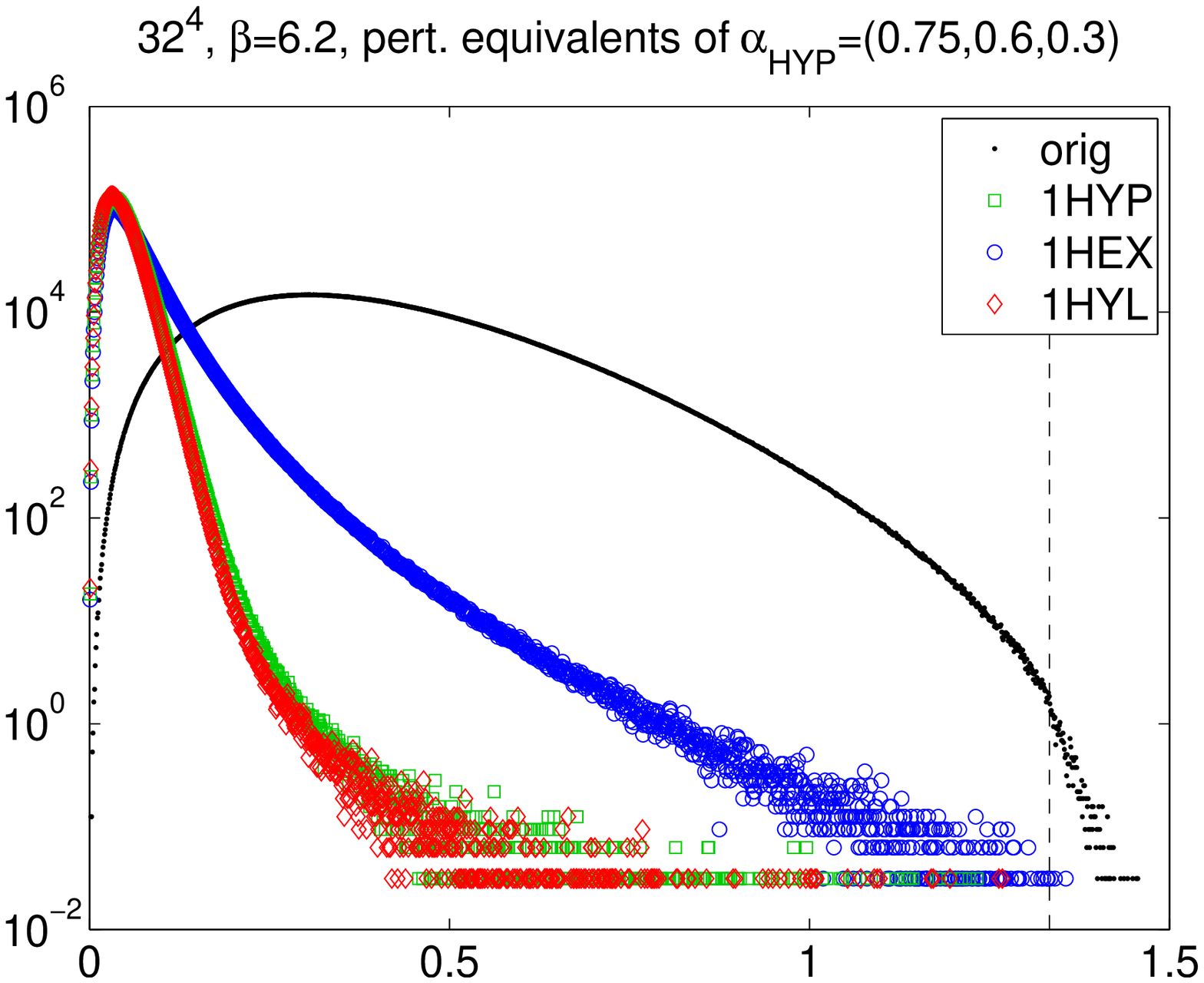,width=8.2cm}
\caption{\sl Distribution of the plaquette on a $32^4$ lattice at $\be=5.8$
(top) and $\be=6.2$ (bottom) after one step of APE/EXP/LOG smearing (left) or
HYP/HEX/HYL smearing (right). To the right of the dashed line at $s=4/3$ the
non-principal definition of the log may prove relevant. Throughout perturbative
equivalents of $\al_\mr{APE}=0.6$ and $\al_\mr{HYP}=(0.75,0.6,0.3)$ have been
used.}
\label{fig1}
\end{figure}

\begin{figure}[t]
\epsfig{file=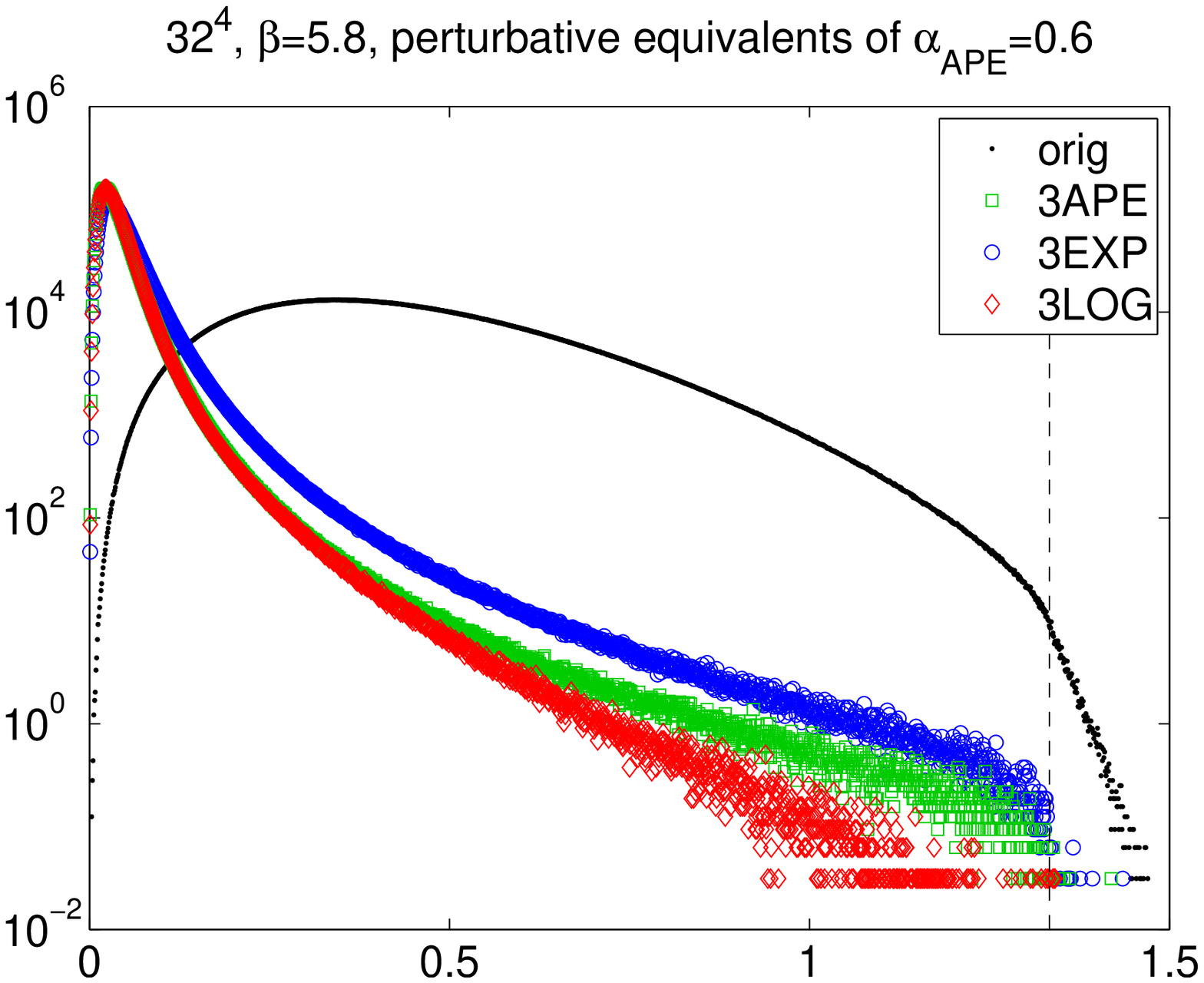,width=8.2cm}\hfill
\epsfig{file=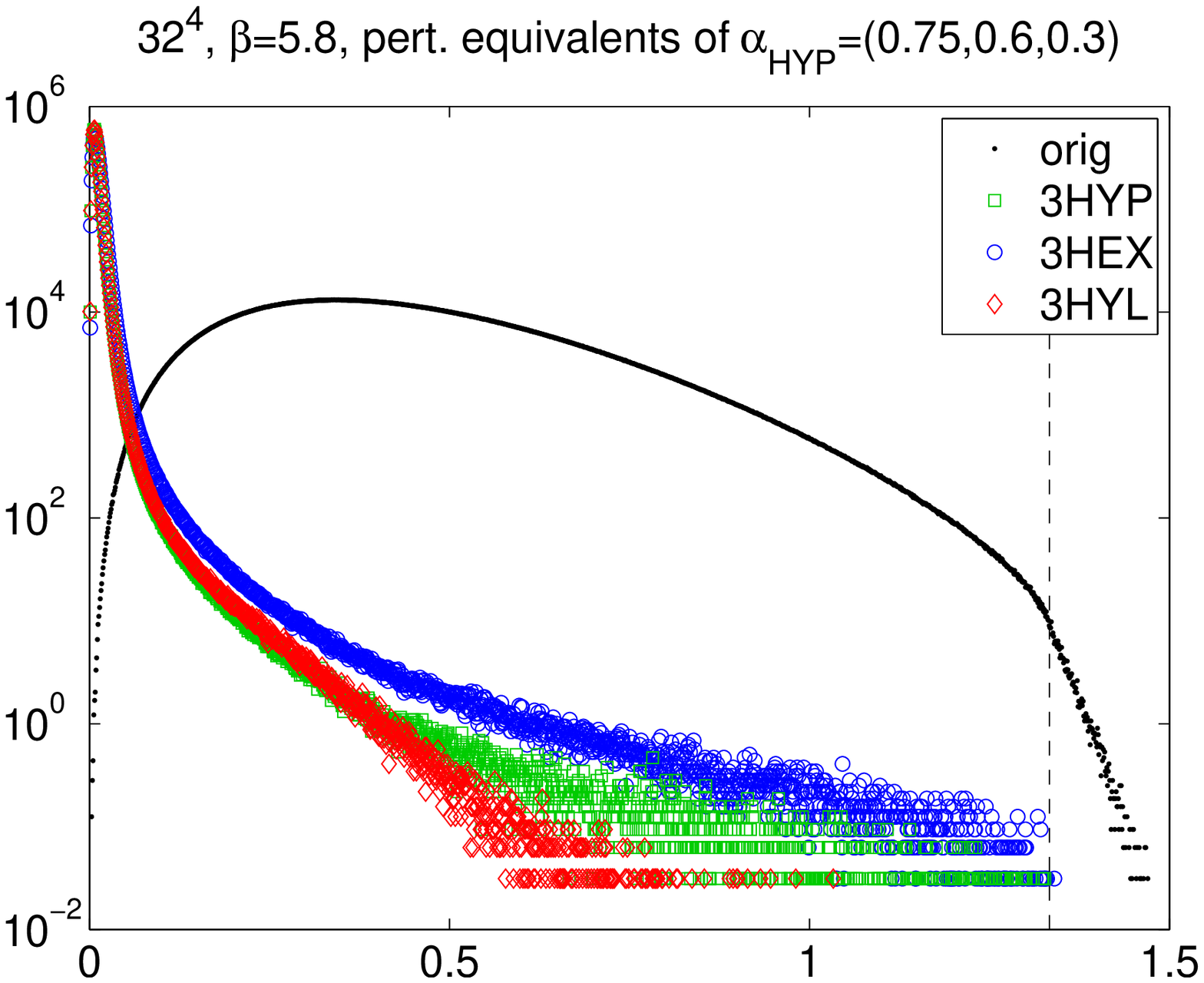,width=8.2cm}\\
\epsfig{file=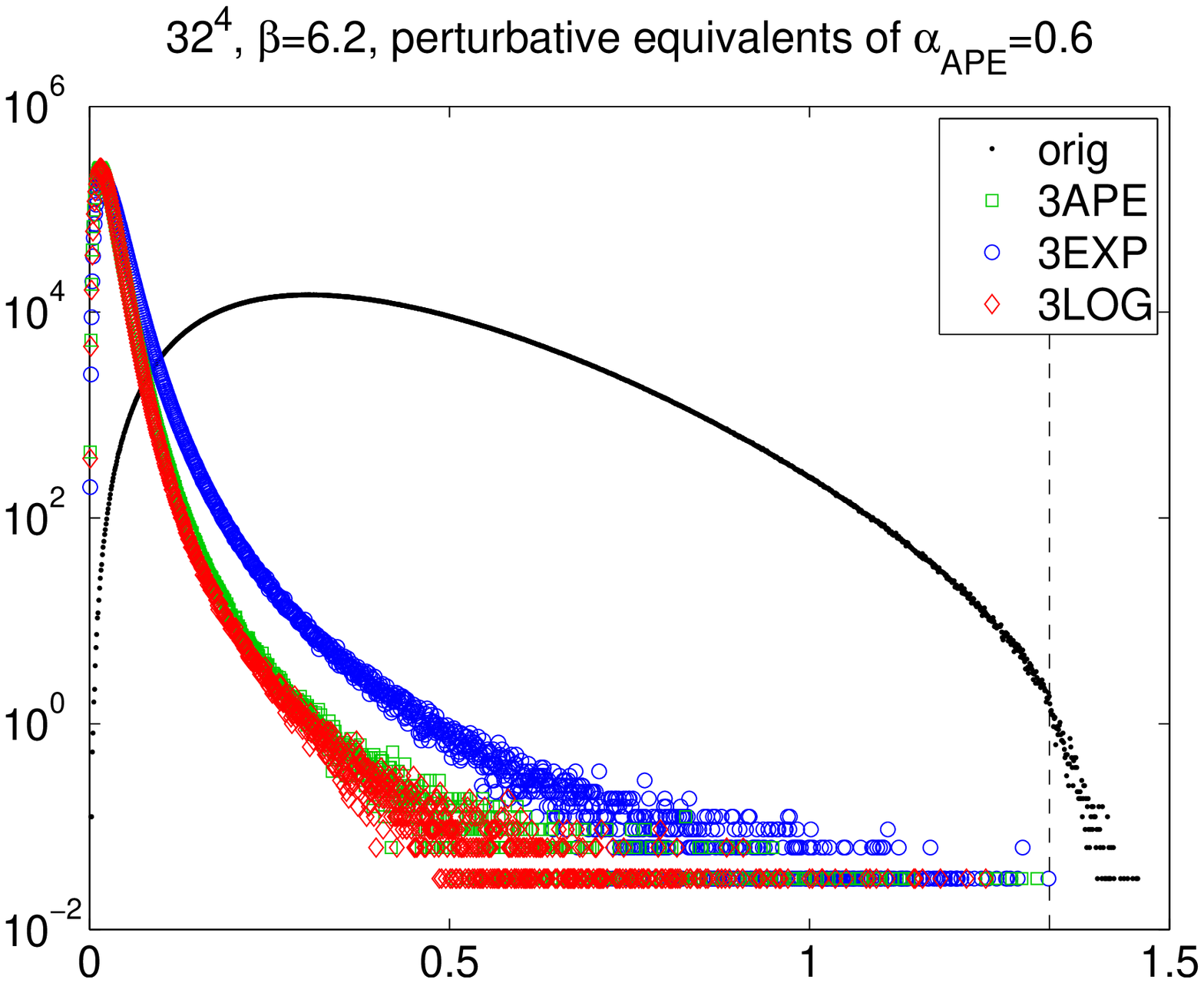,width=8.2cm}\hfill
\epsfig{file=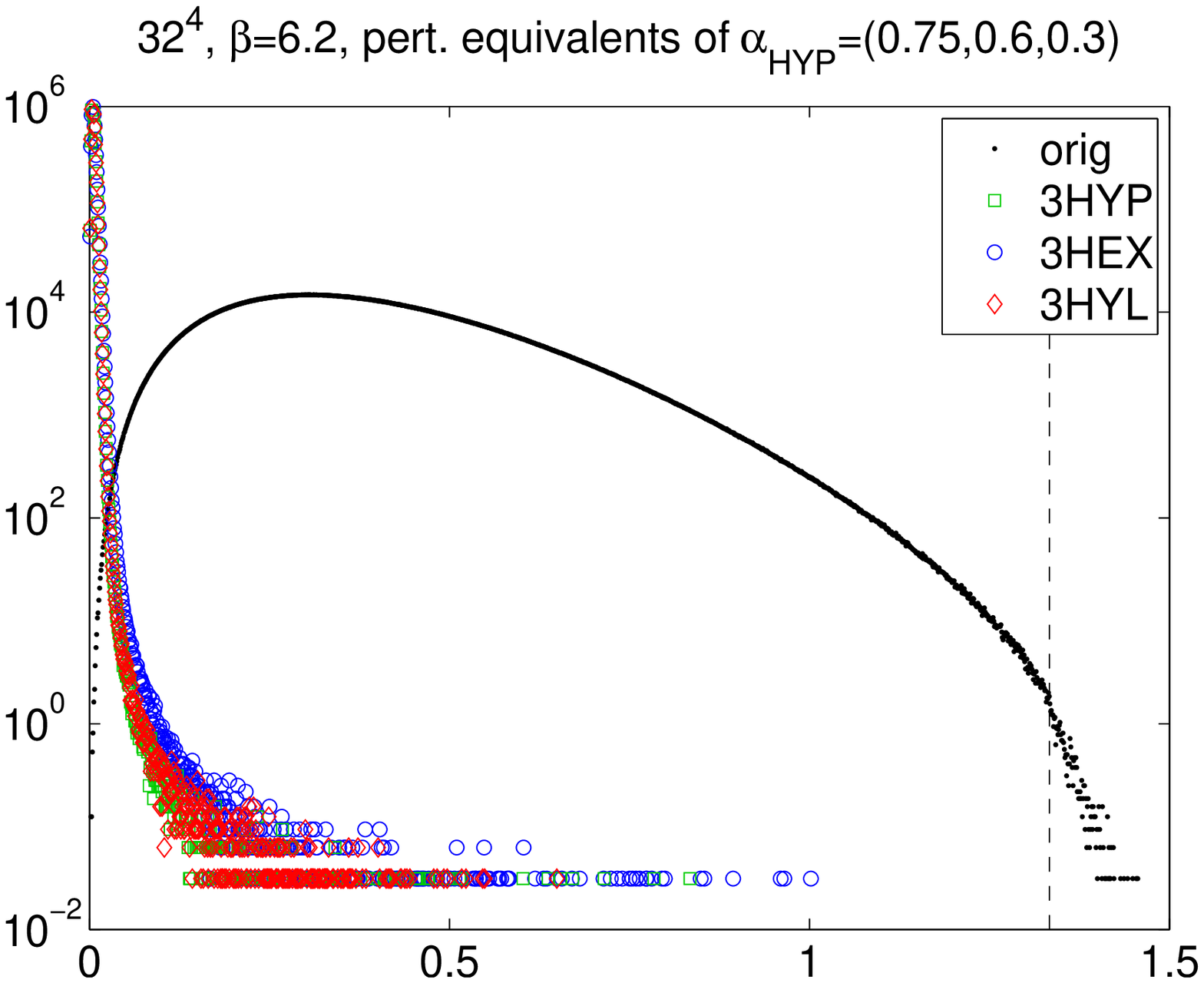,width=8.2cm}
\caption{\sl The same as in Fig.\,1, but after three iterations of the smearing
have been applied.}
\label{fig2}
\end{figure}

Fig.\,\ref{fig1} shows the distribution of the plaquette variable
$s_{\mu\nu}(x)=1-\mr{Re\,Tr}(U_{\mu\nu}(x))/3$ on a $32^4$ grid
in a pure gauge ensemble with the Wilson action at $\be=5.8$ and $\be=6.2$.
The distribution vaguely resembles a black body radiation curve -- it starts
out with a power law and ends with a more-or-less exponential tail which, at
not-so-large $\be$ is affected by the fact that $s\leq1.5$.
Smearing shifts the distribution towards a ``colder'' temperature, in
particular the probability of large or extremal plaquettes is suppressed.
Taking APE smearing as a standard, EXP (alias ``stout'') is somewhat less
effective, while the new LOG smearing is at least as effective.
In this comparison perturbatively equivalent parameters have been used, that
is $\al_\mr{APE}=\al_\mr{LOG}=0.6$ and $\al_\mr{EXP}=0.1$ (see \cite{CDH} for
details).
In the right panel a similar comparison is given for one step of HYP, HEX or
HYL smearing.
Again the perturbatively equivalent parameter set 
$\al_\mr{HYP}=\al_\mr{HYL}=(0.75,0.6,0.3)$ and $\al_\mr{HEX}=(0.125,0.15,0.15)$
has been used, and with this specific choice HYP and HYL are more efficient
than HEX.
The two lower panels show that the difference between the smearings is reduced
as $\be$ increases, in particular with the hypercubic nesting trick in place.

Fig.\,\ref{fig2} shows the same distributions as Fig.\,\ref{fig1}, but this
time after three iterations of the smearing.
Again, APE and LOG are more effective than EXP, and HYP and HYL are better than
HEX, albeit the difference seems less pronounced than in the previous figure.

In these figures the ``trace-free'' logarithm has been used.
It turns out that the principal logarithm (\ref{def_1}) yields an even better
suppression of extremal plaquettes, but for reasons discussed in Sect.\,2 the
``trace-free'' version seems more promising for an application in full QCD, and
I have chosen to consistently show the results for this non-principal
logarithm.

\begin{figure}
\epsfig{file=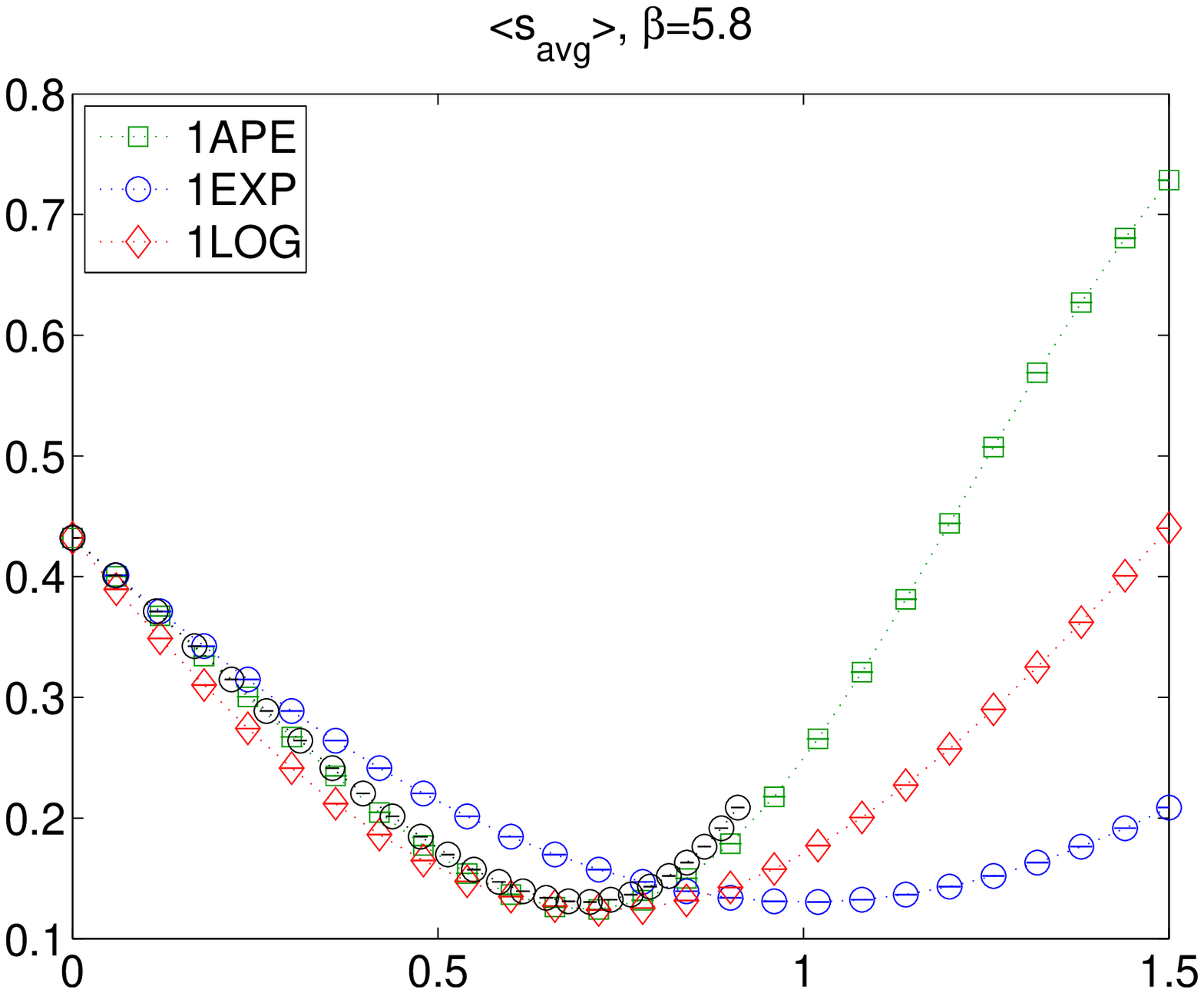,width=8.2cm}\hfill
\epsfig{file=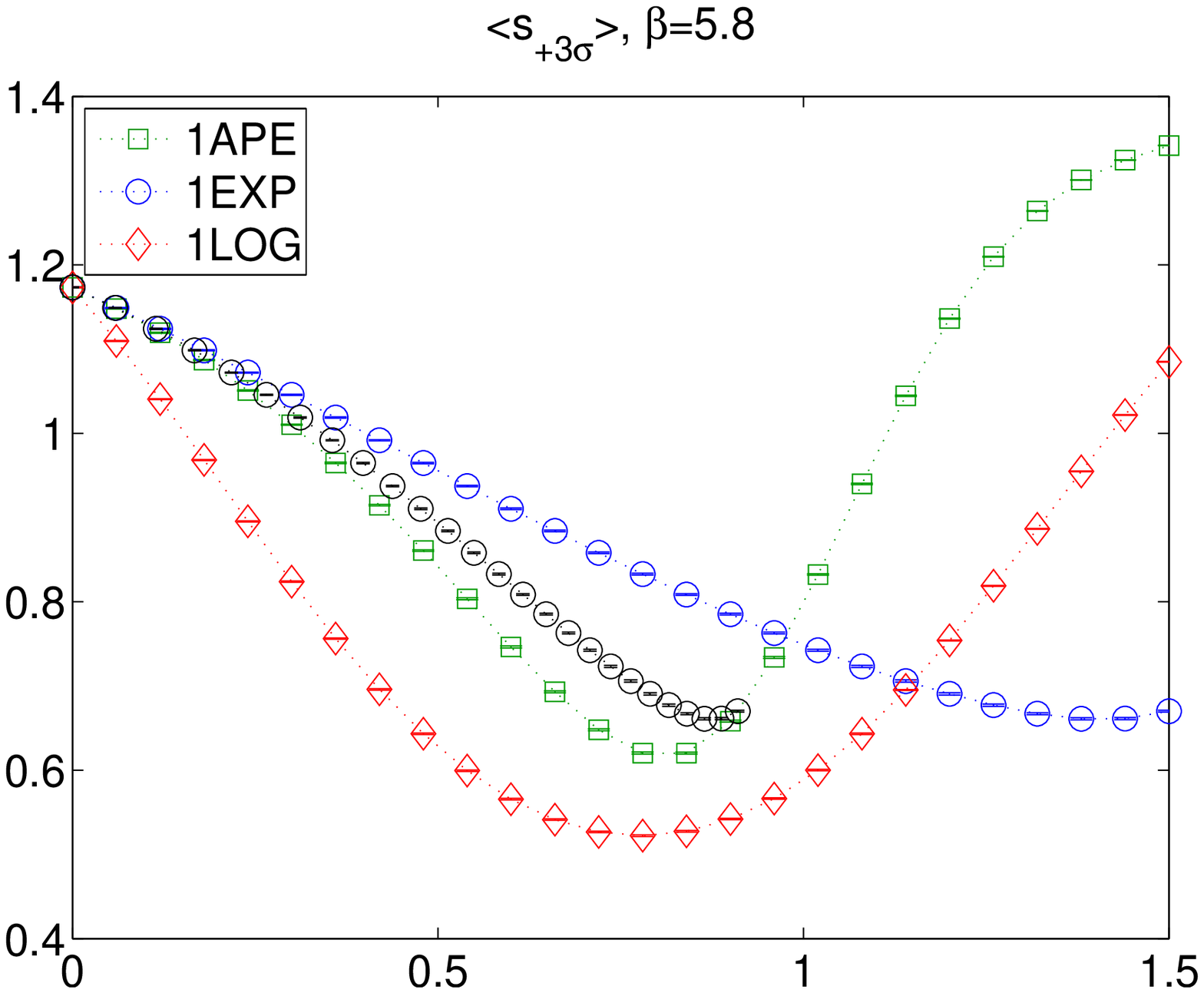,width=8.2cm}
\\[2mm]
\epsfig{file=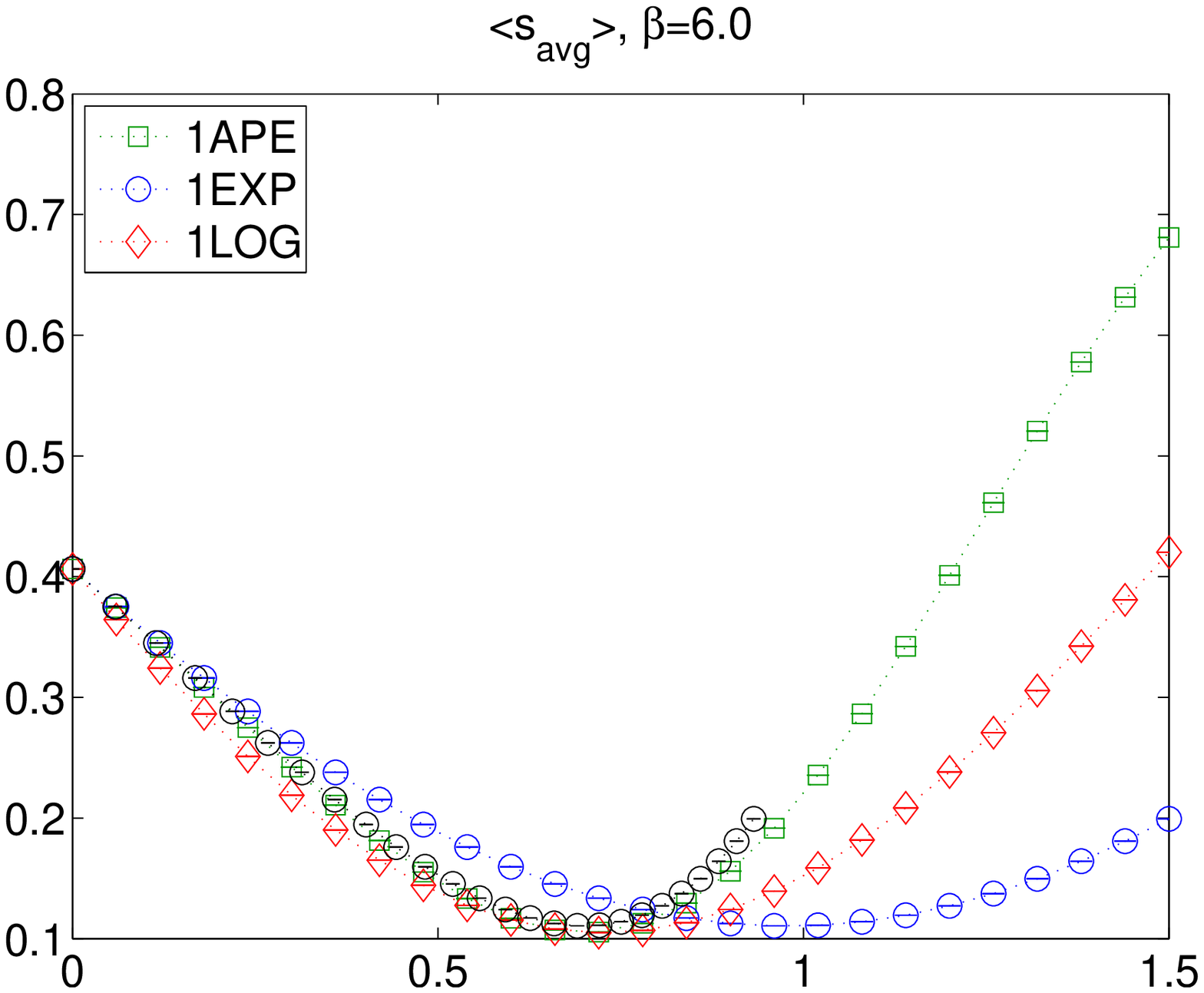,width=8.2cm}\hfill
\epsfig{file=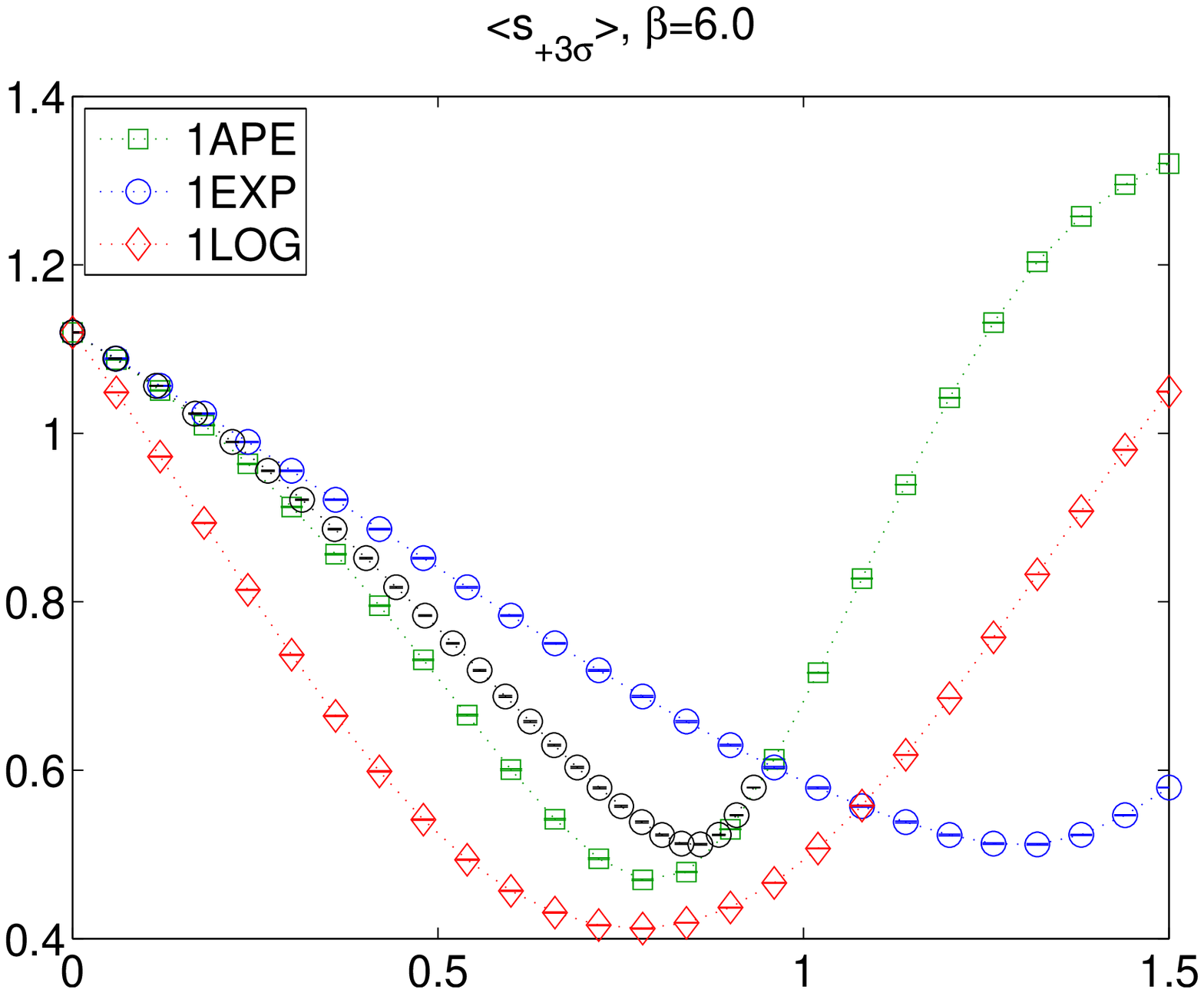,width=8.2cm}
\\[2mm]
\epsfig{file=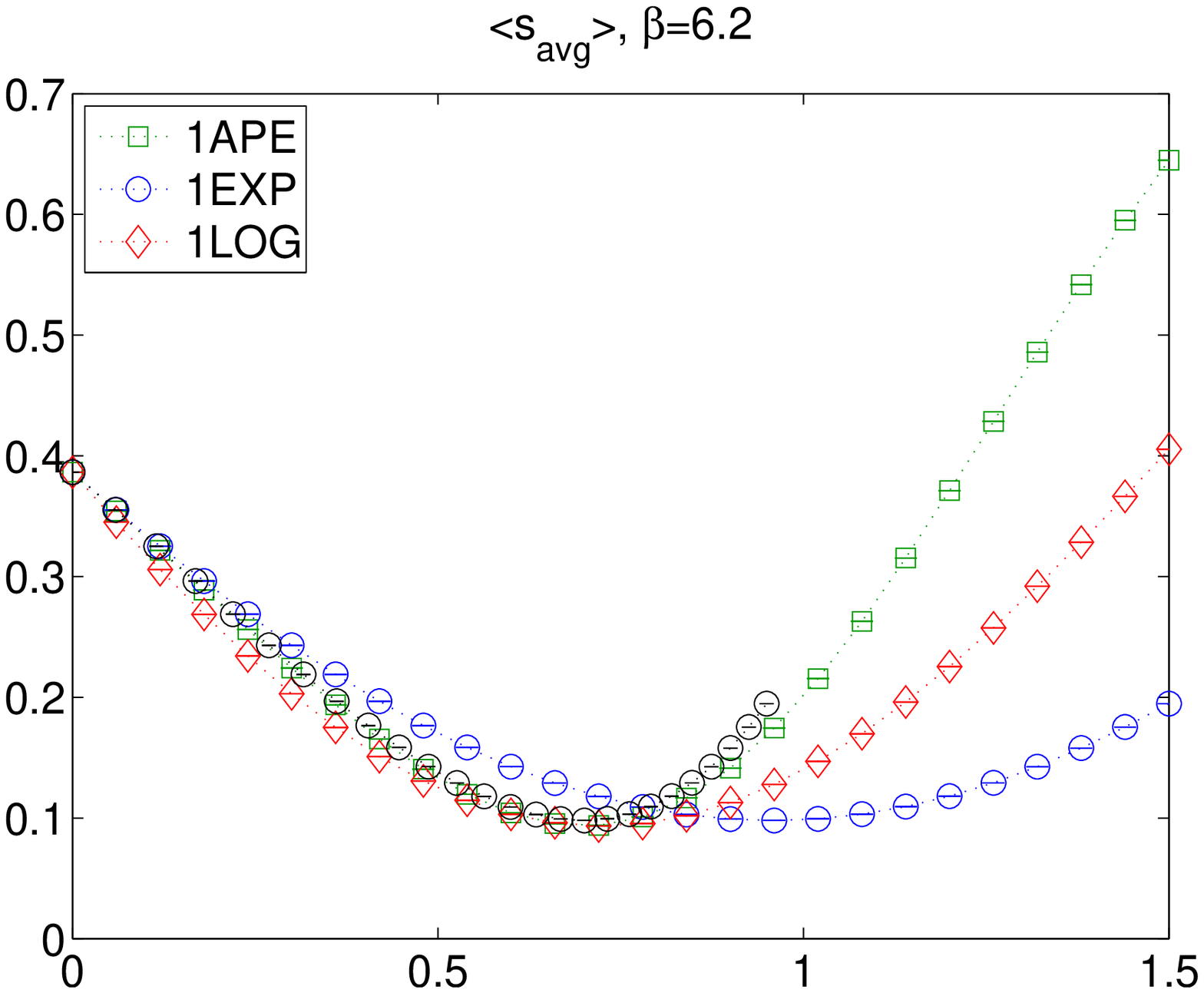,width=8.2cm}\hfill
\epsfig{file=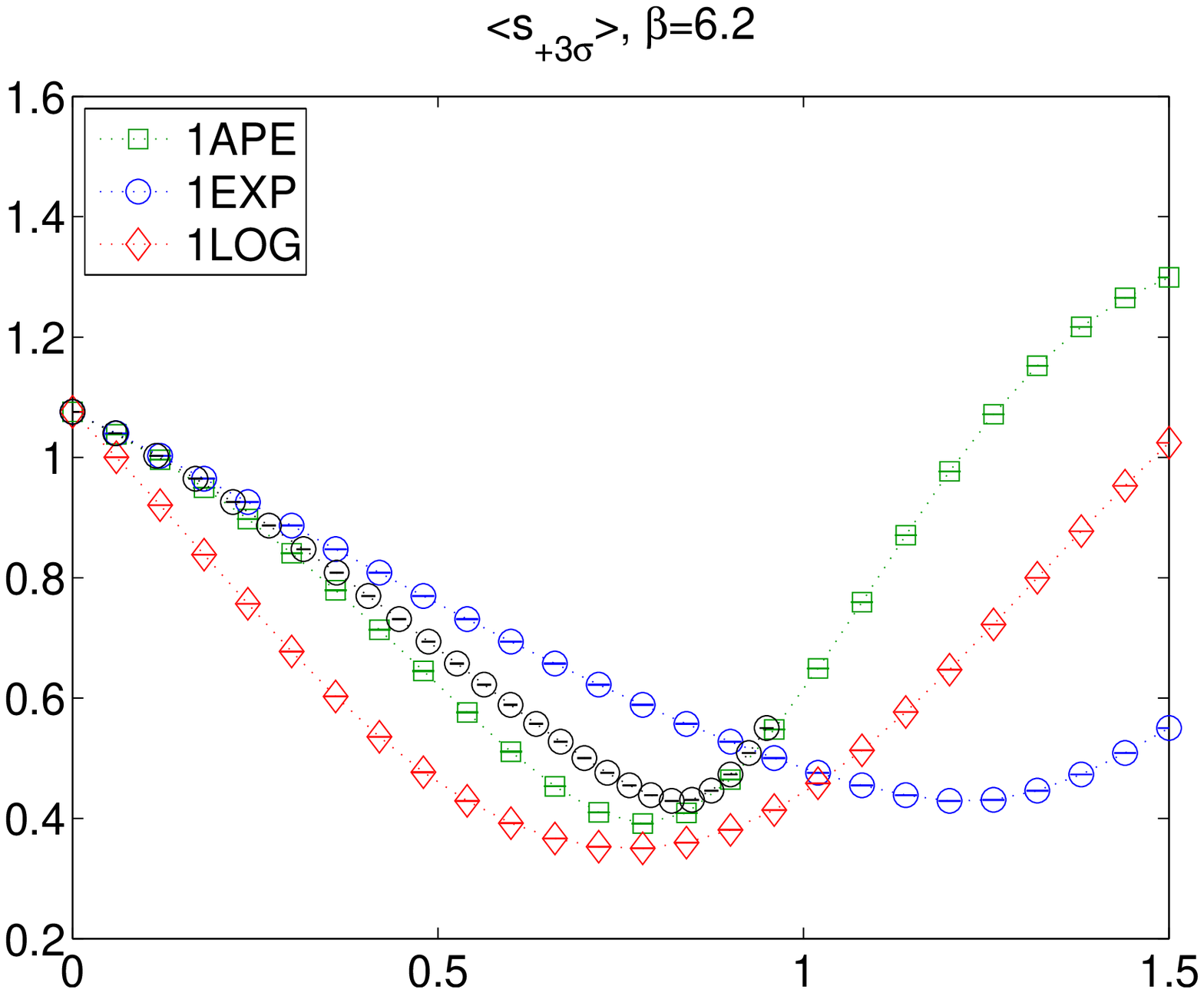,width=8.2cm}
\caption{\sl Average action $s$ (left) and median-plus-$3\si$-percentile of $s$
(right) at $\be=5.8,6.0,6.2$ (from top to bottom) versus smearing parameter
$\al$, after 1 iteration of APE/EXP/LOG smoothing. To make the three $\al$
commensurate, the perturbative equivalence rule has been used, but the 1\,EXP
data have also been redrawn as a function of $\al_\mr{eff}$ as defined in
(\ref{al_eff}).}
\label{fig3}
\end{figure}

\begin{figure}
\epsfig{file=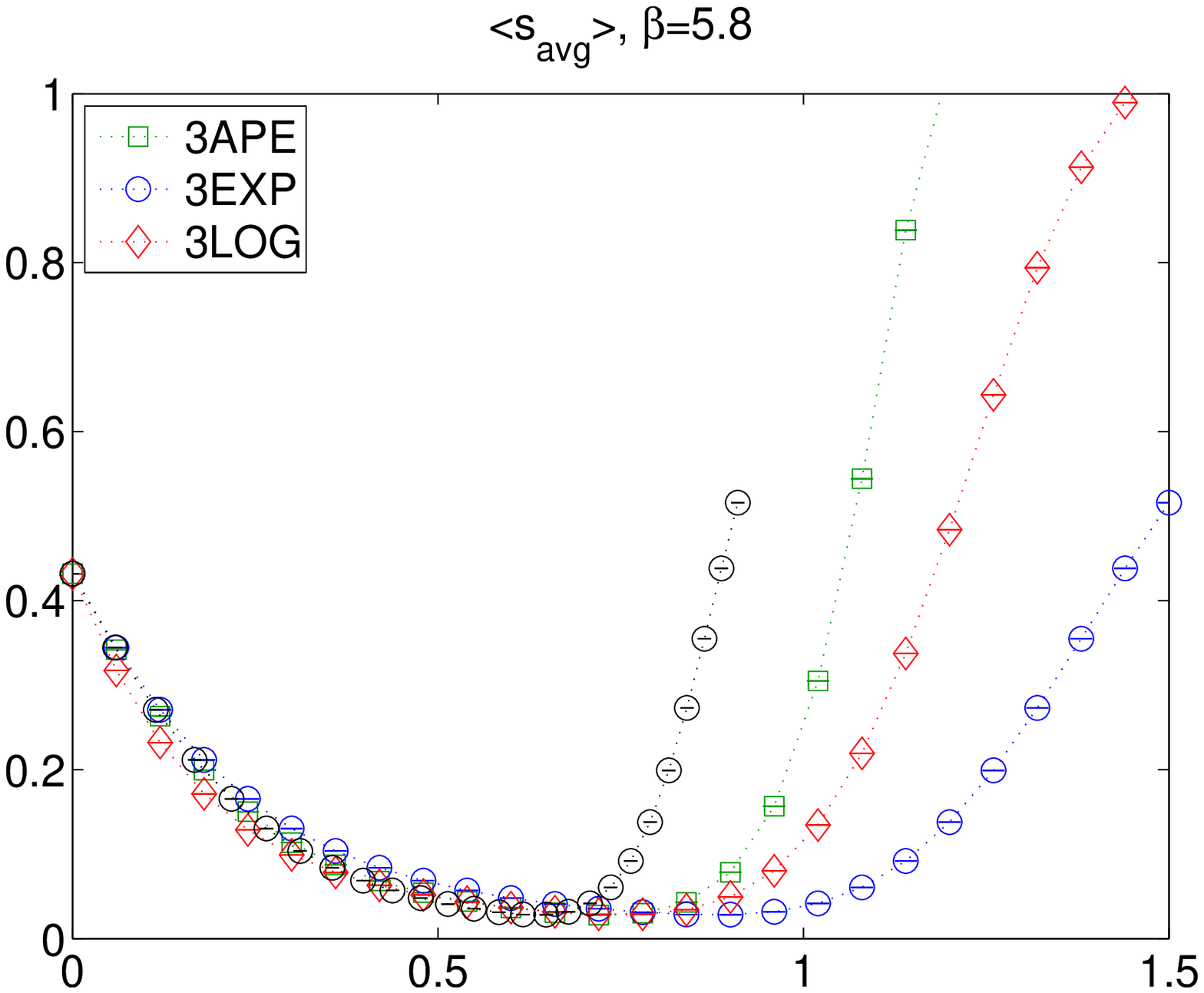,width=8.2cm}\hfill
\epsfig{file=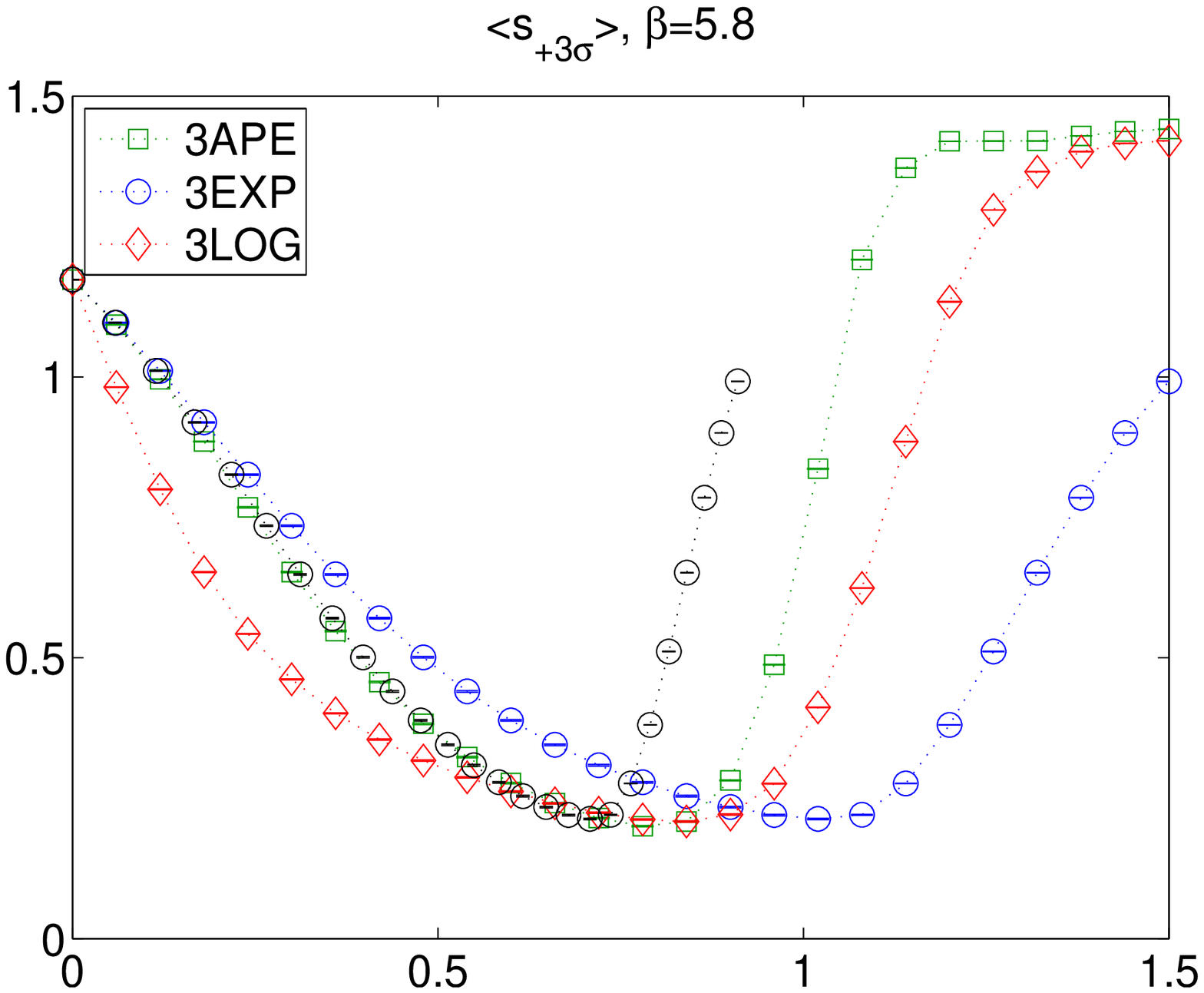,width=8.2cm}
\\[2mm]
\epsfig{file=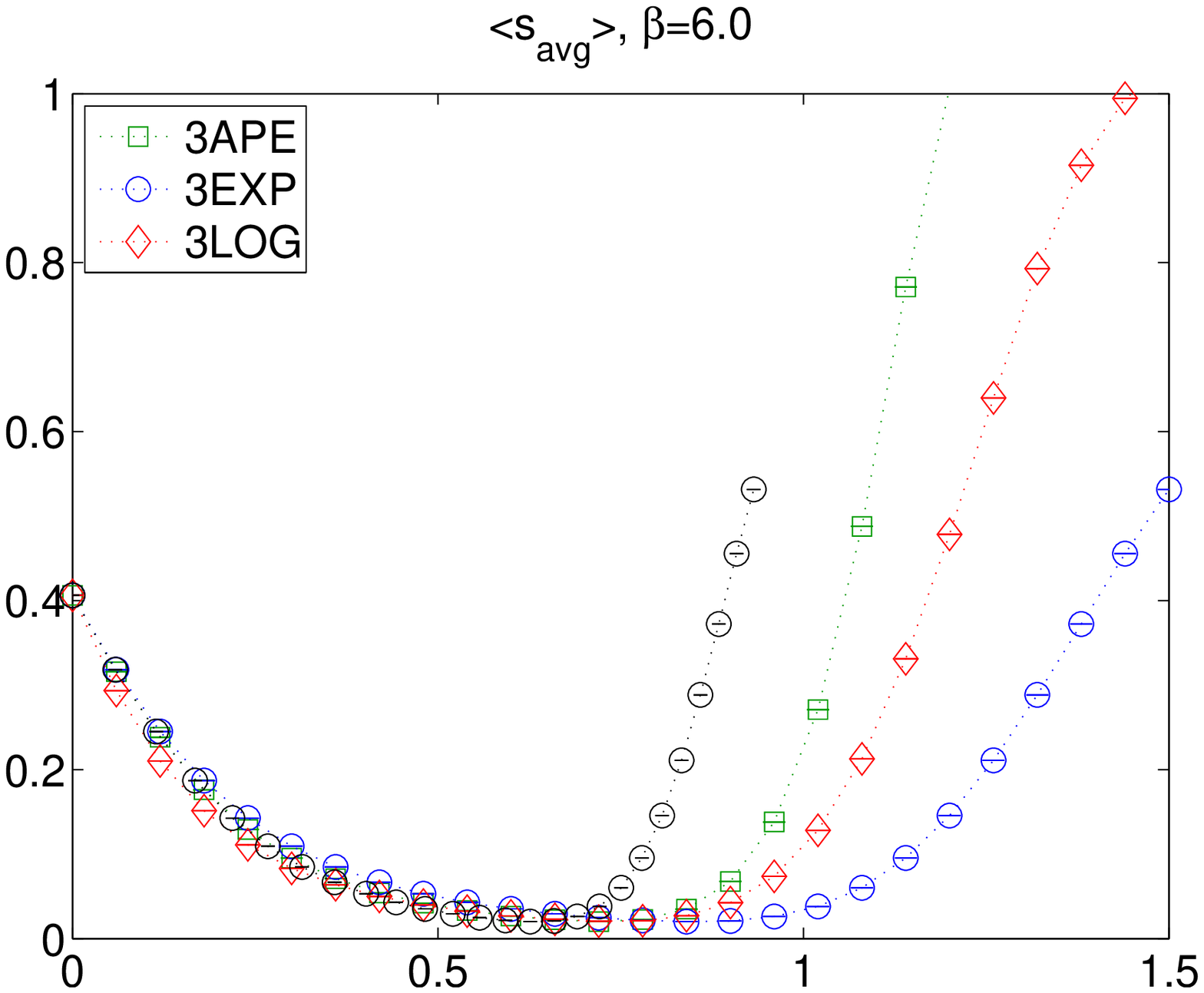,width=8.2cm}\hfill
\epsfig{file=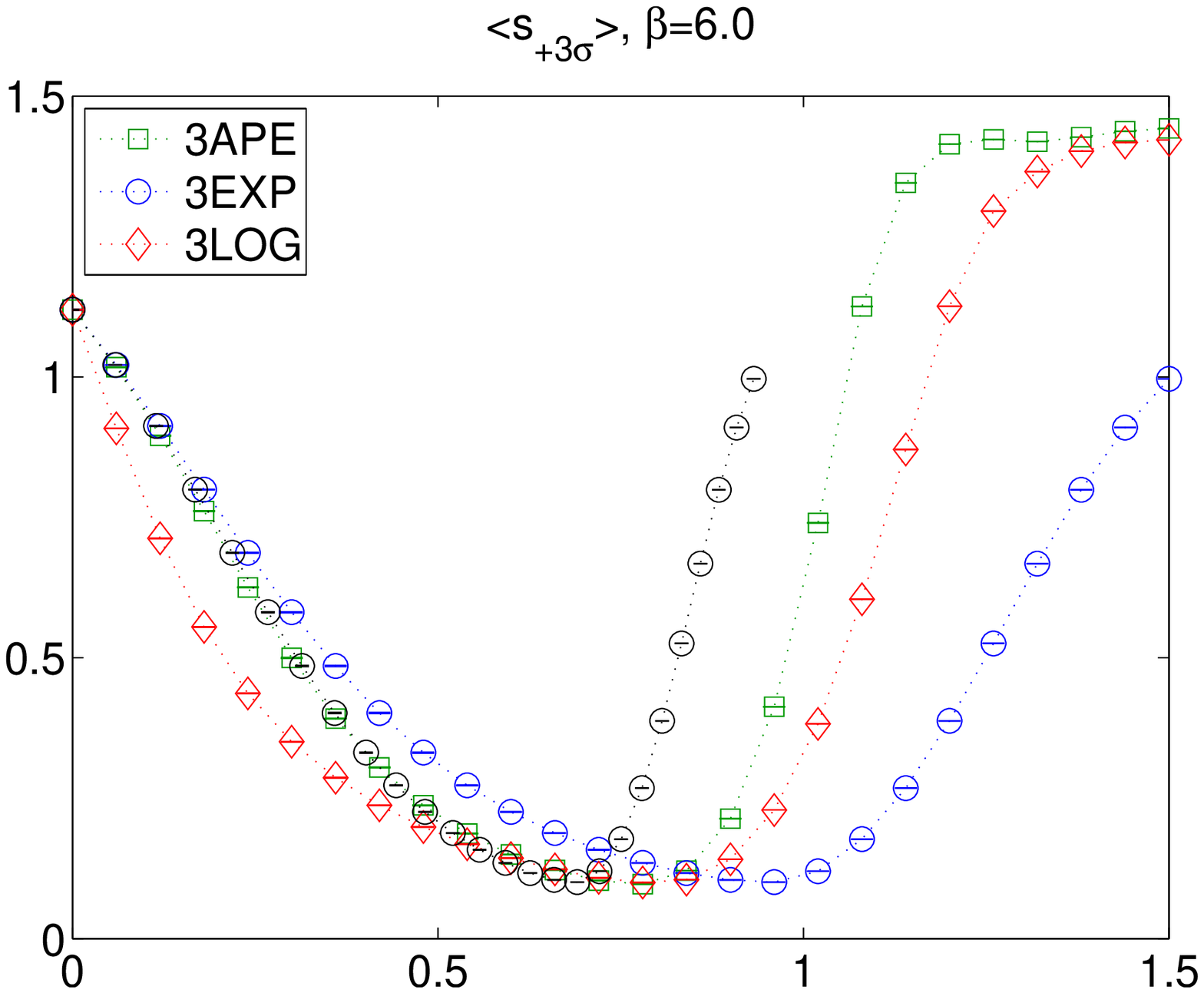,width=8.2cm}
\\[2mm]
\epsfig{file=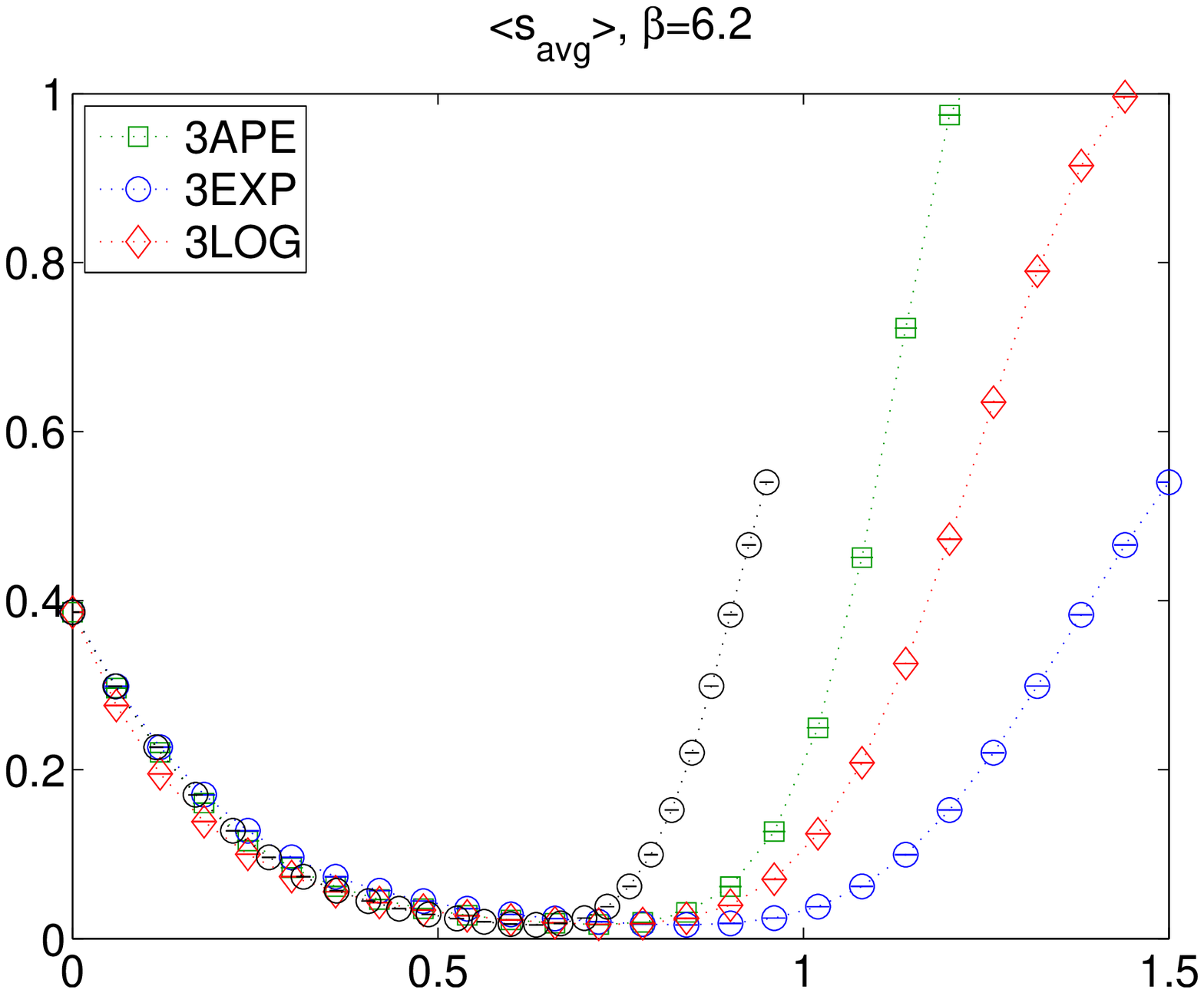,width=8.2cm}\hfill
\epsfig{file=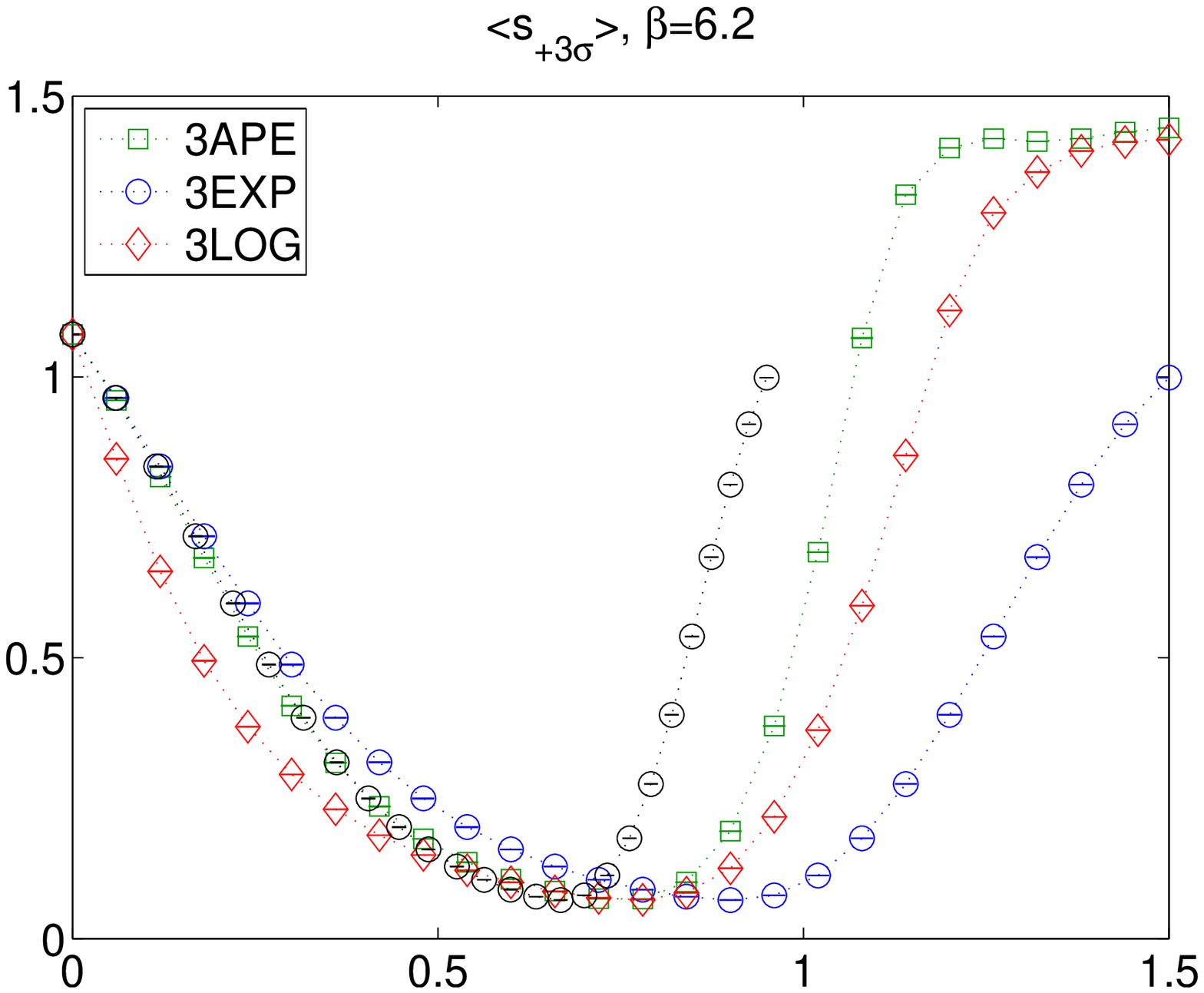,width=8.2cm}
\caption{\sl Same as Fig.\,\ref{fig3}, except that the effect of 3 iterations
of APE/EXP/LOG smoothing is shown. The 3\,EXP data have been redrawn as a
function of $\al_\mr{eff}$, as defined in (\ref{al_eff}), with $\<P\>$ denoting
the plaquette of the original (unsmeared) configuration.}
\label{fig4}
\end{figure}

Evidently, a fair comparison should use ``optimized'' values for each smearing.
However, it is not so clear which quantity sould be used in the optimization.
For instance, statements about the effect of smearing on both the average
plaquette and extremal plaquettes are found in the literature \cite{HYP}.
Moreover, once a specific quantity to optimize for has been selected, at least
two characteristics should enter the comparison:
\begin{itemize}
\itemsep-2pt
\vspace*{-1mm}
\item[({\sl i})]
The value of the selected observable in the extremum.
\item[({\sl ii})]
The width of the ``near-minimal'' region, i.e.\ whether the optimal smearing
effect requires a fine-tuning of the parameters or whether
any ``near-optimal'' choice will do fine, too.
\vspace*{-1mm}
\end{itemize}

To address these questions a scan over different smearing parameters in the
APE/EXP/LOG recipes has been performed.
The focus has been on the average plaquette $s_\mr{avg}$ and an ``extremal''
plaquette, here defined through $s_{+3\si}$ which is meant to denote the
$0.99865$ percentile, rather than $s_\mr{max}$ which is rather noisy.
Data for $\be=5.8,6.0,6.2$ are collected in Figs.\,\ref{fig3} and \ref{fig4}
for 1 and 3 iterations, respectively.
All results are plotted as a function of $\al_\mr{APE}$, $6\al_\mr{EXP}$,
$\al_\mr{LOG}$, but following the suggestion of \cite{n-APE} the EXP/stout
data are also plotted as a function of
\beq
\al_\mr{eff}={6\al_\mr{EXP}\ovr1+6\al_\mr{EXP}(1-\<P\>)}
\label{al_eff}
\eeq
where $\<P\>$ denotes the plaquette (i.e.\ $1-s$) of the original (unsmeared)
configuration.

Regarding point ({\sl i}), from Fig.\,\ref{fig3} it is evident that the minimum
of the average plaquette action is almost the same with all three smearings.
Only the position at which it is assumed is somewhat different with EXP
(it seems $\al\simeq0.7$ is optimal for APE and LOG, and $6\al\simeq1$ is
a good choice for EXP), but most of this effect is gone if the latter data are
plotted as a function of $\al_\mr{eff}$ \cite{n-APE}.
On the other hand, the ``extremal'' plaquette is quite sensitive to the details
of the smearing recipe (right panel).
With this criterion LOG smearing is significantly better than APE, and the
latter is slightly more effective than EXP.
However, this advantage diminishes towards large $\be$, and also for higher
iteration counts as Fig.\,\ref{fig4} shows.

Regarding point ({\sl ii}), we learn from Fig.\,\ref{fig3} (right panel) that
again LOG smearing has a slight advantage over the other two in showing a
broader shape in the ``near-minimal'' region.
In other words, this recipe seems more robust w.r.t.\ the details of the
smearing parameter, in particular if an observable is chosen which is sensitive
to near-extremal plaquettes.
Again, larger $\be$-values and higher iteration counts tend to diminish the
effect.



\section{Effect on selected fermionic observables}


The main rationale for the LOG/HYL smearing (\ref{def_4}, \ref{def_HYL}) has
been to create a smearing that may be used as an ingredient in a fermion action
suitable for full QCD simulations with the HMC algorithm.
Therefore it is important to test a fat-link fermion with this kind of smearing
and to compare it to similar actions where APE/HYP or EXP/HEX smearing has been
used.
This may be done with Wilson or with staggered quarks.
In the former case the focus should be on chiral symmetry breaking, in the
latter case on taste symmetry violation.
For definiteness, I concentrate on Wilson fermions (with a clover term) and I
choose the simplest observable sensitive to chiral symmetry violation:
I measure the residual mass $m_\mr{res}$, defined as the PCAC mass at bare mass
$m_0=0$.
Lacking the CPU power needed to simulate full QCD, I choose the quenched theory
as testing ground.
In fact, from an engineering viewpoint this choice is likely to be more
indicative of the quality of the smearing, since there is no determinant which
would mitigate the effect of spurious almost-zero modes.

With standard conventions the ($r\!=\!1$) Wilson operator takes the form
\beq
D_\mr{W}(x,y)={1\ovr2}\sum_\mu
\Big\{
(\ga_\mu-I) U_\mu(x)\de_{x+\hat\mu,y}-
(\ga_\mu+I) U_\mu\dag(x-\hat\mu)\de_{x-\hat\mu,y}
\Big\}
+{1\ovr2\ka}\de_{x,y}
\label{def_wils}
\eeq
with $I$ a $4\times4$ spinor matrix and $1/(2\ka)=4+m_0$.
The Sheikholeslami-Wohlert clover operator follows by subtracting a hermitean
contribution proportional to the gauge field strength
\cite{Sheikholeslami:1985ij}
\beq
D_\mr{SW}(x,y)=D_\mr{W}(x,y)
-{c_\mr{SW}\ovr2}\sum_{\mu<\nu}\si_{\mu\nu}F_{\mu\nu}\;\de_{x,y}
\label{def_clov}
\eeq
with $\si_{\mu\nu}\!=\!{\ri\ovr2}[\ga_\mu,\ga_\nu]$ and $F_{\mu\nu}$ the
hermitean clover-leaf operator.
The same kind of UV-filtering is applied to the covariant derivative and to the
clover term.
In other words, the gauge link $U_\mu(x)$ in (\ref{def_wils}) is replaced, for
instance, by $U_\mu^\mr{LOG}(x)$ and the field-strength tensor $F_{\mu\nu}(x)$
in (\ref{def_clov}) is built from such links, too (see \cite{CDH} for
references to other options).
Throughout, the tree-level improvement coefficient $c_\mr{SW}\!=\!1$ is used.
The goal is to test the smearings and to compare the non-perturbative data to
the perturbative 1-loop prediction for $m_\mr{res}$.

\begin{table}[t]
\centering
\begin{tabular}{|c|ccccc|}
\hline
($\be,L/a$)  &($5.6,08$)&($5.8,12$)&($6.0,16$)&($6.2,22$)&($6.4,28$)\\
$L/r_0$      &   3.48   &   3.27   &   2.98   &   2.98   &   2.87   \\
$a^{-1}$[GeV]&  0.908   &   1.45   &   2.12   &   2.91   &   3.84   \\
n\_conf      &    256   &    128   &    64    &    32    &   16     \\
\hline
1\,APE       &0.4367(20)&0.2607(11)&0.1931(06)&0.1589(04)&0.1371(03)\\
1\,HYP       &0.1914(13)&0.0937(09)&0.0615(05)&0.0490(02)&0.0421(02)\\
1\,EXP       &0.5470(24)&0.3400(14)&0.2572(08)&0.2129(06)&0.1838(03)\\
1\,HEX       &0.3146(18)&0.1583(10)&0.1030(05)&0.0784(03)&0.0641(02)\\
1\,LOG       &0.4309(20)&0.2570(11)&0.1905(06)&0.1569(04)&0.1355(03)\\
1\,HYL       &0.1923(13)&0.0929(08)&0.0606(05)&0.0482(02)&0.0414(02)\\
\hline
3\,APE       &0.1949(14)&0.0850(10)&0.0489(06)&0.0354(03)&0.0285(02)\\
3\,HYP       &0.0681(11)&0.0242(09)&0.0109(06)&0.0069(03)&0.0053(02)\\
3\,EXP       &0.2368(16)&0.1075(10)&0.0636(06)&0.0462(03)&0.0369(02)\\
3\,HEX       &0.0950(13)&0.0305(10)&0.0128(07)&0.0078(03)&0.0059(02)\\
3\,LOG       &0.1972(14)&0.0858(10)&0.0492(06)&0.0357(03)&0.0287(03)\\
3\,HYL       &0.0696(11)&0.0247(09)&0.0110(07)&0.0068(03)&0.0053(02)\\
\hline
7\,HYL       &0.0303(08)&0.0109(06)&0.0040(05)&0.0020(02)&0.0014(01)\\
\hline
\end{tabular}
\caption{\sl Residual mass $am_\mr{res}$, defined as the PCAC quark mass at
zero bare mass, for various couplings and smearings. The box geometry is
$L^3\!\times\!T$ with $T\!=\!2L$, and errors are statistical.}
\label{tab1}
\end{table}

The details of the gauge configurations can be read off from the heading of
Tab.\,\ref{tab1}.
The box geometry is $L^3\!\times\!T$ with $T\!=\!2L$.
The five couplings between $\be=5.6$ and $\be=6.4$ and the grid sizes have been
chosen such that the physical box volumes are approximately equal, aiming for
$L/r_0\simeq3$ by the formula for $r_0(\be)$ from \cite{Necco:2001xg}.
The values in the $a^{-1}[\GeV]$ line are based on the assumption $r_0=0.5\fm$
and indicate that the lattice spacing varies by a factor 4.
A point- or $U(1)$ wall source has been used, and the point-sink has been
averaged over the time-slice.
The inversion of the clover operator has been performed with an even-odd
preconditioned version of the biconjugate gradient $\gaf$ algorithm
(BCG$\gaf$), with details given in App.\,B.

\begin{figure}[t]
\hspace*{-2mm}
\epsfig{file=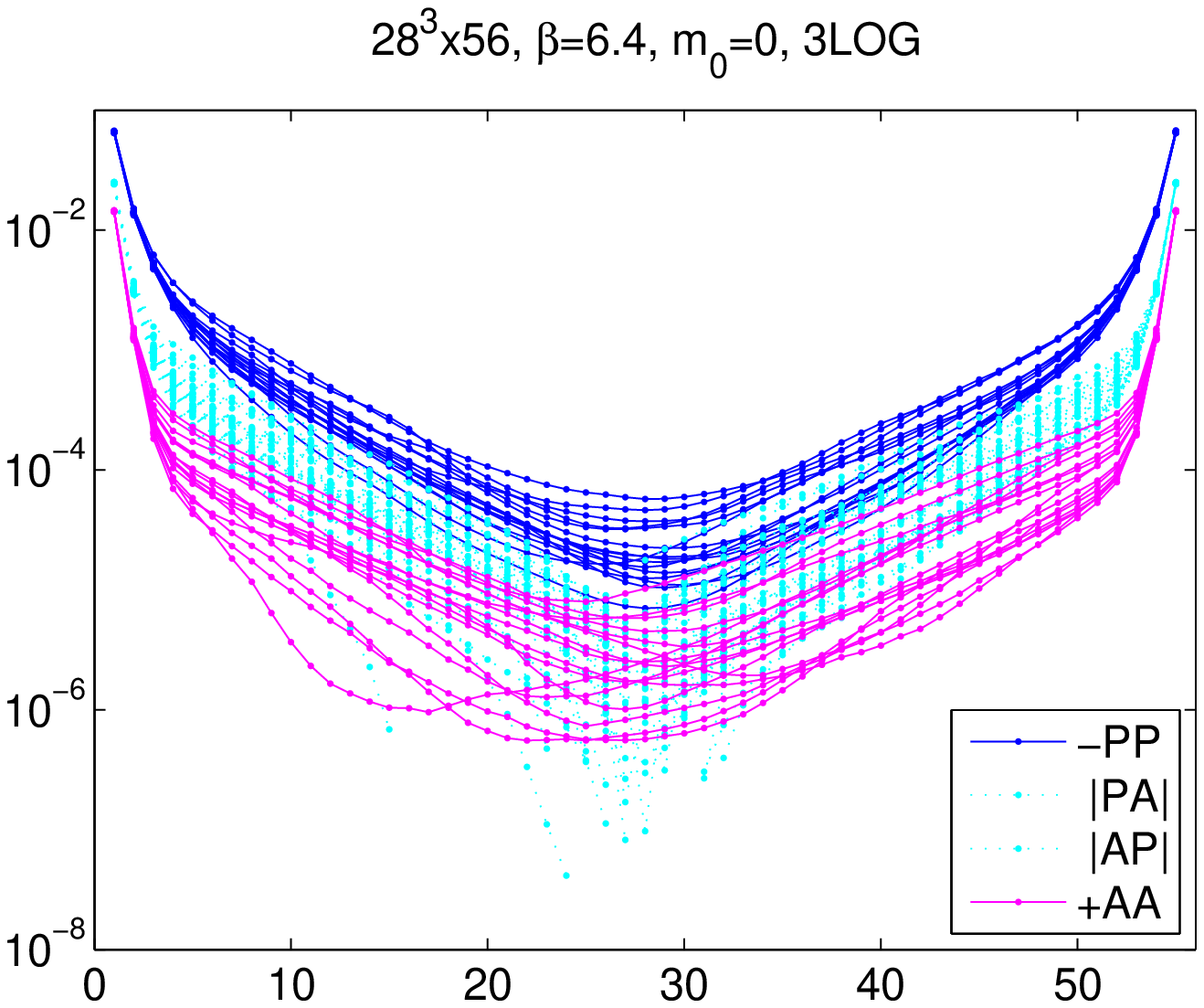,width=8.6cm}
\hspace*{-3mm}
\epsfig{file=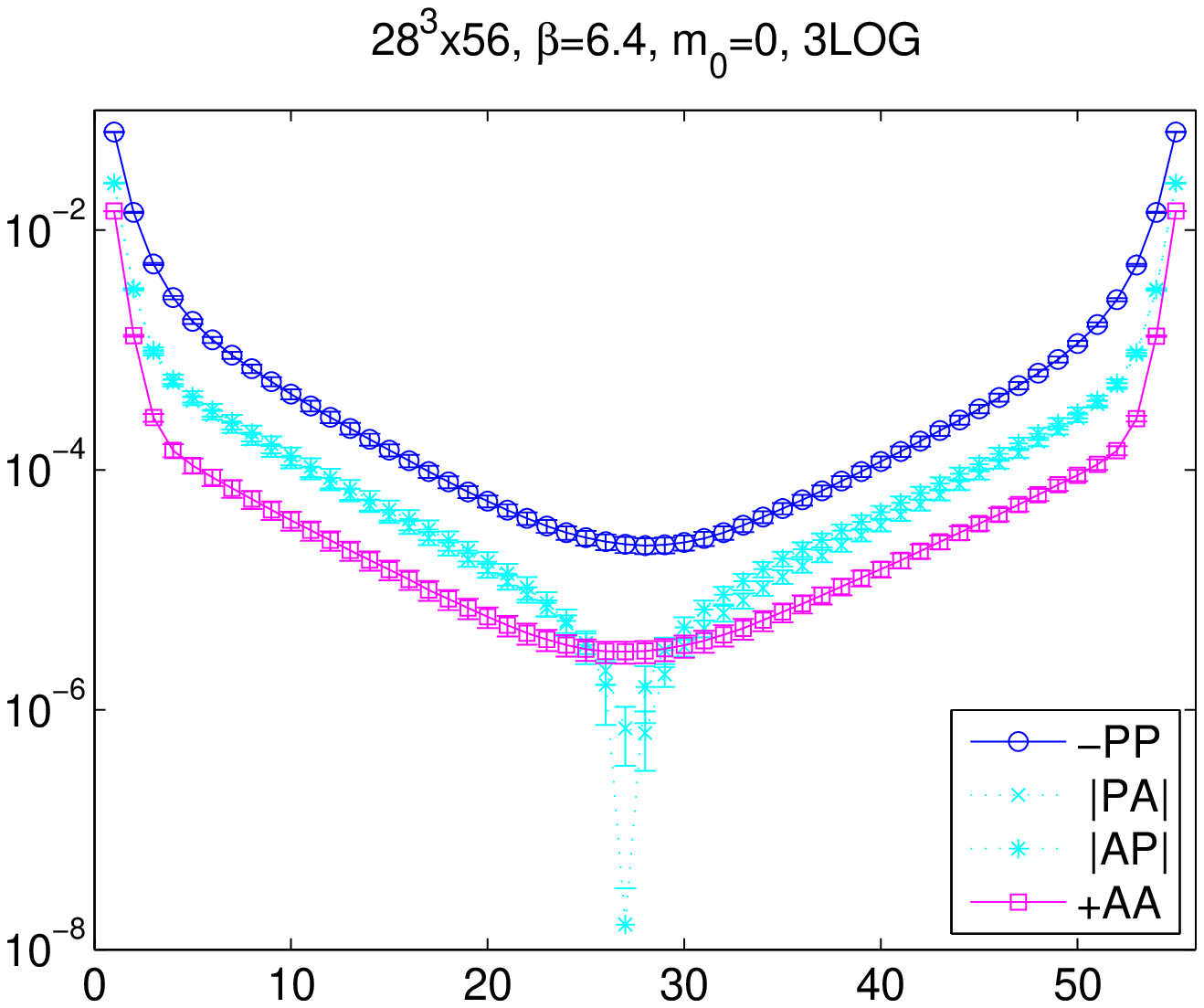,width=8.6cm}
\hspace*{-4mm}
\\
\hspace*{-2mm}
\epsfig{file=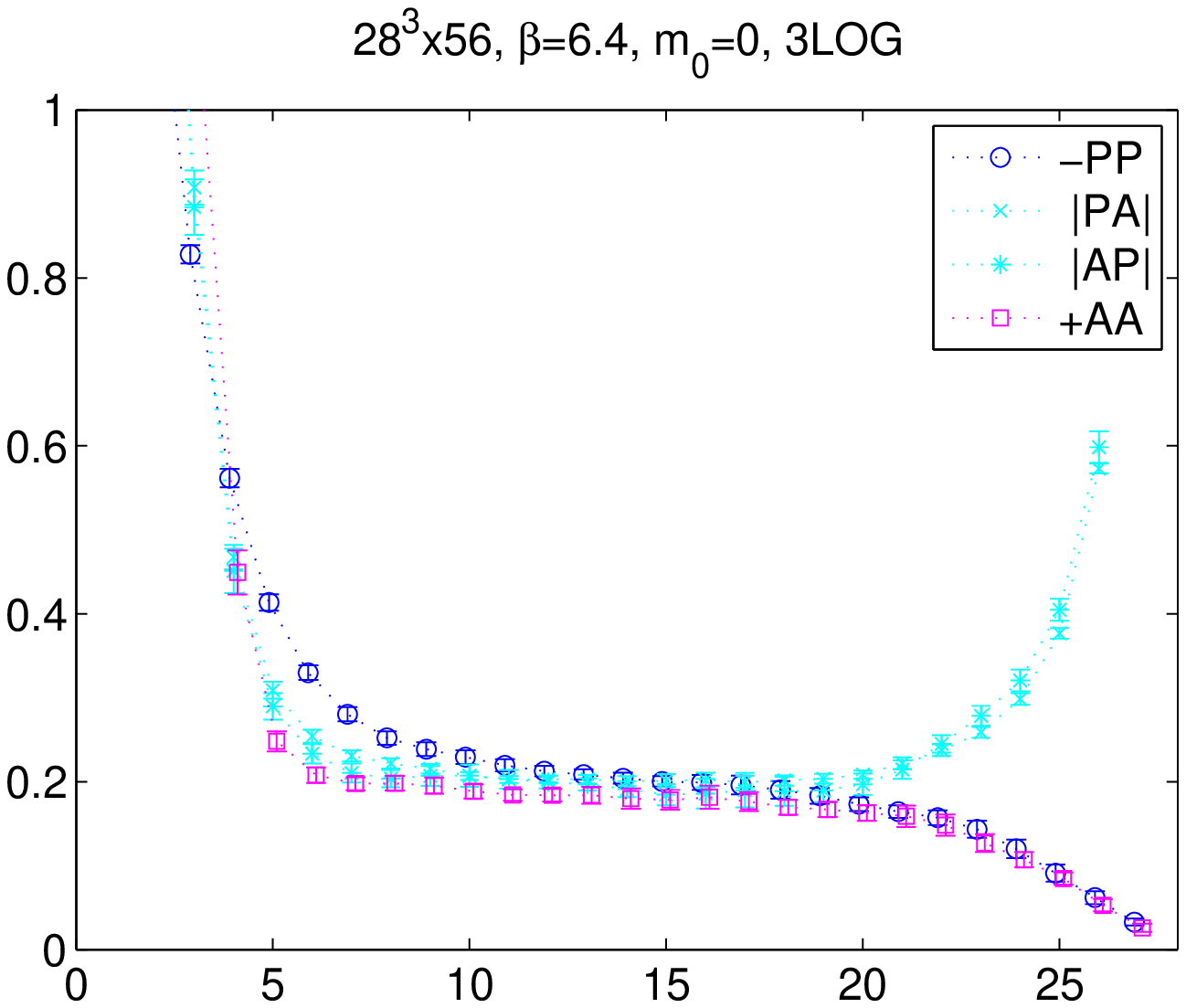,width=8.6cm}
\hspace*{-3mm}
\epsfig{file=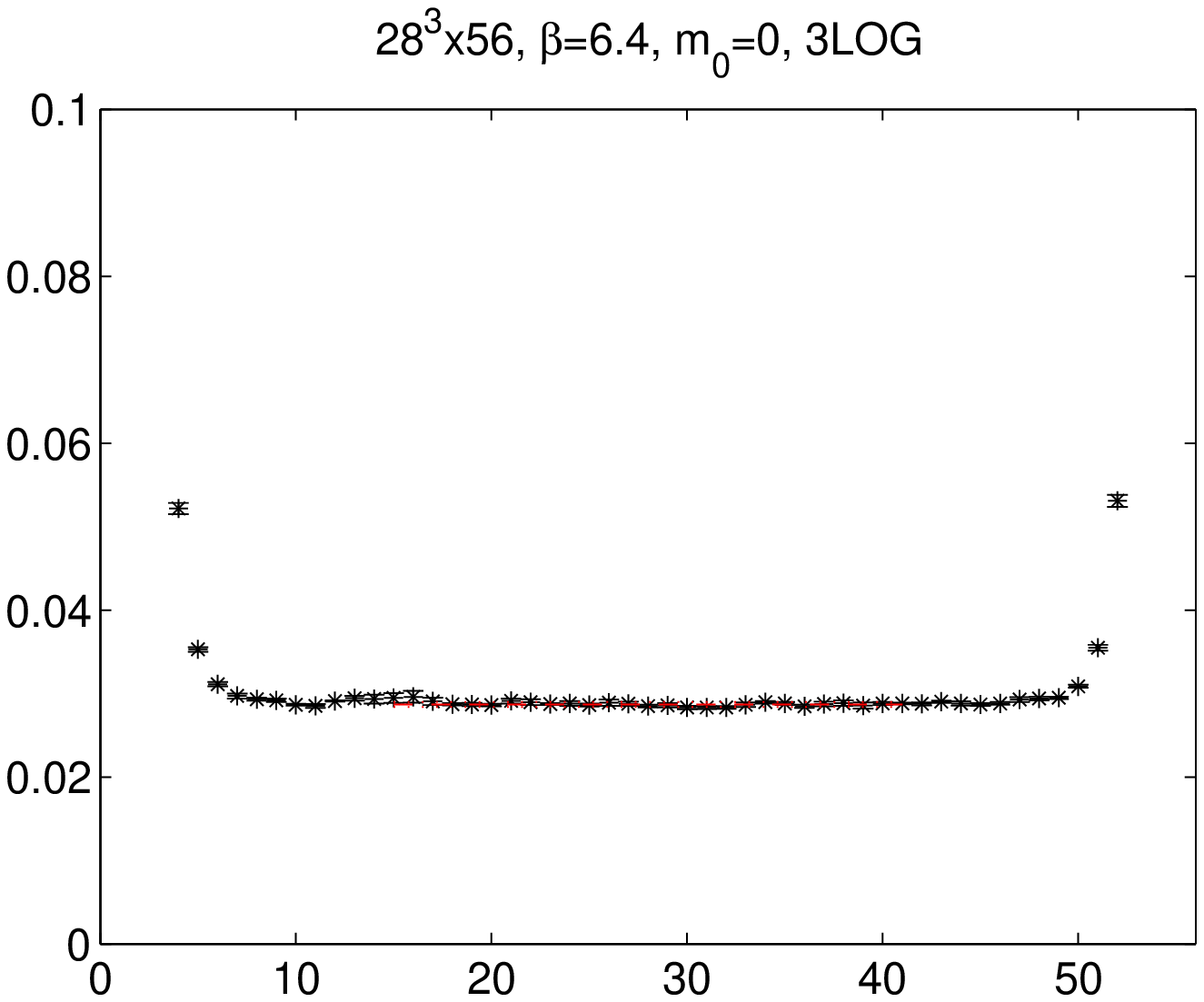,width=8.6cm}
\hspace*{-4mm}
\vspace*{-4mm}
\caption{\sl Individual and averaged pion correlators at $\be=6.4$ and $am_0=0$
with 3\,LOG steps, together with resulting pion mass and PCAC mass (everything
in lattice units).}
\label{fig5}
\end{figure}

\begin{figure}[t]
\hspace*{-2mm}
\epsfig{file=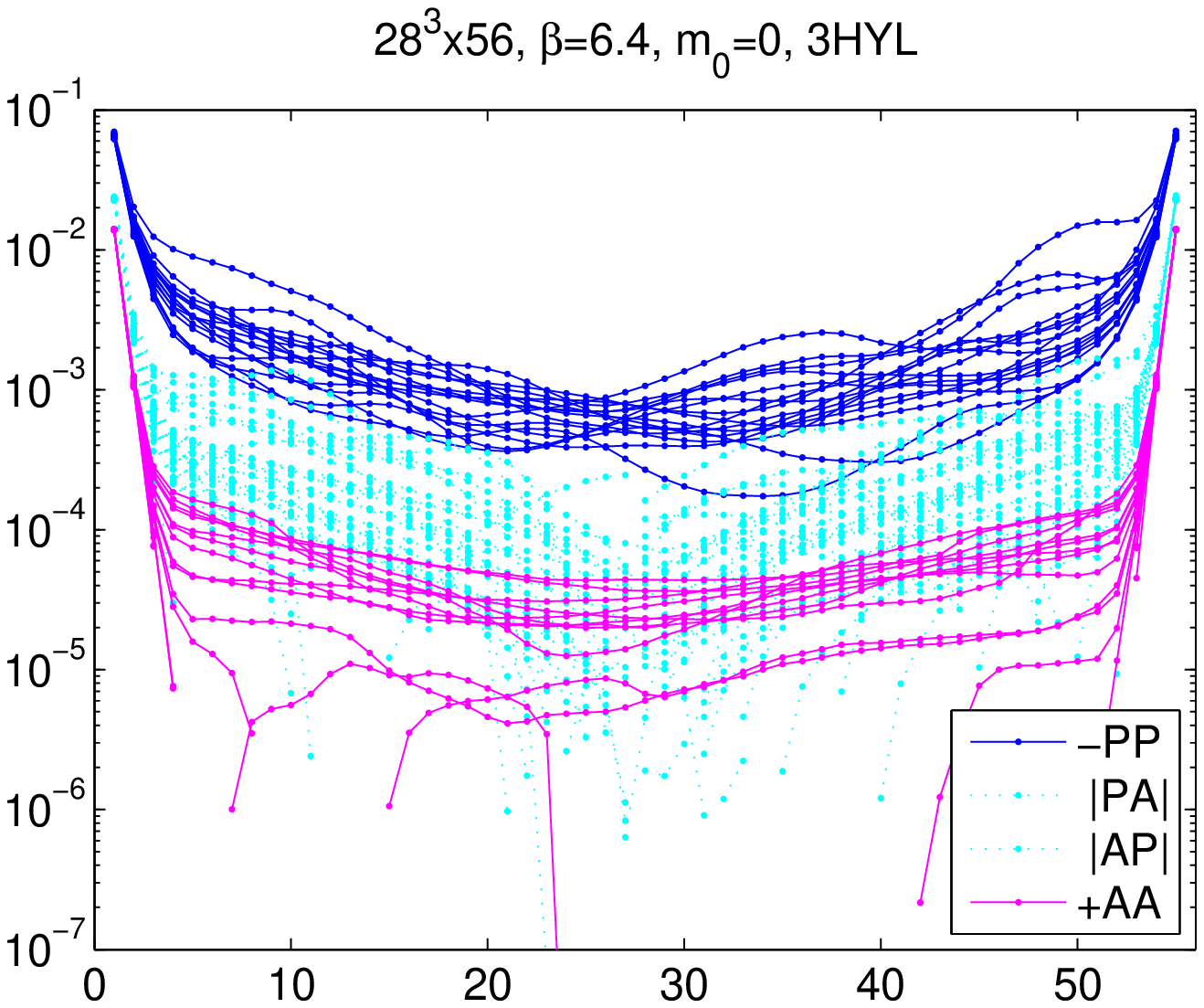,width=8.6cm}
\hspace*{-3mm}
\epsfig{file=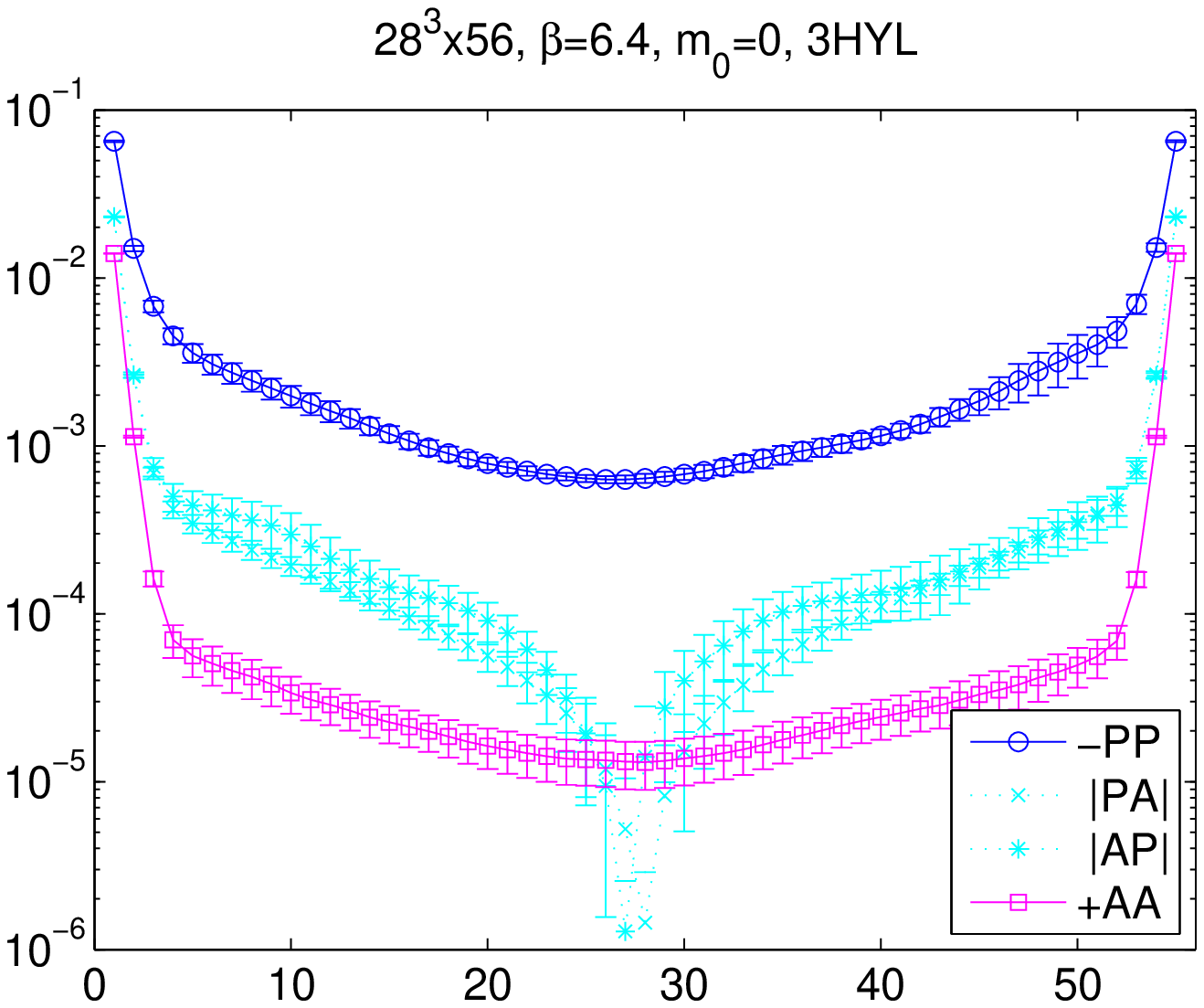,width=8.6cm}
\hspace*{-4mm}
\\
\hspace*{-2mm}
\epsfig{file=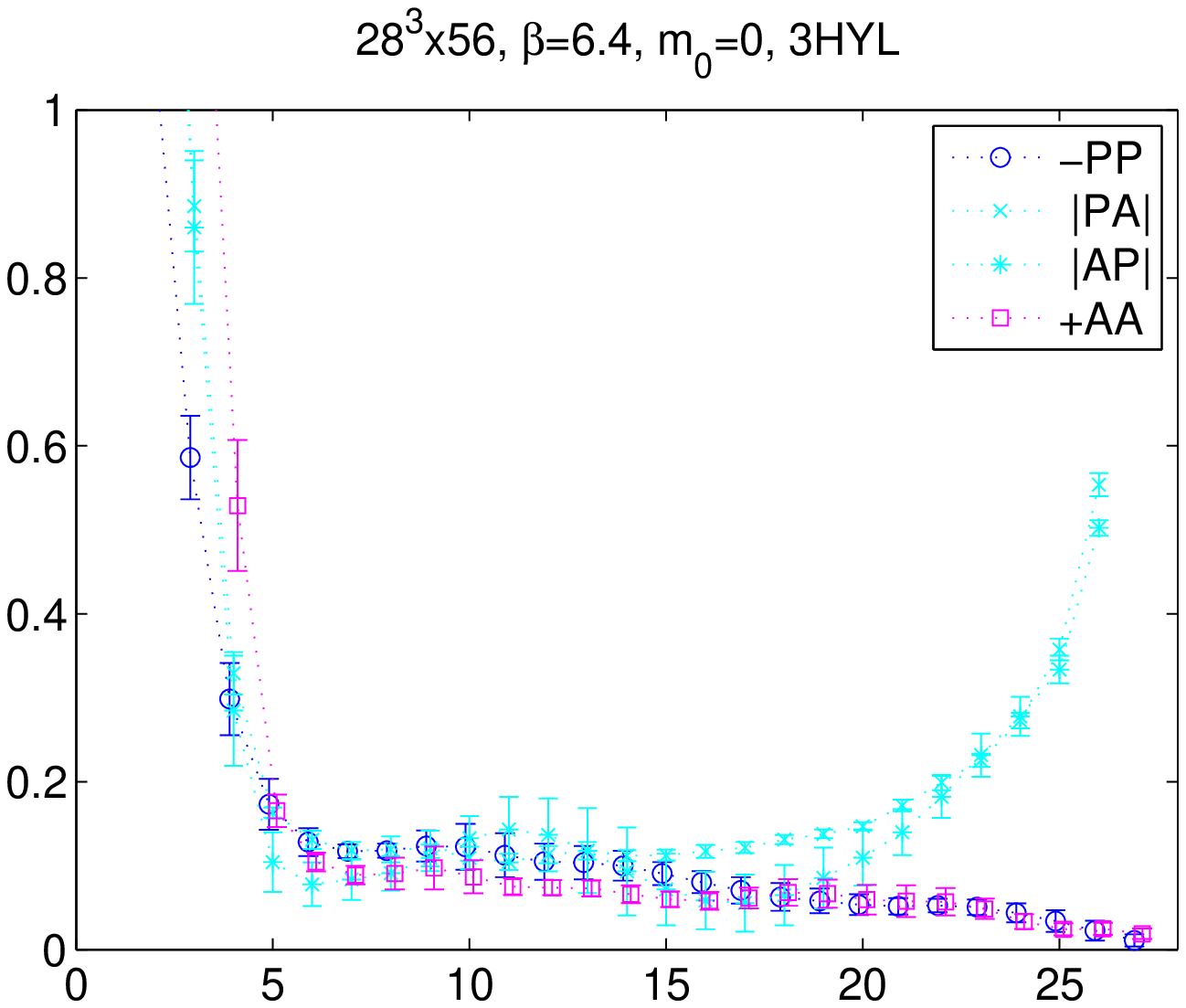,width=8.6cm}
\hspace*{-3mm}
\epsfig{file=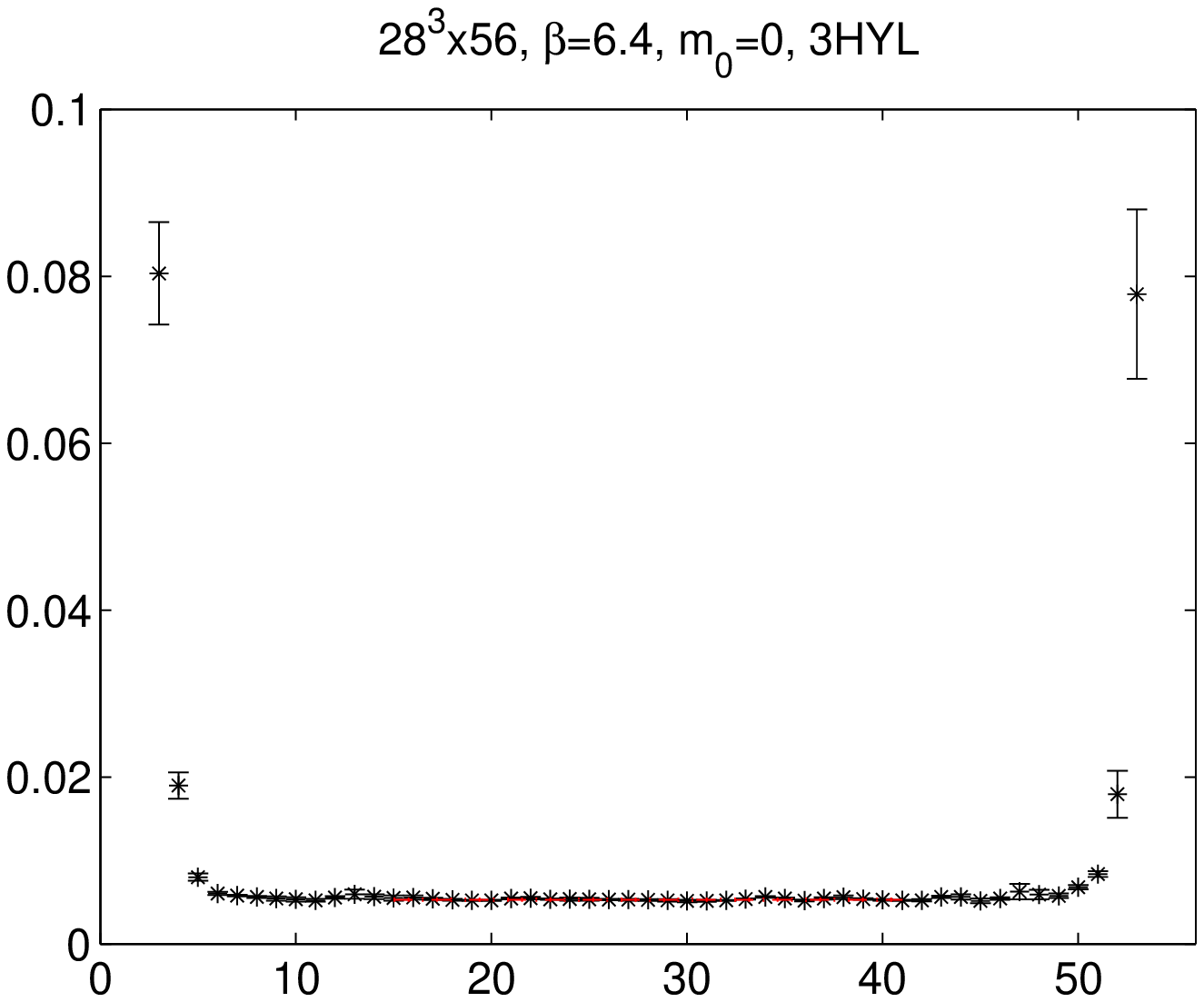,width=8.6cm}
\hspace*{-4mm}
\vspace*{-4mm}
\caption{\sl Same as in Fig.\,\ref{fig5}, but with 3\,HYL smearing steps.}
\label{fig6}
\end{figure}

An illustration of the observables studied is given in Fig.\,\ref{fig5},
based on the run at $\be=6.4$ with the 3\,LOG action.
Using both $\psb\gaf\ps$ and $\psb\ga_4\gaf\ps$ as interpolating fields the
direct and crossed correlators have been calculated, labeled as ``PP'', ''AA''
and ``PA/AP'', respectively.
The individual correlators and the pertinent ensemble averages are shown in the
first and second panel.
The third plot indicates the resulting effective masses for $\Mpi$.
The pion mass seems to be around $\Mpi\simeq0.17a^{-1}=650\MeV$, but the box is
not sufficiently long to see a good plateau.
Finally, the fourth plot contains the plateau of the PCAC mass, the latter
being defined as
\beq
am^\mr{PCAC}={(\pad_4^{}\!+\!\pad_4^*)\<A(x)P(0)\>\ovr4\<P(x)P(0)\>}
\;.
\eeq
In Fig.\,\ref{fig6} the same observables are shown (again at $\be=6.4$) for the
action with 3\,HYL steps.
Evidently, with $am_0=0$ kept fixed, the pion is much lighter now.
By comparing the first two plots to their counterparts in Fig.\,\ref{fig5} one
sees that the correlators are much fuzzier now.
In consequence, the effective mass plot for $\Mpi$ does not show a good plateau
at all.
Luckily, the PCAC mass is still easy to determine, and this is sufficient for
the present investigation.

In Tab.\,\ref{tab1} the values of the residual mass $am^\mr{PCAC}$ at $am_0=0$
are summarized for 13 fermion actions at 5 couplings.
The same (perturbatively equivalent) smearing parameters have been used as in
Figs.\,\ref{fig1} and \ref{fig2}, since we want to match on a common
perturbative prediction (see below); results should not be interpreted as to
give a fair comparison of the smearing recipes per se.
Whether actions with $n_\mr{iter}=3...7$ suffer from ``strong delocalization''
effects (read: have bad scaling properties in some observables) is, at this
moment, not clear.
In state-of-the-art phenomenological studies one has several lattice spacings
and thus means to check.
The 7\,HYL line has been included to indicate that there is no sign of a
saturation of $am_\mr{res}$ versus $n_\mr{iter}$ in the range studied.
The main message from Tab.\,\ref{tab1} is rather encouraging; the LOG
smearing defines a clover action with surprisingly good chiral symmetry
properties.
So far residual masses $am_\mr{res}\!\simeq\!10^{-3}$ have only been
achieved with domain-wall fermions \cite{Allton:2007hx}.


\begin{figure}[t]
\epsfig{file=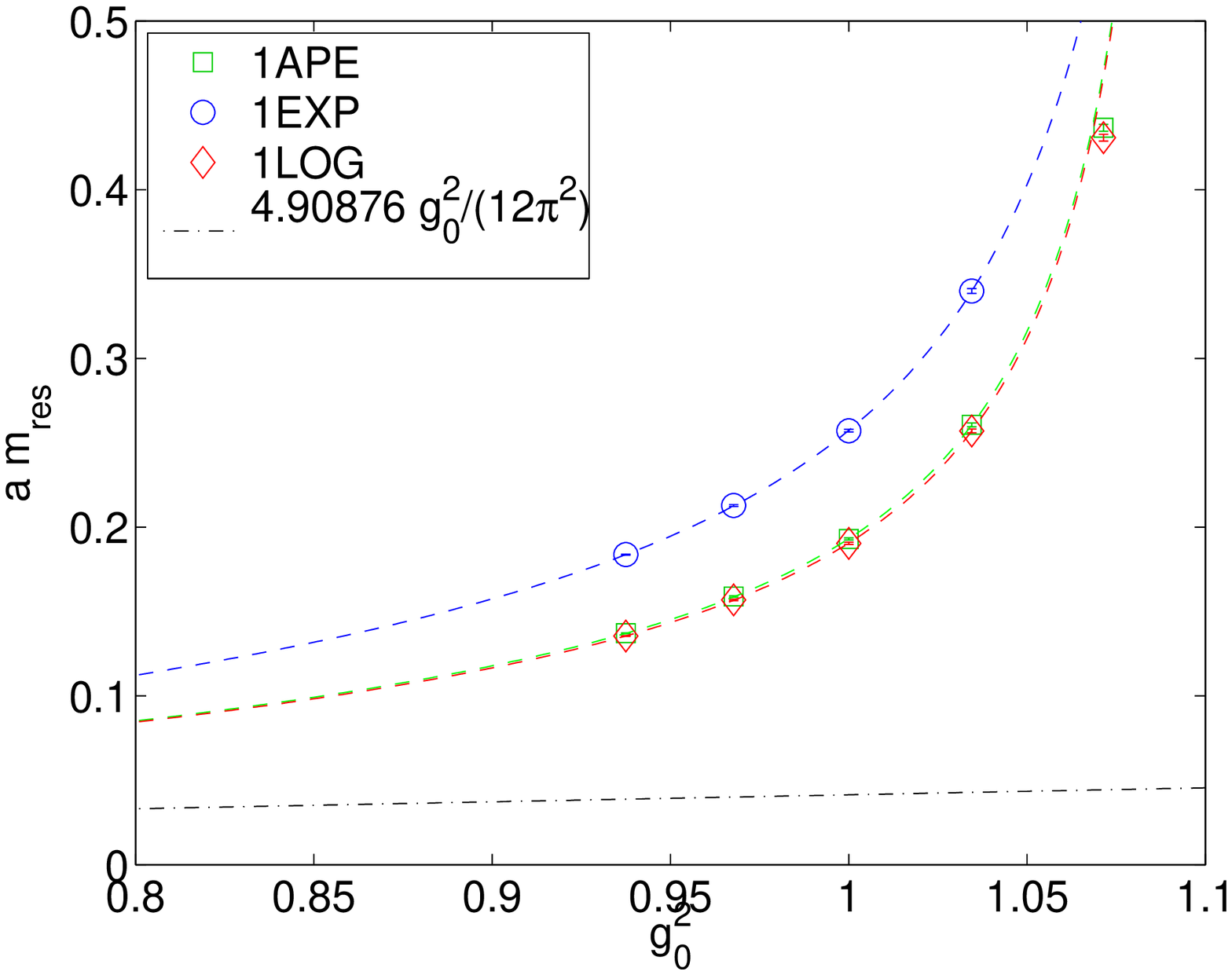,width=8.2cm}\hfill
\epsfig{file=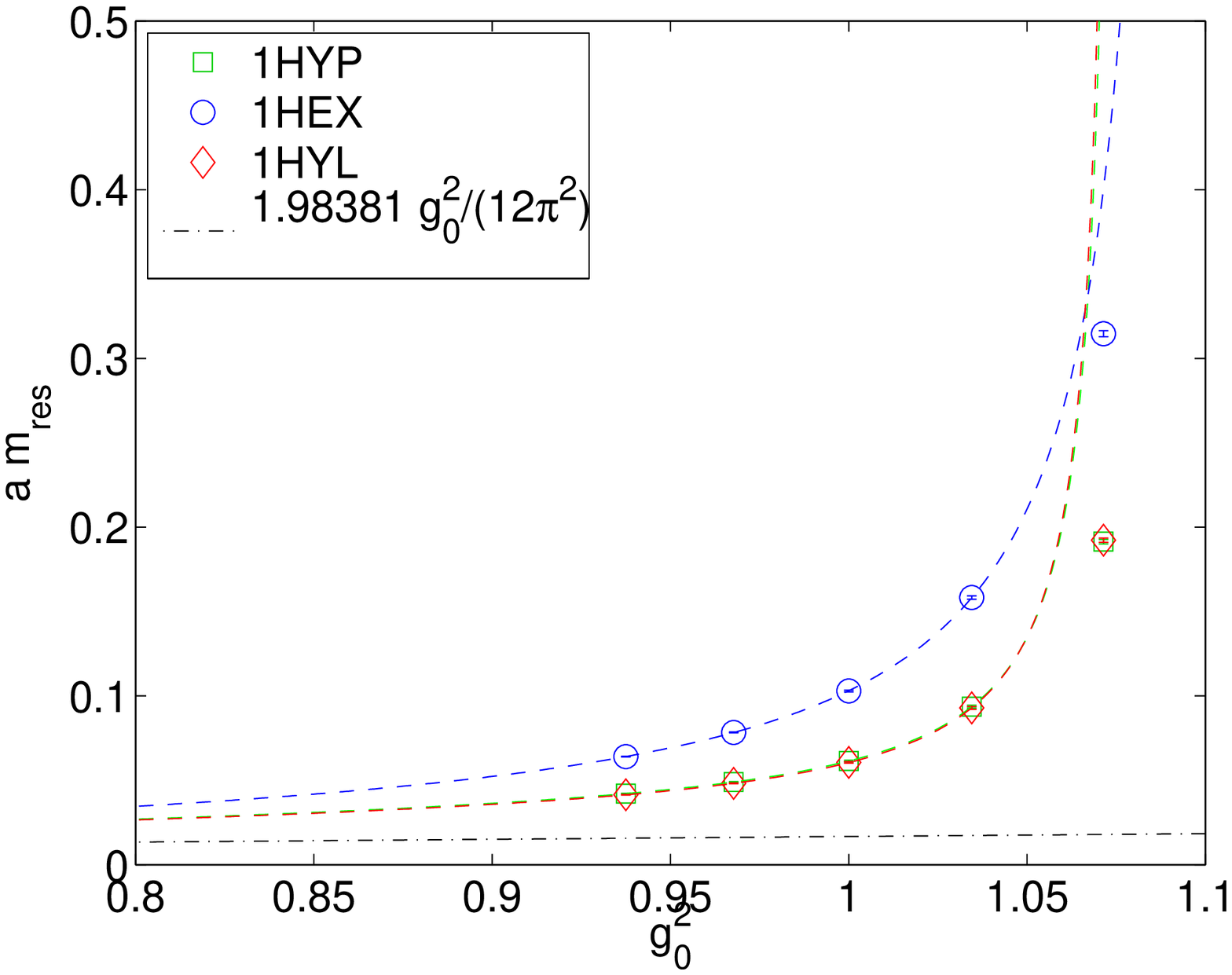,width=8.2cm}
\vspace*{-4mm}
\caption{\sl Graphical representation of the results of Tab.\,\ref{tab1} for
$n_\mr{iter}=1$ without (left) and with (right) hypercubic nesting. The common
1-loop prediction for each panel is shown as a straight dash-dotted line and is
used as a constraint in fitting the data with $\be\geq5.8$ to the ansatz
(\ref{m_res_fit}).}
\label{fig7}
\end{figure}

\begin{table}[t]
\centering
\begin{tabular}{|c|crcc|}
\hline
       &  $c_0$  &$c_1\quad$&$c_2$ &  $c_3$ \\
\hline
1\,APE &[4.90876]& 0.40140&-0.95201&-0.90361\\
1\,HYP &[1.98381]&-0.05780&-0.68221&-0.92911\\
1\,EXP &[4.90876]& 1.15872&-1.51175&-0.89583\\
1\,HEX &[1.98381]&-0.03521&-0.43517&-0.91397\\
1\,LOG &[4.90876]& 0.39395&-0.95340&-0.90421\\
1\,HYL &[1.98381]&-0.09322&-0.65284&-0.92974\\
\hline
3\,APE &[0.77096]& 0.27440&-0.74716&-0.92960\\
3\,EXP &[0.77096]& 0.64259&-0.90602&-0.92456\\
3\,LOG &[0.77096]& 0.25769&-0.72616&-0.92955\\
\hline
\end{tabular}
\caption{\sl Summary of the coefficients in the fit (\ref{m_res_fit}) for the 9
actions where $c_0$ is known in PT.}
\label{tab2}
\end{table}

With Tab.\,\ref{tab1} the main test has been completed, but it is still
interesting to compare the data to the prediction from 1-loop fat-link
perturbation theory.
To lowest order the residual mass $am_\mr{res}$ relates to the frequently
quoted critical mass $am_\mr{crit}$ through
\beq
m_\mr{res}={|m_\mr{crit}|\ovr Z_A}\,{Z_P\ovr Z_S}
\eeq
but the perturbative expansion $Z_X=1+O(g_0^2)$ means that at leading order
there is no difference between $am_\mr{res}$ and $am_\mr{crit}$.
Thus we may directly take the result from \cite{CDH} for our parameters
\beq
am_\mr{res}={g_0^2\ovr16\pi^2}C_F S
\;=\;
{g_0^2\ovr12\pi^2}
\left\{
\begin{array}{cc}
4.90876&(\mbox{1\,APE/EXP/LOG},\;c_\mr{SW}\!=\!1)\\
1.98381&(\mbox{1\,HYP/HEX/HYL},\;c_\mr{SW}\!=\!1)\\
0.77096&(\mbox{3\,APE/EXP/LOG},\;c_\mr{SW}\!=\!1)
\end{array}
\right.
\label{m_res_pt}
\eeq
and confront it with the data.
In Fig.\,\ref{fig7} the data with $n_\mr{iter}=1$ are plotted against
$g_0^2=6/\be$.
Since we are outside the regime where (\ref{m_res_pt}) is adequate it is
natural to use the rational ansatz
\beq
am_\mr{res}={g_0^2\ovr12\pi^2}
\,c_0\,
{1+c_1g_0^2+c_2g_0^4\ovr1+c_3g_0^2}
\label{m_res_fit}
\eeq
to fit the data, where the perturbative constraint (\ref{m_res_pt}) is built
in by setting $c_0=S$.
The results of the fits to the data at $\be\geq5.8$ are summarized in
Tab.\,\ref{tab2}.
With these values in hand, the residual mass for any of these 9 actions can be
accurately predicted in the range $\be\geq5.8$.

\begin{table}[t]
\centering
\begin{tabular}{|c|ccccc|}
\hline
($\be,L/a$)&($5.6,08$)&($5.8,12$)&($6.0,16$)&($6.2,20$)&($6.4,28$)\\
\hline
1\,LOG     &0.9914(46)&0.9421(41)&1.0223(34)&1.1584(32)&1.3203(28)\\
1\,HYL     &0.4425(30)&0.3406(30)&0.3254(26)&0.3559(17)&0.4039(17)\\
\hline
3\,LOG     &0.4538(33)&0.3146(35)&0.2642(30)&0.2632(20)&0.2794(29)\\
3\,HYL     &0.1601(25)&0.0906(33)&0.0590(36)&0.0502(19)&0.0515(14)\\
\hline
7\,HYL     &0.0698(18)&0.0401(22)&0.0215(25)&0.0150(18)&0.0135(14)\\
\hline
\end{tabular}
\caption{\sl Same as Tab.\,\ref{tab1}, but converted to $r_0$ units, based on
the formula for $r_0/a$ given in \cite{Necco:2001xg}.}
\label{tab3}
\vspace*{5mm}
\begin{tabular}{|c|ccccc|}
\hline
($\be,L/a$)&($5.6,08$)&($5.8,12$)&($6.0,16$)&($6.2,20$)&($6.4,28$)\\
\hline
1\,LOG     &  388(18) &  381(17) &  424(14) &  489(13) &  565(12) \\
1\,HYL     &  165(11) &  131(12) &  129(10) &  144(07) &  166(07) \\
\hline
3\,LOG     &  166(12) &  119(13) &  103(12) &  105(08) &  113(11) \\
\hline
\end{tabular}
\caption{\sl Same as Tab.\,\ref{tab3}, but converted to MeV and
$(\overline\mr{MS},\mu\!=\!2\GeV$), based on $Z_m$ from (\ref{Z_m}).}
\label{tab4}
\end{table}

Finally it is interesting to convert the residual masses into physical units
and to $(\overline{\mr MS},2\GeV)$ conventions.
In the first step we use again the formula from \cite{Necco:2001xg}; the result
is shown in Tab.\,\ref{tab3}.
Assuming $r_0=0.5\fm$, the second step is performed by means of
$m^{\overline{\mr MS}}(\mu)=Z_m(a\mu)m^\mr{PCAC}(a^{-1})$, and we use the
1-loop perturbative prediction [as usual, up to $O(g_0^4)$ contributions]
\beq
Z_m=Z_S^{-1}=
\Big( 1-{g_0^2\ovr3\pi^2}[{z_S\ovr3}-\log(a^2\mu^2)] \Big)^{-1}=
1+{g_0^2\ovr3\pi^2}[{z_S\ovr3}-\log(a^2\mu^2)]
\label{Z_m}
\eeq
with the ingredients $z_S=4.11106$ for 1\,APE/EXP/LOG, $z_S=-1.43930$ for
3\,APE/EXP/LOG and $z_S=-0.03678$ for 1\,HYP/HEX/HYL from \cite{CDH}.
The results are shown in Tab.\,\ref{tab4}.
Assuming that $\Mpi=140\MeV$ corresponds to
$m^{\overline{\mr MS}}(2\GeV)=4\MeV$ we can now estimate the pion mass by
multiplying $140\MeV$ with the square-root of
$m^{\overline{\mr MS}}(2\GeV)/(4\MeV)$.
In fact, since $Z_m$ is so close to 1, even an estimate based on the bare PCAC
mass might be accurate to 10\% or so.
In case of the 3\,HYL action it predicts $\Mpi\simeq320\MeV$ for $\be=6.4$, in
fair agreement with the result from the effective mass plot.
In case of the 7\,HYL action the effective mass plot cannot be used as a check,
but the estimate $\Mpi\simeq160\MeV$ might still be accurate within $20\MeV$.
Note finally that -- with $m_0\!=\!0$ or $\ka\!=\!1/8$ fixed -- all of this has
been achieved in a regime (to be precise: at the edge of the regime) where we
are safe against ``exceptional configurations''.


\section{Towards full QCD with the RHMC algorithm}


With the tests of the LOG/HYL smearing in the quenched case completed, it seems
worth while to spend a first thought one how one would use this smearing in
full QCD.

The driving ingredient in a HMC algorithm is the molecular dynamics evolution
which is determined by the so-called HMC force.
Let $D_m$ be an arbitrary undoubled fermion action, implicitly dependent on the
``thin'' gauge links $U$.
The pseudofermion action is defined as
\beq
S_\mr{pf}=\<\ph|\,M^{-1/2}\,|\ph\>
=\int\ph\dag(x)\,M^{-1/2}(x,y)\,\ph(y)\;d^4\!x\,d^4\!y
\;,\quad
M=D_m\dag D_m^{}>0
\eeq
where $\ph$ denotes a boson field with the spinor and color components of a
standard Dirac flavor.
We choose $\Nf\!=\!1$ for full generality; the version for even $\Nf$ is even
simpler.
The HMC force, finally, is defined as minus the derivative of $S_\mr{pf}$ with
respect to the thin links \cite{Clark:2006wq}.
The standard way to proceed is to make use of the fact that the $p$-th order
diagonal rational approximation of $x^{-1/2}$ over the relevant spectral range
admits a partial fraction formulation \cite{Clark:2006wq}
\beq
x^{-1/2}\simeq\al_0+\sum_{k=1}^p{\al_k\ovr x+\be_k}
\eeq
with $\al_k\!>\!0$ ($k\!=\!1...p$) and $0\!<\!\be_1\!<\!...\!<\!\be_p$.
As a result, the pseudo-fermion force is given by
\beq
F_\mr{pf}=-S_\mr{pf}'=\sum_{k=1}^p\al_k\,
\<\ph|(M+\be_k)^{-1}\,M'\,(M+\be_k)^{-1}|\ph\>
\label{fat1}
\eeq
where the prime denotes the variation w.r.t.\ a single element of the gauge
field $A_\mu^a(x\!+\!\hat\mu/2)$, defined as a Gell-Mann component of
$\log(U_\mu(x))$.
In other words, what is needed to work out $S_\mr{pf}'$ is the derivative of
$M$ w.r.t.\ the gauge field.
In explicit terms this is
\beq
M'=
{dM\ovr dA_\mu^\mr{a}}=
{dD_m\dag\ovr dA_\mu^\mr{a}} D_m^{} + D_m\dag {dD_m\ovr dA_\mu^\mr{a}}=
(D_m\dag)'D_m^{}+D_m\dag(D_m^{})'
\label{fat2}
\eeq
where the derivatives $(D_m\dag)'$ and $(D_m)'$ follow as a product of the
variation of the chosen fermion action under a change of the ``fat'' links,
times the inner derivative \cite{Kamleh:2004xk}.
In the very end, the variation $dS_\mr{pf}/dA_\mu^\mr{a}(x\!+\!\hat\mu)$
combines both types%
\footnote{We use the standard definition (with $\til U$ denoting a generic fat
link)
\bdm
{d(.)\ovr dA_\mu^\mr{a}}={\pad (.)\ovr\pad A_\mu^\mr{a}}+
{\pad (.)\ovr\pad U_\nu}\,{dU_\nu\ovr dA_\mu^\mr{a}}+
{\pad (.)\ovr\pad U_\nu\dag}\,{dU_\nu\dag\ovr dA_\mu^\mr{a}}
\;,\;
{\pad (.)\ovr\pad U_\mu}=
{\pad (.)\ovr\pad\til U_\nu}\,{\pad\til U_\nu\ovr\pad U_\mu}+
{\pad (.)\ovr\pad\til U_\nu\dag}\,{\pad\til U_\nu\dag\ovr\pad U_\mu}
\;,\;
{\pad (.)\ovr\pad U_\mu\dag}=
{\pad (.)\ovr\pad\til U_\nu}\,{\pad\til U_\nu\ovr\pad U_\mu\dag}+
{\pad (.)\ovr\pad\til U_\nu\dag}\,{\pad\til U_\nu\dag\ovr\pad U_\mu\dag}
\edm
in which the partial derivative w.r.t.\ a given link picks up only that link
and not its hermitean conjugate, that is $U_\mu$ and $U_\mu\dag$ are considered
independent with regard to partial differentiation.}
of derivatives.
The former is standard, except that in the result a replacement
$U\to U^\mr{LOG}$ is needed.
The latter is specific to the chosen smearing recipe and hence deserves a
closer look.

We restrict ourselves to the ``trace-free'' logarithm (\ref{def_4}).
The product rule (\ref{product}) yields
\bea
{\pad U_\mu^\mr{LOG}(x)^\mr{c}\ovr \pad U_\nu(y)^\mr{c}}\!\!&\!\!=\!\!&\!\!
(U_\mu(x)'\!\otimes\!I_3)\!\cdot\!
{\pad\ovr\pad U_\nu(y)}
\Big\{
\exp\!\Big(
{\al\ovr2(d\!-\!1)}\!\sum_{\pm\rh\neq\mu}\!
\mr{tf\/log}[U_\rh(x)U_\mu(x\!+\!\hat\rh)U_\rh\dag(x\!+\!\hat\mu)U_\mu\dag(x)]
\Big)^\mr{c}
\Big\}
\nonumber
\\
\!&\!+\!&\!
I_3\!\otimes\!
\exp\!\Big(
{\al\ovr2(d\!-\!1)}\!\sum_{\pm\rh\neq\mu}\!
\mr{tf\/log}[U_\rh(x)U_\mu(x\!+\!\hat\rh)U_\rh\dag(x\!+\!\hat\mu)U_\mu\dag(x)]
\Big)\cdot
I_9\,\de_{\mu,\nu}\de_{x,y}
\label{deri1}
\\
{\pad U_\mu^\mr{LOG}(x)^\mr{c}\ovr \pad U_\nu\dag(y)^\mr{c}}\!\!&\!\!=\!\!&\!\!
(U_\mu(x)'\!\otimes\!I_3)\!\cdot\!
{\pad\ovr\pad U_\nu\dag(y)}
\Big\{
\exp\!\Big(
{\al\ovr2(d\!-\!1)}\!\sum_{\pm\rh\neq\mu}\!
\mr{tf\/log}[U_\rh(x)U_\mu(x\!+\!\hat\rh)U_\rh\dag(x\!+\!\hat\mu)U_\mu\dag(x)]
\Big)^\mr{c}
\Big\}
\quad\!
\label{deri2}
\eea
and analogously for the daggered fat-links.
Here we use the tensor notation where the l.h.s.\ is a $9\times9$ matrix with
$\pad U_\mu^\mr{LOG}(x)_{ij}/\pad U_\nu(y)_{kl}$ in the
$3(j\!-\!1)\!+\!i$-th row and $3(l\!-\!1)\!+\!k$-th column.
In other words, $U_\mu^\mr{LOG}(x)$ and $U_\nu(y)$ are ``reshaped'' into
$9\times1$ column vectors, and then the derivative of the $p$-th element
of the first with respect to the $q$-th element of the second appears in
position $(p,q)$.
Likewise, on the r.h.s.\ the derivative is a $9\times9$ matrix, and $U_\mu(x)$
has been ``blown up'' by tensoring with the identity such that the two
$9\times9$ matrices would multiply according to standard matrix multiplication
rules.
Some details of the underlying formalism have been collected in App.\,C.
The task is now to collect the various pieces that come from the derivative of
the exponential.
Upon applying the chain rule (\ref{chain}) for matrix functions, I obtain
\bea
{\pad\ovr\pad U_\nu(y)^\mr{c}}
\Big\{
\exp\!\Big(
{\al\ovr2(d\!-\!1)}\!\sum_{\pm\rh\neq\mu}\!
\mr{tf\/log}[U_\rh(x)U_\mu(x\!+\!\hat\rh)U_\rh\dag(x\!+\!\hat\mu)U_\mu\dag(x)]
\Big)^\mr{c}
\Big\}
&=&F_1 \cdot F_2 \cdot F_3
\\
{\pad\ovr\pad U_\nu\dag(y)^\mr{c}}
\Big\{
\exp\!\Big(
{\al\ovr2(d\!-\!1)}\!\sum_{\pm\rh\neq\mu}\!
\mr{tf\/log}[U_\rh(x)U_\mu(x\!+\!\hat\rh)U_\rh\dag(x\!+\!\hat\mu)U_\mu\dag(x)]
\Big)^\mr{c}
\Big\}
&=&F_1 \cdot F_2 \cdot F_4
\eea
where each factor
\bea
F_1&=&{\pad\exp(X)^\mr{c}\ovr\pad X^\mr{c}}
\quad\mbox{evaluated at}\quad
X={\al\ovr2(d\!-\!1)}\sum_{\pm\rh\neq\mu}
\mr{tf\/log}[U_\rh(x)U_\mu(x\!+\!\hat\rh)U_\rh\dag(x\!+\!\hat\mu)U_\mu\dag(x)]
\nonumber
\\[1mm]
F_2&=&
{\al\ovr2(d\!-\!1)}\sum_{\pm\rh\neq\mu}
{\pad\,\mr{tflog}(X)^\mr{c}\ovr\pad X^\mr{c}}
\quad\mbox{evaluated at}\quad
X=U_\rh(x)U_\mu(x\!+\!\hat\rh)U_\rh\dag(x\!+\!\hat\mu)U_\mu\dag(x)
\nonumber
\\[1mm]
F_3&=&
[U_\rh\dag(x\!+\!\hat\mu)U_\mu\dag(x)]'\!\otimes\!I_3
\cdot
{\pad(U_\rh(x)U_\mu(x\!+\!\hat\rh))^\mr{c}\ovr\pad U_\nu(y)^\mr{c}}
\nonumber
\\[1mm]
&=&
[U_\mu(x\!+\!\hat\rh)U_\rh\dag(x\!+\!\hat\mu)U_\mu\dag(x)]'\!\otimes\!I_3
\,\de_{\nu,\rh}\de_{x,y} +
[U_\rh\dag(x\!+\!\hat\mu)U_\mu\dag(x)]'\!\otimes\!U_\rh(x)
\,\de_{\mu,\nu}\de_{x+\hat\rh,y}
\nonumber
\\[1mm]
F_4&=&\,
I_3\!\otimes\![U_\rh(x)U_\mu(x\!+\!\hat\rh)]
\cdot
{\pad(U_\rh\dag(x\!+\!\hat\mu)U_\mu\dag(x))^\mr{c}\ovr\pad U_\nu\dag(y)^\mr{c}}
\nonumber
\\[1mm]
&=&\,
U_\mu\dag(x)'\!\otimes\![U_\rh(x)U_\mu(x\!+\!\hat\rh)]
\,\de_{\nu,\rh}\de_{x+\hat\mu,y} +
I_3\!\otimes\![U_\rh(x)U_\mu(x\!+\!\hat\rh)U_\rh\dag(x\!+\!\hat\mu)]
\,\de_{\mu,\nu}\de_{x,y}
\nonumber
\eea
is a $9\times9$ matrix, and similarly for the daggered fat-links.
If the LOG smearing is iterated, several factors (\ref{deri1}, \ref{deri2})
are lined up, each one evaluated with the appropriate argument.

For completeness, let us consider fat-link clover fermions, although this
has been done before \cite{Kamleh:2004xk}.
Introducing, for each term in (\ref{fat1}), $|\et_k\>=(M+\be_k)^{-1}|\ph\>$ and
$|\ze_k\>=D_m|\et_k\>$ we have
\beq
F_\mr{pf}=-S_\mr{pf}'=
\sum_{k=1}^p\al_k\Big[\<\et_k|(D_m\dag)'|\ze_k\>+\<\ze_k|(D_m^{})'|\et_k\>\Big]
\eeq
where we enjoy the benefits of dealing with a scalar function of a scalar
argument (such that the full apparatus of App.\,C is not needed).
From a glimpse at (\ref{def_wils}) it follows that
\beq
{\pad S_\mr{pf, W}\ovr \pad U_\mu^\mr{LOG}(x)}=
{1\ovr2}\sum_{k=1}^p\al_k
\mr{Tr}_\mr{spin}\Big[
\et_k\dag(x\!+\!\hat\mu)(\ga_\mu+I)\ze_k(x)-
\ze_k\dag(x)(\ga_\mu-I)\et_k(x\!+\!\hat\mu)
\Big]
\eeq
where the layout of the color structure on the r.h.s.\ is just adapted to
whichever convention on the l.h.s.\ is chosen.
Similarly, from a glimpse at (\ref{def_clov}) it follows that
\beq
{\pad S_\mr{pf, SW}\ovr \pad U_\mu^\mr{LOG}(x)}={c_\mr{SW}\ovr2}
\sum_y
\sum_{k=1}^p\al_k
\mr{Tr}_\mr{spin}\Big[
\et\dag(y)\si_{\ka\la}{\pad F_{\ka\la}^\mr{LOG}(y)\ovr\pad U_\mu^\mr{LOG}(x)}
\ze(y)+
\ze\dag(y)\si_{\ka\la}{\pad F_{\ka\la}^\mr{LOG}(y)\ovr\pad U_\mu^\mr{LOG}(x)}
\et(y)
\Big]
\;.
\eeq

Upon putting the various expressions together, one has an analytic expression
for the HMC force of a LOG-filtered clover (staggered, overlap, etc.) action.
The generalization to the HYL smearing (\ref{def_HYL}) follows by lining up
several factors of (\ref{deri1}, \ref{deri2}) [with restricted sums].


\section{Modified gauge action and topological charge density}


In the numerical investigations of this note the traditional Wilson gauge
action
\beq
S_G={2\Nc\ovr g_0^2}\sum_{x,\mu<\nu}
\Big\{
1-{1\ovr\Nc}\mr{Re}\,\mr{Tr}(U_{\mu\nu}(x))
\Big\}
\label{def_sold}
\eeq
has been used.
However, with the technical means to compute the matrix logarithm of the
plaquette $U_{\mu\nu}(x)=U_\mu(x)U_\nu(x\!+\!\hat\mu)
U_\mu\dag(x\!+\!\hat\nu)U_\nu\dag(x)$ in hand, two modifications are possible.

The first one concerns a constraint, to be added to whichever gauge action is
used.
Clearly, the discussion around (\ref{def_1}\,-\,\ref{def_4}) would have been
much simpler, if we could be sure that there is no lattice in our ensemble
with a single $U_{\mu\nu}(x)$ which has a non-trace-free principal logarithm.
It can be shown that a sufficient condition for $\mr{Tr}\log(U_{\mu\nu}(x))=0$
is $\mr{Re}\,\mr{Tr}(U_{\mu\nu}(x))>-1$.
This suggests adding a piece which penalizes all plaquettes with
$\mr{Re}\,\mr{Tr}(U_{\mu\nu}(x))<0$, without affecting those with
$\mr{Re}\,\mr{Tr}(U_{\mu\nu}(x))\geq0$.
Alternatively, one may directly test for
$\mr{Tr}\log(U_{\mu\nu}(x))=\pm2\pi\ri$
and assign, in such a case, a prohibitive extra action.
This point is summarized through
\beq
S_G\;\longrightarrow\;S_G+
\left\{
\begin{array}{ll}
+{1\ovr\ep}\sum_{x,\mu<\nu}\th(-r)\exp(1/r)
&\quad\mr{with}\quad r=\mr{Re}\,\mr{Tr}(U_{\mu\nu}(x))\\[2mm]
-{1\ovr\ep}\sum_{x,\mu<\nu}[\mr{Tr}\log(U_{\mu\nu}(x))]^2
&\quad\mr{with}\quad \ep\ll1\;.
\end{array}
\right.
\eeq

The second one concerns a modification of the ``bulk'' piece of the gauge
action.
Recalling that the rationale behind Wilson's choice (\ref{def_sold}) is to come
up with a simple recipe to produce a term $F_{\mu\nu}^2$ from
$U_{\mu\nu}=\exp(\ri g_0a F_{\mu\nu})=
1+\ri g_0a^2 F_{\mu\nu}-{1\ovr2}g_0^2a^4 F_{\mu\nu}^2-...$, it is tempting to
generate $F_{\mu\nu}$ directly from the logarithm of the plaquette and square
it explicitly.
With $L_{\mu\nu}(x)=\mr{tflog}\,U_{\mu\nu}(x)$ and
$\bar{L}_{\mu\nu}(x)={1\ovr4}[L_{\mu\nu}(x)+L_{\mu\nu}(x\!-\!\hat\mu)
+L_{\mu\nu}(x\!-\!\hat\nu)+L_{\mu\nu}(x\!-\!\hat\mu\!-\!\hat\nu)]$ at hand,
it appears that
\beq
S_G[U]=a^4\sum_{x,\mu<\nu}
\mr{Tr}[F_{\mu\nu}(x)F_{\mu\nu}(x)]
\doteq
-{1\ovr g_0^2}\sum_{x,\mu<\nu}
\mr{Tr}[\bar{L}_{\mu\nu}(x)\bar{L}_{\mu\nu}(x)]
\label{def_snew}
\eeq
would be a rather natural choice.
Likely this gauge action defines a theory with good scaling properties, good
rotational symmetry and nice overlap with the perturbative regime, as it is
rather similar to the Manton action \cite{Manton:1980ts}.
On the other hand, tunneling of the topological charge needs to be
investigated, and clearly (\ref{def_snew}) is not as easy to simulate as the
Wilson action (\ref{def_sold}).

Of course, the same modification may be applied to $F\tilde{F}$, too.
This yields the expression
\beq
q[U]=
{a^4\ovr32\pi^2}\sum_{x,\mu\nu\rh\si}
\ep_{\mu\nu\rh\si}
\mr{Tr}[F_{\mu\nu}(x)F_{\rh\si}(x)]
\doteq
-{1\ovr32\pi^2 g_0^2}\sum_{x,\mu\nu\rh\si}
\ep_{\mu\nu\rh\si}
\mr{Tr}[\bar{L}_{\mu\nu}(x)\bar{L}_{\rh\si}(x)]
\eeq
for the bare topological charge density.
Unlike in the previous case, we do not expect this to be much better than the
usual clover-leaf expression for the topological charge density, since in that
case it is standard to perform an averaging of the anti-hermitean part of the
plaquette.

In the same spirit, it seems natural to define the clover action through
\beq
D_\mr{SW}(x,y)=D_\mr{W}(x,y)
-{c_\mr{SW}\ovr2\ri g_0}\sum_{\mu<\nu}\si_{\mu\nu}\bar{L}_{\mu\nu}\;\de_{x,y}
\eeq
instead of (\ref{def_clov}) [where $F_{\mu\nu}$ follows from the anti-hermitean
part of the same 4 plaquettes which enter $\bar{L}_{\mu\nu}$].
Again, at least with some filtering, this change will be rather insignificant.


\section{Summary and outlook}



The purpose of this paper has been to define and test a smearing which is
applicable, in the context of the HMC algorithm, to studies of full QCD.
The testing has been restricted to pure gauge theory observables and to clover
fermions in quenched QCD, but it is clear that the main lesson concerns the
fermion formulation per se.

The key idea behind this LOG smearing is to do the ``averaging'' in the Lie
algebra, such that one would end up with a smeared (``fat'') link which is
naturally in the original gauge group.
This is a clear advantage if one intends to use such fat-links in observables
which are sensitive to the structure of the gauge group (e.g.\ Polyakov loops),
but it might also simplify calculations in fermionic systems (e.g.\ by
providing a more direct connection to perturbation theory).
Whenever a single LOG smearing (\ref{def_4}) is not enough, it seems advisable
to use the hypercubic nesting to define the HYL smeared links (\ref{def_HYL}).

Tests in the pure gauge theory have shown that the new LOG smearing is at least
as effective at suppressing UV fluctuations as known alternatives.
From a practical viewpoint the LOG smearing seems to have an advantage in
having a rather broad ``near-optimal'' region, i.e.\ good results do not
require any fine-tuning of the smearing parameter.


An important point is that the LOG/HYL smearing yields a fat-link which is
differentiable with respect to the thin links it is built from.
This makes it a useful ingredient in defining a UV-filtered fermion action
which can be used to study full QCD with the HMC algorithm.
The tests in the quenched theory have shown quite convincing results, though
most of the good properties of such actions are not specific to the LOG/HYL
recipes.
These smearings are efficient in the sense that modest parameters and/or
iteration counts lead to clover-fermions with quite acceptable chiral
properties.
It seems that, upon iterating the smearing, arbitrarily small residual masses
can be attained.
The smallest value found, $am_\mr{res}=0.0014(1)$ with 7\,HYL steps at
$\be=6.4$, is in a regime which, without smearing, can only be accessed by
domain-wall fermions \cite{Allton:2007hx}.
It seems noteworthy that these results have been obtained with non-negative
bare mass where one is immune against ``exceptional configurations''.
An obvious continuation of the current work would be to compare the scaling
properties of such actions against mildly filtered or unfiltered (``thin
link'') varieties.
From Fig.\,\ref{fig6} one sees that the
correlators of highly smeared actions at small quark mass become sensitive to
individual instantons and/or other topological objects and get rather fuzzy.
Likely, this is the flipside of any fermion action with good chiral properties.
The present tests have focused on the residual mass, as this is the simplest
observable sensitive to chiral symmetry breaking.
Perhaps a study of the spectral gap of the hermitean Wilson/clover operator
with LOG/HYL-filtering would be useful, too.

Wilson fermions have been out of fashion for over a decade, since in the
original formulation chiral symmetry is broken in a rather severe way, and
standard Symanzik improvement alone does not bring a major change in this
respect.
It seems almost guaranteed that, upon combining a clover term with a modest
amount of link-fattening (e.g.\ two HYL steps), the clover action is rendered a
competitive choice for pushing towards $\Mpi=140\MeV$ in full QCD.

\bigskip

\noindent
{\bf Acknowledgments}:
\newline
I would like to thank Anna Hasenfratz for raising my interest in differentiable
fat-link recipes and Ferenc Niedermayer for a series of most enjoyable
discussions.
Computations have been performed on stand-alone PCs with 4GB memory.
I'm indebted to Markus Moser and Matthias Nyfeler for creating temporary swap
space.
This work was supported by the Swiss NSF.

\clearpage

\appendix


\section{Matrix exponential, matrix logarithm and more}


Among the main concepts for computing matrix valued functions on the computer,
viz.
\begin{itemize}
\vspace{-2pt}
\itemsep-4pt
\item[({\sl i}\,)]
eigenvalue and eigenfunction based routines
\item[({\sl ii}\,)]
clever use of the Cayley-Hamilton theorem
\item[({\sl iii}\,)]
iterative methods
\vspace{-2pt}
\end{itemize}
a combination of ({\sl i}\,) and ({\sl ii}\,) often yields the most efficient
implementation.
Nonetheless, it turns out that iterative methods --~though originally designed
to deal with medium-size full and huge sparse systems~-- represent a good
choice already for $3\!\times\!3$ matrices, since they are easy to implement,
numerically stable, and still reasonably fast in terms of CPU time.
Below a robust implementation of the matrix exponential and the matrix
logarithm is described.


\subsection{Matrix exponential with iterative methods}

The matrix exponential of an arbitrary $m\!\times\!m$ matrix $A$ is defined
through (with $A^0\!=\!I$)
\beq
\exp(A)=\sum_{k=0}^\infty{A^k\ovr k!}
\eeq
since this series has infinite convergence range.
A general recipe to speed up the convergence is known as the ``scaling and
squaring'' method.
It is based on the trivial observation
\beq
\exp(A)=[\exp(A/2^n)]^{2^n}
\eeq
which is paraphrased as ``take $\exp(A/2)$ and square it, and nest this $n$
times''.
The point is, of course, that $A/2$ has smaller eigenvalues (or singular
values) than $A$, making the power series for $A/2$ converge faster than the
one for $A$.
Clearly, theorems may be derived for the optimal choice of $n$.
However, in practice it is often sufficient to start with a reasonable
$n_\mr{min}$, and to increase it until the result is stable.
An implementation of the estimator $E_n(.)$, in which also the precision of the
rational approximation $R_n(.)$ is gradually increased, reads
\begin{tabbing}
abstand\=for \=for \kill
       \>$E_{n_\mr{min}-1}=I$\\
       \>for $n=n_\mr{min}:\infty$\\
       \>    \>$S_n=A/2^n$,
               $R_n=[\sum_{k=0}^{2n}{1\ovr k!}(S_n/2)^k]
                    [\sum_{k=0}^{2n}{1\ovr k!}(-S_n/2)^k]^{-1}$,
               $E_n=(R_n)^{2^n}$\\
       \>    \>if $||E_n-E_{n-1}||< \ep$ exit\\
       \>end \> \hspace{14cm} \stepcounter{equation}(\theequation)
       {\newcounter{alg_exp}\setcounter{alg_exp}{\theequation}}
\end{tabbing}
where $||.||$ is any matrix norm and $\ep$ is usually chosen slightly larger
than the machine precision.
An aside: In a HMC algorithm one needs to calculate $\exp(A)$ for
$A\!\in\!su(3)$.
In order to guarantee reversibility, one demands that $E(-A)$ is the
exact inverse of $E(A)$, even with a low grade $E(.)$.
Our representation $R_n(A)=[I+A/2+...][I-A/2+...]^{-1}$, and thus $E_n(.)$, is
of this form.


\subsection{Matrix logarithm with iterative methods}

The matrix logarithm of a non-singular $m\!\times\!m$ matrix $A$ is, in
general, not unique.
However, if $A$ has no eigenvalues on the closed negative real axis, then
there is one solution to the equation $\exp(X)=A$ for which all eigenvalues
$x_i$ of $X$ satisfy $-\pi<\mr{Im}(x_i)<\pi$ ($i=1,...,m$), and this solution
is called the principal logarithm of $A$ and denoted $\log(A)$.
Since there is no power series representation of $\log(A)$ which converges in
the entire cut plane, the usefulness of an ``inverse scaling and squaring''
approach \cite{KenneyLaub}, based on the identity
\beq
\log(A)=\log(A^{1/2^n})\,\,2^n
\eeq
is evident.
It may be paraphrased as ``take the square root, compute the logarithm, double
it, and nest this $n$ times''.
Hence this approach leaves us with the problem to compute, with high precision,
the principal square root of an arbitrary matrix $A$, but the gain is that the
argument of the innermost (``reduced'') logarithm is, for large enough $n$,
close to the identity.

For the square root it is convenient to use an iterative method, too.
The Newton iteration
\beq
X_{k+1}={1\ovr2}(X_k+AX_k^{-1})\;,\qquad X_0=A
\eeq
with $\lim_{k\to\infty}X_k=A^{1/2}$ has good theoretical properties, but its
poor numerical stability renders it useless in practice \cite{Higham1}.
Fortunately, the Denman-Beavers iteration \cite{DenmanBeavers} for $A$ with a
spectrum contained in the cut complex plane (i.e.\ with no non-positive real
eigenvalue)
\bea
Y_{k+1}&=&{1\ovr2}(Y_k+Z_k^{-1})\;,\qquad Y_0=A
\nonumber\\
Z_{k+1}&=&{1\ovr2}(Z_k+Y_k^{-1})\;,\qquad Z_0=I
\label{DB_orig}
\eea
has the quadratic convergence pattern
\bdm
\lim_{k\to\infty}Y_k=A^{1/2}\;,\qquad\lim_{k\to\infty}Z_k=A^{-1/2}
\edm
and is stable against round-off errors \cite{Higham1}.
Throughout this appendix it is understood that the inverse is determined via a
Gauss elimination (with pivoting whenever needed) and hence exact to machine
precision.
In particular, if one is interested only in $A^{1/2}$ (or $A^{-1/2}$), the
product form of the Denman-Beavers iteration \cite{Higham1}
\bea
Y_{k+1}&=&{1\ovr2}Y_k(I+M_k^{-1})
\;,\qquad Y_0=A
\nonumber\\
Z_{k+1}&=&{1\ovr2}(I+M_k^{-1})Z_k
\;,\qquad Z_0=I
\nonumber\\
M_{k+1}&=&{1\ovr4}(M_k+2I+M_k^{-1})
\;,\qquad M_0=A
\label{DB_prod}
\eea
with the line for $Z_{k+1}$ (or for $Y_{k+1}$) omitted and the quadratic
convergence pattern from above (plus $\lim_{k\to\infty}M_k=I$, and hence
$||M_k-I||<\ep$ as a natural exit criterion) proves superior.
The reason is that only one inverse is required and the matrix to be inverted
is, after a few steps, so close to the identity that no pivoting is needed
(which is primarily useful when dealing with large matrices on massively
parallel systems).
In general, one will also use the scaling idea to accelerate this product form
of the Denman-Beavers iteration (see \cite{Higham1} for details), but in our
case $\det(A)\!=\!1$ implies that no scaling is needed in this step.
Note that (\ref{DB_orig}, \ref{DB_prod}) are non-standard in the sense that the
matrix $A$ does not show up in the iteration step.

For the second ingredient, the logarithm of a matrix close to the identity, one
option is to utilize a diagonal rational approximation of $\log(1-x)$, for
instance one of
\bea
r_{11}(x)&=&{-2x\ovr2-x}
\nonumber\\
r_{22}(x)&=&{-6x+3x^2\ovr6-6x+x^2}
\nonumber\\
r_{33}(x)&=&{-60x+60x^2-11x^3\ovr60-90x+36x^2-3x^3}
\nonumber\\
r_{44}(x)&=&{-2940x+4410x^2-1820x^3+175x^4\ovr2940-5880x+3780x^2-840x^3+42x^4}
\nonumber
\eea
with $X=1-A$.
Here it is understood that the numerator and denominator are at least evaluated
by means of the Horner scheme, but the larger $m$ in $r_{mm}(.)$, the more it
pays to use a sophisticated method \cite{Higham2}.
An alternative representation, which is easier to implement, is
\beq
\log(Z)=
2\Big\{
         (Z\!-\!1)(Z\!+\!1)^{-1}+
{1\ovr3}[(Z\!-\!1)(Z\!+\!1)^{-1}]^3+
{1\ovr5}[(Z\!-\!1)(Z\!+\!1)^{-1}]^5+...
\Big\}
\label{def_logred}
\eeq
which is appropriate for $\mr{Re}(z_i)\!\geq\!0, z_i\!\neq\!0$ $(\forall i)$.
In our situation this condition is met after the first square root has been
evaluated.
An aside: If the inversion is exact, upon feeding (\ref{def_logred}) with
$Z^{-1}$ one obtains exactly the negative of what one gets with $Z$, in spite
of the truncation.

Putting things together, one may either opt for the more elaborate algorithm of
Ref.\,\cite{Higham1}
\begin{tabbing}
abstand\=for \=for \=for \kill
       \>$Y^{(0)}=A$\\
       \>for $i=1:\infty$\\
       \>    \>choose $k_i$ according to Theorem\,5.1 of Ref.\,\cite{Higham1}\\
       \>    \>$M_0=Y^{(i-1)}$, $Y_0=Y^{(i-1)}$\\
       \>    \>for $k=0:k_i-1$\\
       \>    \>    \>$Y_{k+1}=Y_k(I+M_k^{-1})/2$,
                     $M_{k+1}=(2I+M_k+M_k^{-1})/4$\\
       \>    \>end\\
       \>    \>$M^{(i)}=M_{k_i}$, $Y^{(i)}=Y_{k_i}$, $n=i$\\
       \>    \>if $||I-Y^{(k)}||\leq{1\ovr2}$ and (7.5) of Ref.\,\cite{Higham1}
               satisfied with $m_n\leq8$ exit\\
       \>end\\
       \>form the Pade approximant $X=r_{m_n m_n}(I-Y^{(n)})$\\
       \>rescale and correct through $X=X2^n-\sum_{i=1}^n(M^{(i)}-I)2^{i-1}$
             \> \hspace{14cm} \stepcounter{equation}(\theequation)
\end{tabbing}
which uses a Pade approximant of maximum order [8/8] for the reduced logarithm,
or one may decide to stay with the simpler algorithm (starting again with a
fixed $n_\mr{min}\!\geq\!2$)
\begin{tabbing}
abstand\=for \=for \=for \kill
       \>$L_{n_\mr{min}-1}=0$\\
       \>let $S_{n_\mr{min}-1}$ be the result of $n_\mr{min}\!-\!1$ nestings of
         (\ref{DB_prod}), each one run to machine precision\\[1mm]
       \>for $n=n_{\mr{min}}:\infty$\\
       \>    \>let $S_n$ be the result of (\ref{DB_prod}) with $A=S_{n-1}$,
               again run to machine precision\\
       \>    \>$R_n=2\sum_{k=1}^{4n}{1\ovr2k-1}[(S_n-I)(S_n+I)^{-1}]^{2k-1}$\\
       \>    \>$L_n=2^nR_n$\\
       \>    \>if $||L_n-L_{n-1}||< \ep$ exit\\
       \>end \> \hspace{14cm} \stepcounter{equation}(\theequation)
       {\newcounter{alg_log}\setcounter{alg_log}{\theequation}}
\end{tabbing}
which is similar in spirit to (\arabic{alg_exp}) for the matrix exponential.
In the second case one may choose to increment $n$ by more than one unit.
In these implementations no property of $SU(3)$ matrices has been exploited.
Accordingly, if the argument is known to be a special unitary matrix, it is
useful to check that the matrix logarithm is anti-hermitean and traceless
modulo $2\pi\ri$.


\subsection{Higher matrix roots with iterative methods}

A problem in our approach of nesting $n$ inverse scaling and squaring steps is
that $n\!-\!1$ times we use the output of a square-root operation as input for
the next one.
In this form round-off errors will accumulate and the approximant of the
$2^n$-fold root will deviate from $A^{1/2^n}$ by an error which grows
exponentially with $n$.
It then is natural to look for a post-iteration which renders the result of the
root-cascade exact.
We thus face the question how to compute the fourth (eighth, etc.) root of a
matrix, if a relatively good initial guess is already known.

Having a reasonable guess for $A^{1/p}$, we have, via a simple inversion, also
a guess for  $A^{-1/p}$, and vice versa.
The simplest strategy is to run the (stable) Newton postiteration
\beq
X_{k+1}={p+1\ovr p}X_k-{1\ovr p}X_k^{1+p}A
\;,\qquad X_0\simeq A^{-1/p}
\eeq
with the approximate inverse of the $p$-th root as a starting value, and then
invert the result.
Note that this recursion is expensive; it requires $n+2=\log_2(p)+2$ matrix
multiplications per step.
Still, since in a post-iteration typically just 1 or 2 steps are needed, this
is acceptable.


\subsection{Post-iteration of the matrix logarithm}

Alternatively, one might stay with the uncorrected $n$-fold square root
cascade, and correct, instead, the final logarithm.
The standard approach for this is to use the Newton iteration
\beq
X_{n+1}=X_n-I+{1\ovr2}(e^{-X_n}A+Ae^{-X_n})
\label{log_postiter}
\eeq
where we have opted for the symmetric version.
Again, the individual step is expensive (a new exponential is needed in each
step), but for a post-iteration this is acceptable.


\subsection{Matrix logarithm for unitary arguments}

For unitary argument the principal matrix logarithm is purely anti-hermitean,
$\log(U)=\ri H$ with $H=H\dag$ and $\mr{spec}(H)\in\;]\!-\!\pi,\pi[$.
Based on the experience with (and some of the ingredients from) the general
algorithm (\arabic{alg_log}) it is straight-forward to devise an algorithm
tailored to yield the matrix logarithm of a unitary argument $U$
\cite{middleages}.

At the beginning one has $\cos(H)$ and $\ri\sin(H)$, defined as the hermitean
and anti-hermitean parts of $U$.
One may think of accumulating knowledge about subsequent half-angle sine and
cosine functions, such that in the end one has
$\sin(H/2^n)/\cos(H/2^n)=\tan(H/2^n)$ and by means of an arctan-representation
which is valid for small arguments one gets $H/2^n$ and thus $H$.

In fact, two tricks ease our task.
The first one is the identity
\beq
\tan(H/4)=\sin(H)\Big[1+\cos(H)+\sqrt{2}[1+\cos(H)]^{1/2}\Big]^{-1}
\label{tan_quarter}
\eeq
which allows us to directly jump to the quarter-angle operator, at the expense
of a single Denman-Beavers iteration (\ref{DB_prod}).
The second one is again based on $H/4$ having a spectrum in the open interval
$]\!-\!\pi/4,\pi/4[$; now $Z=\tan(H/4)$ is just in the range of convergence of
\beq
\mr{arctan}(Z)=Z-{1\ovr3}Z^3+{1\ovr5}Z^5-...=
\sum_{k=0}^\infty{(-1)^k\ovr2k+1}Z^{2k+1}
\qquad(||Z||<1)
\;.
\label{def_arctan}
\eeq
As a result, no nested square-root scheme is needed, and the only exit
criterion is concerned with the absolutely convergent series (\ref{def_arctan}).
In practice it proves useful to monitor the numerical convergence of the
series, and to post-iterate with (\ref{log_postiter}) if needed.


\subsection{Unitary projection with iterative methods}

The ability to calculate $A^{-1/2}$ is also useful for the polar decomposition
that is needed in the projection step of the traditional APE or HYP smearing
procedure.
Given a non-unitary $V_\mu(x)$ [the linear combination inside the wavy bracket
in (\ref{def_APE})], one proceeds in two steps.
Usually, one thinks of first projecting to $U(3)$, and then adjusting the
determinant
\beq
W_\mu(x)=V_\mu(x)[V_\mu\dag(x)V_\mu(x)]^{-1/2}
\;,\qquad
U_\mu(x)=W_\mu(x)/\det\nolimits^{1/3}[W_\mu(x)]
\label{proj1}
\eeq
where $W_\mu(x)$ is the unitary part of $V_\mu(x)$, which is unique if
$V_\mu(x)$ is nonsingular.
In practice it is often a better choice to reverse the order, i.e.\ to compute
\beq
W_\mu(x)=V_\mu(x)/\det\nolimits^{1/3}[V_\mu(x)]
\;,\qquad
U_\mu(x)=W_\mu(x)[W_\mu\dag(x)W_\mu(x)]^{-1/2}
\label{proj2}
\eeq
to speed up the iterative polar decomposition, if the latter is done without
rescaling.

One way to compute the unitary part is via the quadratically convergent Newton
iteration
\beq
X_{k+1}={1\ovr2}(\ga_k X_k+\ga_k^{-1}(X_k\dag)^{-1})
\;,\qquad X_0=A
\eeq
with $\lim_{k\to\infty}X_k=A(A\dag A)^{-1/2}$, where
$\ga_k=|\det(X_k)|^{-1/2}$,
$\ga_k=(\si_\mr{min}(X_k)\si_\mr{max}(X_k))^{-1/2}$, or
$\ga_k=(||X_k^{-1}||/||X_k||)^{1/2}$ (with the Frobenius norm) represent
typical choices for the scaling parameter.
Upon first adjusting the determinant to 1, i.e.\ upon using
$A=V_\mu(x)/\det^{1/3}[V_\mu(x)]$, the choice $\ga_k=1\;(\forall k)$ becomes
quite efficient.

Another option is to use the product form of the Denman-Beavers iteration
(\ref{DB_prod}) for $A^{-1/2}$ with
$A=V_\mu\dag(x)V_\mu(x)/\det^{1/3}[V_\mu\dag(x)V_\mu(x)]$ and leftmultiply the
result with $V_\mu(x)/\det^{1/3}(V_\mu(x))$.
Again, due to $\det(A)\!=\!1$, this is efficient without further rescaling.

Finally, it is worth pointing out that, for arbitrary given $V$, the matrix
$P_{U(N)}=V(V\dag V)^{-1/2}$ maximizes $\mr{Re}\,\trc(V\dag U)$ over all
unitary $U$, but the $SU(N)$ matrix $P_{\det=1}P_{U(N)}V=P_{U(N)}P_{\det=1}V$
does not maximize $\mr{Re}\,\trc(V\dag U)$ over all special unitary matrices.
This is of interest in the context of a direct overrelaxation in $SU(N)$, as
discussed in \cite{deForcrand:2005xr}.


\subsection{Eigenvalue based logarithm of a unitary matrix}

For an arbitrary $3\times3$ matrix $A$, the Vieta theorem for the
characteristic polynomial yields
\beq
\la_1+\la_2+\la_3=\mr{Tr}(A)\;,\quad
\la_1\la_2+\la_1\la_3+\la_2\la_3=\det(A)\mr{Tr}(A^{-1})\;,\quad
\la_1\la_2\la_3=\det(A)
\eeq
and for a unitary matrix this simplifies to
\beq
\la_1+\la_2+\la_3=\mr{Tr}(U)\;,\quad
\la_1\la_2+\la_1\la_3+\la_2\la_3=\mr{Tr}(U)^*\;,\quad
\la_1\la_2\la_3=\det(U)
\eeq
where the star denotes complex conjugation.

As a result, the eigenvalue based method for computing the matrix logarithm of
a unitary $3\times3$ matrix $U$ starts with the coefficients of the
characteristic polynomial $x^3+ax^2+bx+c=0$, that is with $a=-\mr{Tr}(U)$,
$b=\mr{Tr}(U)^*=-a^*$, $c=-\det(U)$.
Next, form the complex quantities $Q=a^2/9-b/3$ and
$R=a^3/27-ab/6+c/2=aQ/3-ab/18+c/2$.
If $Q$ and $R$ are both real and $R^2<Q^3$, then the cubic equation has three
real roots.
In the $SU(3)$ case, this means either $\la_1=\la_2=\la_3=1$, or one eigenvalue
$+1$ and two $-1$.
Otherwise, form $A=-[R+\sqrt{R^2-Q^3}]^{1/3}$ [with the complex square-root
chosen such that $\mr{Re}(R^*\sqrt{R^2-Q^3})>0$] and $B=Q/A$.
With these at hand, the solution (Cardano, 1545)
\beq
x_1=(A+B)-{a\ovr3}\;,\;
x_2=-{1\ovr2}(A+B)-{a\ovr3}+\ri{\sqrt{3}\ovr2}(A-B)\;,\;
x_3=-{1\ovr2}(A+B)-{a\ovr3}-\ri{\sqrt{3}\ovr2}(A-B)
\eeq
is written in such a way as to minimize round-off errors \cite{PTVF}.

Having the eigenvalues, the pertinent eigenvectors may be found through solving
a linear system; the eigenvector $v^{(i)}$ ($i=1..3$) is simply the solution of
$(U-\la_iI)v^{(i)}=0$.
Building on the fact that $S=U-\la_iI$ is singular, a convenient choice for the
components of $v^{(i)}$ is the minors of, e.g., the first row:
$v^{(i)}_1=S_{22}S_{33}-S_{23}S_{32}$,
$v^{(i)}_2=S_{23}S_{31}-S_{21}S_{33}$,
$v^{(i)}_3=S_{21}S_{32}-S_{22}S_{31}$.
Here, it is assumed that the second and third row are linearly independent
[but, of course, any row is in the span of the other two].
Whenever the maximum degeneracy is two-fold, there is at least one such choice
which yields a non-zero eigenvector.

With the normalized eigenvectors at hand, the principal logarithm of $U$ would
be $\log(U)=\sum_{i=1}^3\log(\la_i)v^{(i)}v^{(i)\,\dagger}$.
The virtue of the eigenvalue based approach is that it gives us the means to
identify the ``troublesome'' $\la_i$ whenever the principal logarithm has a
non-zero trace.
Assume the trace is $+2\pi\ri$; in that case
$\log(\la_1)+\log(\la_2)+\log(\la_3)=2\pi\ri$.
Identify the $\log(\la_i)$ with the largest imaginary part [they all lie on the
imaginary axis].
Set $\mu_i=\log(\la_i)-2\pi\ri$ for that $i$ and $\mu_j=\log(\la_j)$ for the
other two.
The trace-free logarithm is then
$\mr{tf\/log}(U)=\sum_{i=1}^3\mu_i v^{(i)}v^{(i)\,\dagger}$, with an obvious
generalization to the case where the trace of the principal logarithm is
$-2\pi\ri$.
In mathematical terms this procedure is equivalent to shifting the cut of the
logarithm such that one of the eigenvalues lies on the second Riemann sheet.

Finally, note that this procedure is in marked contrast to the naive approach
of subtracting one third of the trace from the principal logarithm, as is
done in (\ref{def_2}, \ref{def_3}).
This naive recipe amounts to shifting each $\log(\la_i)$ by $\pm2\pi\ri/3$.


\subsection{Appendix summary}

While faster algorithms exist to compute the matrix exponential, the principal
logarithm and the projection to $SU(3)$, the iterative methods described in
this appendix yield numerically stable results with 64-bit machine precision
after $O(<10)$ steps.
Moreover, the possibility to warrant exact HMC reversibility, even if one opts
for a lower precision, seems attractive.
For the non-principal logarithm used in (\ref{def_4}), the eigenvalue based
method seems most convenient.


\section{Tricks for EO-preconditioned BCG$\gaf$ solver}


In (full or quenched) QCD one solves the Dirac equation $Dx=b$ for a given
right-hand side $b$.
A key feature of the clover $D$ is that it couples nearest neighbors,
apart from a mass and a clover contribution which do not hop at all.
Accordingly, upon labeling the sites in a checkerboard (``even-odd'') fashion,
the problem takes a block-offdiagonal form, except for the generalized mass
contribution which remains site-diagonal.
Hence $D$ can be block-$LU$-factorized \cite{DeGrand:1988vx}
\beq
D=
\Big(\begin{array}{cc}D_\mr{ee}&D_\mr{eo}\\D_\mr{oe}&D_\mr{oo}\end{array}\Big)=
L\tilde{D}U=
\Big(\begin{array}{cc}1&0\\D_\mr{oe}^{}D_\mr{ee}^{-1}&1\end{array}\Big)
\Big(\begin{array}{cc}\star&0\\0&\star\end{array}\Big)
\Big(\begin{array}{cc}1&D_\mr{ee}^{-1}D_\mr{eo}^{}\\0&1\end{array}\Big)
\eeq
where the Schur complements $L$ and $U$ are easily inverted (by flipping the
sign of the block-offdiagonal piece) and the block-diagonalized $\tilde{D}$
takes the form
\beq
\tilde{D}=
\Big(\begin{array}{cc}1&0\\-D_\mr{oe}^{}D_\mr{ee}^{-1}&1\end{array}\Big)
\Big(\begin{array}{cc}D_\mr{ee}&D_\mr{eo}\\D_\mr{oe}&D_\mr{oo}\end{array}\Big)
\Big(\begin{array}{cc}1&-D_\mr{ee}^{-1}D_\mr{eo}^{}\\0&1\end{array}\Big)=
\Big(
\begin{array}{cc}
D_\mr{ee}^{}&0\\
0&D_\mr{oo}^{}\!-\!D_\mr{oe}^{}D_\mr{ee}^{-1}D_\mr{eo}^{}
\end{array}
\Big)
\;.
\eeq
As a result, the original problem is split into three parts:
$(i)$ left-multiply the source with $L^{-1}$, that is define the new source $c$
with $c_\mr{e}=b_\mr{e}$ and
$c_\mr{o}=b_\mr{o}-D_\mr{oe}^{}D_\mr{ee}^{-1}b_\mr{e}$;
$(ii)$ solve the non-trivial block problem $\tilde{D}_\mr{oo}y_\mr{o}=c_\mr{o}$
and the trivial one $y_\mr{e}=\tilde{D}_\mr{ee}^{-1}c_\mr{e}$;
$(iii)$ left-multiply the solution with $U^{-1}$, that is obtain $x$ from $y$
through
$x_\mr{o}=y_\mr{o}$ and $x_\mr{e}=y_\mr{e}-D_\mr{ee}^{-1}D_\mr{eo}y_\mr{o}$.

It is worth pointing out two peculiarities of the clover case.
First, contrary to the Wilson case, where $D_\mr{oo}$ and $D_\mr{ee}$ are
constants, they now happen to be site diagonal, that is they are composed of
$V$ little $12\times12$ matrices [for $SU(3)$ gauge group, with $V$ the number
of sites].
In the chiral representation, the latter are further reduced to two
$6\times6$ matrices.
Still, this has a severe impact on the memory requirement.
In order to have a fast forward-application routine, it is customary to
allocate an array which contains all $6\times6$ matrices needed.
This amounts to a vector of $72V$ complex entries, to be compared to the
$36V$ complex entries of the gauge field and the $6V$ complex entries of a
half-vector.

Second, the non-trivial problem in step $(ii)$ may be traded
for the more symmetric version
$(1-D_\mr{oo}^{-1}D_\mr{oe}^{}D_\mr{ee}^{-1}D_\mr{eo}^{})y_\mr{o}=
D_\mr{oo}^{-1}c_\mr{o}$ with a slightly lowered condition number
\cite{Jansen:1996yt,Chiarappa:2006hz}.
On the other hand, the reduced operator
$D_\mr{red}={1\ovr2}(D_\mr{oo}-D_\mr{oe}^{}D_\mr{ee}^{-1}D_\mr{eo}^{})$ is
$\gaf$-hermitean; thus the original version can be mapped into
$H_\mr{red}y_\mr{o}={1\ovr2}\gaf c_\mr{o}$ with the hermitean,
indefinite $H_\mr{red}=\gaf D_\mr{red}$.
In addition, it implies that the operator
$D_\mr{sym}=2(1-D_\mr{oo}^{-1}D_\mr{oe}^{}D_\mr{ee}^{-1}D_\mr{eo}^{})$
enjoys a $\gaf D_\mr{oo}^{}$-hermiticity, that is
$D_\mr{sym}\dag=
2(1-\gaf D_\mr{oe}^{}D_\mr{ee}^{-1}D_\mr{eo}^{}D_\mr{oo}^{-1}\gaf)=
\gaf D_\mr{oo}^{} D_\mr{sym} D_\mr{oo}^{-1} \gaf$.
Similarly, the other symmetric operator, that follows from $D_\mr{sym}$ by
eo-flipping the suffixes, is $\gaf D_\mr{ee}^{}$-hermitean (and has an
identical spectrum).
In all these considerations it is assumed that the little $6\times6$
matrices are inverted beforehand, so that $D_\mr{ee}^{-1}$ (and, if needed,
$D_\mr{oo}^{-1}$) are known.

The operator $D_\mr{red}$ being $\gaf$-hermitean, it is tempting to use the
BCG$\gaf$-algorithm \cite{deForcrand:1995bs} to solve the reduced system in
step $(ii)$ above.
As is well known, with finite-precision arithmetics BCG$\gaf$ is prone to
suffer from instabilities, and this holds true after EO-preconditioning, too.
The instability is mainly due to the indefinite scalar products [with a $\gaf$
between the two vectors] genuine to this algorithm.
Such an object may happen to be quite small in absolute magnitude, while the
individual terms in the sum are not.
In practice it is, nonetheless, observed that the algorithm performs quite
convincingly, if the following three ``tricks'' are used:
\begin{enumerate}
\item
perform the summation in the $(.,\gaf.)$-type scalar products in quadruple
precision [the products to be summed over are still computed in standard double
precision]
\item
recompute the true residue much more frequently than one would do in an
algorithm without indefinite scalar products, e.g.\ every 10-th step instead
of every 100-th step
\item
keep track of the vector which gave, so far, the smallest true residue, and
restart, if certain conditions are met, from this vector.
\end{enumerate}
Evidently, devising a good set of conditions is important.
In some of the simulations presented in this article, BCG$\gaf$ was restarted
if the norm of the actual residue was three orders of magnitude above the best
residue encountered so far and, at the same time, the last restart would date
back at least 500 steps.
It turned out that a better condition is to demand that the minimum residue
would persist for 500 steps and the last restart would date back at least 1500
steps.
It is advisable to code the routine such that -- in case the required residual
norm would not be reached within the maximum number of steps -- it would return
the best approximation encountered and another algorithm (e.g.\ CGNE) would
take over.
In the simulations presented in this article this has never happened.


\section{Tensor calculus and chain rule for matrix functions}


In this appendix a convenient notation is introduced whereupon the chain rule
for matrix valued functions looks like the usual chain rule for scalar
functions.
The material is mostly taken from \cite{Pollock}, but restricted to the case of
square matrices, since this is all that we need.

Considering a function $Y(X)$ with $X$ and $Y$ both $N\!\times\!N$ matrices,
the derivative $\pad Y/\pad X$ is conveniently represented as a
$N^2\!\times\!N^2$ matrix, since it is supposed to contain information on how
each element of $Y$ depends on each element of $X$.
In the mathematical literature three conventions regarding the ordering of its
entries may be found, but only one of them allows for a simple-looking
generalization of the chain rule for scalar functions.

For definiteness, let us recall some standard mathematical notation.
A matrix $A$ is repre\-sented as $A=(a_{ij}e_i^j)$ with the usual summation
convention for repeated indices within a bracket.
Here, $e_i^j$ is an $N\!\times\!N$ matrix with zeros everywhere except for the
place in row $i$ and column $j$.
The Kronecker product of $A$ and $B=(b_{kl}e_k^l)$ is then a $N^2\!\times\!N^2$
matrix
\beq
A \otimes B = (\{a_{ij}b_{kl}\} e_{ik}^{jl})
\label{kronecker}
\eeq
which is constructed by blowing up $A$ and plugging $B$ into every partition
(multiplied with the element of the former $A$).
Accordingly, the product $a_{ij}b_{kl}$ is found in position $(p,q)$ of the
new matrix with $p=(i\!-\!1)N+k$ and $q=(j\!-\!1)N+l$.
In fact, this is the reason for the notation used in (\ref{kronecker}), where
only information on the ordering of covariant indices among themselves (and
ditto for contravariant indices) is retained, but no information on the
relative ordering of covariant and contravariant indices.
One is invited to read $(ik)$ as the new row index and $(jl)$ as the new column
index, with the breakdown into row and column number as given above.
When traveling through the new matrix, $i$ and $j$ move slowly, while $k$ and
$l$ move rapidly.
Upon tensoring (\ref{kronecker}) with a matrix $C=(c_{mn}e_m^n)$ the old
layout is again blown up and we arrive at
\beq
A \otimes B \otimes C = (\{a_{ij}b_{kl}c_{mn}\} e_{ikm}^{jln})
\eeq
with the row and column multiindices $(ikm)$ and $(jln)$ translating into
$p=((i\!-\!1)N+k\!-\!1)N+m$ and $q=((j\!-\!1)N+l\!-\!1)N+n$, respectively.
The important point is that these Kronecker products form an associative
algebra, that is
\beq
(A \otimes B \otimes C) (D \otimes E \otimes F) =
(AD \otimes BE \otimes CF)
\;.
\eeq

In the first definition the derivative of $Y$ with respect to $X$ is a
partitioned matrix $[\pad Y/\pad x_{ij}]$ whose $(i,j)$ partition is put
together by taking the derivatives $\pad y_{kl}/\pad x_{ij}$ for all entries of
$Y$.
In other words, the layout is similar to that of $X\!\otimes\!Y$, that is
\beq
\Bigg[{\pad Y\ovr\pad x_{ij}}\Bigg]=
\Bigg({\pad y_{kl}\ovr \pad x_{ij}} e_{ik}^{jl}\Bigg)
\;.
\eeq

In the second definition the derivative of $Y$ with respect to $X$ is a
partitioned matrix $[\pad y_{kl}/\pad X]$ whose $(k,l)$ partition is put
together by taking the derivatives $\pad y_{kl}/\pad x_{ij}$ for all entries of
$X$.
In other words, the layout is similar to that of $Y\!\otimes\!X$, that is
\beq
\Bigg[{\pad y_{kl}\ovr\pad X}\Bigg]=
\Bigg({\pad y_{kl}\ovr \pad x_{ij}} e_{ki}^{lj}\Bigg)
\;.
\eeq

In the third definition the derivative of $Y$ with respect to $X$ is obtained
by first reshaping both matrices into $N^2\!\times\!1$ column vectors, denoted
$X^\mr{c}$ and $Y^\mr{c}$, respectively.
In this step the second column is appended to the first one, and so on.
Then the derivative of the $p$-th element of $X^\mr{c}$ with respect to the
$q$-th element of $Y^\mr{c}$ is stored in position $(p,q)$.
In other words, the layout is similar to that of
$Y^\mr{c}\!\otimes\!(X^\mr{c})'=(\{x_{ij}y_{kl}\}e_{lk}^{ji})$, with $'$
denoting the transposition, that is
\beq
\Bigg[{\pad Y^\mr{c}\ovr\pad X^\mr{c}}\Bigg]=
\Bigg({\pad y_{kl}\ovr \pad x_{ij}} e_{lk}^{ji}\Bigg)
\;.
\label{right}
\eeq

When comparing these definitions one notices that the first two differ only by
the order among the covariant indices and among the contravariant indices.
Therefore, the two can be mapped into each other by means of transpositions.
By contrast, the third definition differs from the previous two in a more
profound manner; for this change covariant indices need to be converted into
contravariant indices and vice versa.
Another notation for this third definition, which sometimes occurs in the
literature, is $\pad\mr{vec}Y/\pad\mr{vec}'X=\pad\mr{vec}Y/\pad(\mr{vec}X)'$.

We start with having a look into the product rule with this third definition.
Let $Y(X)$ and $Z(X)$ be two matrix valued functions which depend on $X$.
According to the standard rules, the derivative of the product
$W=YZ=(w_{kn}e_k^n)$ with respect to $X$ is
\beq
\Bigg[{\pad W^\mr{c}\ovr\pad X^\mr{c}}\Bigg]=
\Bigg({\pad w_{kn}\ovr \pad x_{ij}} e_{nk}^{ji}\Bigg)=
\Bigg({\pad y_{kl}\ovr \pad x_{ij}}z_{ln} e_{nk}^{ji}\Bigg)+
\Bigg(y_{km}{\pad z_{mn}\ovr \pad x_{ij}} e_{nk}^{ji}\Bigg)
\eeq
where $w_{kn}=(y_{ko}z_{on})$ has been used.
On the other hand, the derivative of $Y$ with respect to $X$ is the
$N^2\!\times\!N^2$ matrix (\ref{right}), which cannot be multiplied with $Z$,
unless the latter is tensored with the identity $I$ to have the right
dimension.
However, standard algebra gives
\bea
(Z\otimes I)
\Bigg[{\pad Y^\mr{c}\ovr\pad X^\mr{c}}\Bigg]\!\!&\!\!=\!\!&\!\!
\Bigg(z_{mn}\de_{pq}e_{mp}^{nq}\Bigg)
\de_{nl}\de_{qk}
\Bigg({\pad y_{kl}\ovr \pad x_{ij}} e_{lk}^{ji}\Bigg)=
\Bigg(z_{ml}\de_{pk}e_{mp}^{lk}\Bigg)
\Bigg({\pad y_{kl}\ovr \pad x_{ij}} e_{lk}^{ji}\Bigg)=
\Bigg({\pad y_{kl}\ovr \pad x_{ij}} z_{ml} e_{mk}^{ji}\Bigg)
\nonumber
\\
(I\otimes Y)
\Bigg[{\pad Z^\mr{c}\ovr\pad X^\mr{c}}\Bigg]\!\!&\!\!=\!\!&\!\!
\Bigg(\de_{pq}y_{kl}e_{pk}^{ql}\Bigg)
\de_{qn}\de_{lm}
\Bigg({\pad z_{mn}\ovr \pad x_{ij}} e_{nm}^{ji}\Bigg)=
\Bigg(\de_{pn}y_{km}e_{pk}^{nm}\Bigg)
\Bigg({\pad z_{mn}\ovr \pad x_{ij}} e_{nm}^{ji}\Bigg)=
\Bigg(y_{km} {\pad z_{mn}\ovr \pad x_{ij}} e_{nk}^{ji}\Bigg)
\nonumber
\eea
and therefore the product rule for matrix derivatives is seen to take the form
\beq
\Bigg[{\pad (YZ)^\mr{c}\ovr\pad X^\mr{c}}\Bigg]=
(Z'\otimes I)\Bigg[{\pad Y^\mr{c}\ovr\pad X^\mr{c}}\Bigg]+
(I \otimes Y)\Bigg[{\pad Z^\mr{c}\ovr\pad X^\mr{c}}\Bigg]
\;.
\label{product}
\eeq
In fact, upon generalizing the size of $X$ to $M\!\times\!N$, and that of $Y$
and $Z$ to $P\!\times\!Q$ and $Q\!\times\!R$, respectively, the l.h.s.\ of
(\ref{product}) is an $PR\!\times\!MN$ matrix.
At the same time the first term on the r.h.s.\ is the product of a
$PR\!\times\!PQ$ matrix (with $I=I_P$) and a $PQ\!\times\!MN$ matrix.
And the second term is the product of a $PR\!\times\!QR$ matrix (with $I=I_R$)
and a $QR\!\times\!MN$ matrix.
Hence either term, and the r.h.s.\ in total, is of size $PR\!\times\!MN$, in
perfect agreement with the l.h.s.

The main reason why it pays to choose the layout (\ref{right}) is that this
definition allows for a neat extension of the standard chain rule.
Let $Z=Z(Y)$ be a matrix valued function of $Y$ and $Y=Y(X)$ a matrix
valued function of $X$.
With similar manipulations as above it follows that
\beq
\Bigg[{\pad Z^\mr{c}\ovr\pad X^\mr{c}}\Bigg]=
\Bigg[{\pad Z^\mr{c}\ovr\pad Y^\mr{c}}\Bigg]
\Bigg[{\pad Y^\mr{c}\ovr\pad X^\mr{c}}\Bigg]
\label{chain}
\eeq
where the l.h.s.\ is a $QR\!\times\!MN$ matrix
and the r.h.s.\ is a $QR\!\times\!PQ$ times a $PQ\!\times\!MN$ matrix.
Therefore, also this rule remains valid for arbitrarily shaped source and
target matrices.
Note that, with any of the other definitions of the matrix derivative, one
would be forced to introduce a clumsy ``star product'' which does not obey the
usual rules of matrix multiplication.

It is worth noticing that the similarity to the chain rule for scalar functions
is somehow limited, since the derivative of $\exp(.)$ and $\log(.)$ is not
given by $\exp(.)$ and $(.)^{-1}$, respectively [which would be matrices of
inappropriate size].
Due to the infinite series in the definition of $\exp(X)$, the $(k,l)$ element
of the exponential depends on each $x_{ij}$.
Accordingly, $\pad\exp(X)^\mr{c}/\pad X^\mr{c}$ is a full $N^2\!\times\!N^2$
matrix, and a similar statement holds true for
$\pad\log(X)^\mr{c}/\pad X^\mr{c}$.
Still, with
\beq
{\pad(X^n)^\mr{c}\ovr\pad X^\mr{c}}=
{\pad(X^{n-1} \cdot X)^\mr{c}\ovr\pad X^\mr{c}}=
(X'\!\otimes\!I){\pad(X^{n-1})^\mr{c}\ovr\pad X^\mr{c}}+
I\!\otimes\!X^{n-1}
\eeq
for $I=I_N$ it follows that the derivative of the $n$-th power can be written
in compact form
\beq
{\pad(X^n)^\mr{c}\ovr\pad X^\mr{c}}=
\sum_{\ell=0}^{n-1}(X'\!\otimes\!I)^{n-1-\ell} (I\!\otimes\!X^\ell)=
\sum_{\ell=0}^{n-1}(X')^{n-1-\ell}\!\otimes\!X^\ell
\eeq
and similarly the derivative of the inverse of a matrix $Y=Y(X)$ is given by
\beq
{\pad(Y^{-1})^\mr{c}\ovr\pad X^\mr{c}}=
-(Y^{-1\,\prime}\!\otimes\!Y^{-1}){\pad Y^\mr{c}\ovr\pad X^\mr{c}}
\;.
\label{inverse}
\eeq
Thanks to the exponential being given by a globally convergent power series, it
follows that
\beq
{\pad\exp(X)^\mr{c}\ovr\pad X^\mr{c}}=
\sum_{n=1}^\infty {1\ovr n!}
\sum_{\ell=0}^{n-1}(X')^{n-1-\ell}\!\otimes\!X^\ell
\eeq
and the chain rule (\ref{chain}) says that the inverse of it, if it exists, is
$\pad\log(Y)^\mr{c}/\pad Y^\mr{c}$ at $Y=\exp(X)$.

With the formalism presented in this appendix the HMC force for an arbitrary
fat-link action can be worked out in a way which relies only on the standard
matrix multiplication law and
may, as a result, benefit from optimized linear algebra subroutines.

\clearpage


\end{document}